\documentclass[a4paper,11pt]{article}
\usepackage{jheppub} 
\usepackage{graphicx,dcolumn,bm,epsfig,cancel,hyperref}
\usepackage{multirow}
\usepackage{amsmath}
\usepackage{xcolor}
\usepackage{amssymb, times}

\newcommand{\comment}[1]{{}}

\usepackage{color}
\title{The Benefits of Diligence: How Precise are Predicted  Gravitational Wave Spectra in Models with Phase Transitions?}

\author[a]{Huai-Ke Guo}
\author[a]{Kuver Sinha}
\author[a]{Daniel Vagie}
\author[b]{Graham White}
\affiliation[a]{Department of Physics and Astronomy, University of Oklahoma, Norman, OK 73019, USA}
\affiliation[b]{Kavli IPMU (WPI), UTIAS, The University of Tokyo, Kashiwa, Chiba 277-8583, Japan}
\emailAdd{ghk@ou.edu}
\emailAdd{kuver.sinha@ou.edu}
\emailAdd{Daniel.d.vagie-1@ou.edu}
\emailAdd{graham.white@ipmu.jp}
\begin{document}


\abstract
{Models of particle physics that feature phase transitions typically provide predictions for  stochastic gravitational wave signals at future detectors and such predictions are used to delineate portions of the model parameter space that can be constrained. The question is: how precise are such predictions? Uncertainties enter  in the calculation of the macroscopic thermal parameters and the dynamics of the phase transition itself. We calculate such uncertainties with increasing levels of sophistication in treating the phase transition dynamics. Currently, the highest level of diligence corresponds to careful treatments of the source lifetime; mean bubble separation; going beyond the bag model approximation in solving the hydrodynamics equations and explicitly calculating  the fraction of energy in the fluid from these equations rather than using a fit; and including fits for the energy lost to vorticity modes and reheating effects. The lowest level of diligence incorporates none of these effects. We compute the percolation and nucleation temperatures, the mean bubble separation, the fluid velocity, and ultimately the gravitational wave spectrum corresponding to the level of highest diligence for three explicit examples: SMEFT, a dark sector Higgs model, and the real singlet-extended Standard Model (xSM). In each model, we contrast  different levels of diligence in the calculation and find that the difference in the final predicted signal can be several orders of magnitude. Our results indicate that calculating the gravitational wave spectrum for particle physics models and deducing precise constraints on the parameter space of such models  continues to remain very much a work in progress and warrants care.
}


\maketitle

\section{Introduction}
Even the very early Universe is transparent to gravitational waves, making searches for the gravitational wave background of the Universe a unique probe of the cosmos before big bang nucleosynthesis. Ubiquitous in the literature is the generation of a gravitational wave background from an inhomogeneous transition of the ground state (for a review see \cite{Weir:2017wfa,Mazumdar:2018dfl,Caprini:2019egz}). In the standard model of particle physics, there is no mechanism for such a gravitational wave background to be produced. Specifically, both the QCD and electroweak transitions are predicted to be smooth \cite{Kajantie:1995kf,Kajantie:1996mn,Kajantie:1996qd,Csikor:1998eu,Bazavov:2011nk,Gupta:2011wh}. This implies that any gravitational wave background resulting from a strong first order phase transition is proof that the standard model is incomplete. \par 
The electroweak transition can be made strongly first order through the introduction of new states at around the electroweak scale
\cite{Grojean:2004xa,Grojean:2006bp,Delaunay:2007wb,Cline:2009sn,Carena:2011jy,Gil:2012ya,Carena:2012np,Fairbairn:2013uta,Profumo:2014opa,Kozaczuk:2014kva,Vaskonen:2016yiu,Baldes:2016rqn,Dorsch:2016nrg,Chiang:2017nmu,Cao:2017oez,vonHarling:2017yew,Bian:2017wfv,Chala:2018ari,Alves:2018jsw,Angelescu:2018dkk,Beniwal:2018hyi,Bruggisser:2018mrt,Athron:2019teq,Kainulainen:2019kyp,Ellis:2019flb,Ellis:2019tjf,Papaefstathiou:2020iag,Postma:2020toi,Alves:2020bpi,Baum:2020vfl,Paul:2020wbz,Zhou:2020idp,Zhang:2021alu,Cline:2021iff}. The QCD transition can be catalyzed by changing the number of light fermions \cite{Davoudiasl:2019ugw} or having a very large lepton asymmetry \cite{Schwarz:2009ii,Caprini:2010xv,Barenboim:2017dfq}. Additionally there are strong motivations to believe that the standard model is incomplete and additions to the standard model can also leave cosmic fingerprints. For instance, baryonic matter can only explain a fraction of the matter observed and the missing dark matter can be a part of a hidden sector that undergoes a phase transition \cite{Schwaller:2015tja,Chao:2017vrq,Croon:2018erz,Archer-Smith:2019gzq,Pandey:2020hoq,Mohamadnejad:2019vzg,Helmboldt:2019pan,Croon:2019iuh,Fairbairn:2019xog,Breitbach:2018ddu,Bhoonah:2020oov,Hall:2019rld,Li:2020eun,Ghosh:2020ipy,Huang:2021rrk,Huang:2020mso,Bigazzi:2020avc,Azatov:2021ifm}. Second, the near unification of gauge coupling constants along with conspiracy of gauge anomaly cancellation motivates grand unification which can sequentially break into the standard model gauge group and leave a gravitational wave background \cite{Chao:2017ilw,Croon:2018kqn,Haba:2019qol,Greljo:2019xan,Brdar:2019fur,Okada:2020vvb,Huang:2020bbe}. Finally, the generation of neutrino masses can arise through a $B-L$ breaking transition \cite{Marzola:2017jzl,Ellis:2019oqb,Haba:2019qol,Bian:2019szo,Okada:2018xdh,Hasegawa:2019amx}. In each case, an observed signal not only sheds light on our cosmic history, but on a range of energy scales spanning from sub-GeV to the PeV scale \cite{Dev:2016feu} (even higher scales have been proposed, though technology needs to improve to make the sensitivity cosmologically relevant \cite{Aggarwal:2020olq} with the possible exception of NEMO \cite{Ackley:2020atn}). \par 
Any strong first order transition produces three contributions to a stochastic background (see, e.g., \cite{Weir:2017wfa,Mazumdar:2018dfl,Caprini:2019egz}). 
For a transition in the thermal plasma, only a negligibly small fraction of the energy released will remain in the
scalar field when the bubble wall reaches a constant velocity~\cite{Hindmarsh:2017gnf}, 
or when it ``runs away''~\cite{Bodeker:2017cim}. So the bubble collision contribution is negligible when considering transitions in a thermal plasma, though it can become dominant
for transitions in vacuum, e.g., those in a dark sector without a thermal bath.
In addition, the contribution from magnetohydrodynamic turbulence is subdominant compared
with that from sound waves~\cite{Hindmarsh:2013xza,Hindmarsh:2016lnk,Hindmarsh:2017gnf}, with an efficiency factor that is roughly $(5\sim 10)\%$ of that for sound waves~\cite{Hindmarsh:2015qta}, though its spectrum is highly uncertain as of now~\cite{Caprini:2015zlo,Kahniashvili:2008pf,Kahniashvili:2008pe,Kahniashvili:2009mf,Caprini:2009yp,Kisslinger:2015hua,Pol:2019yex}.
Since most transitions in the early universe are highly likely to be proceeding in the thermal plasma, we will focus here on the gravitational wave production from sound waves.
 This acoustic contribution has been studied both in simulations and a combination of analytic and numerical techniques and there has been much recent progress. \par 

\begin{figure}
    \centering
    \includegraphics[width=0.9\textwidth]{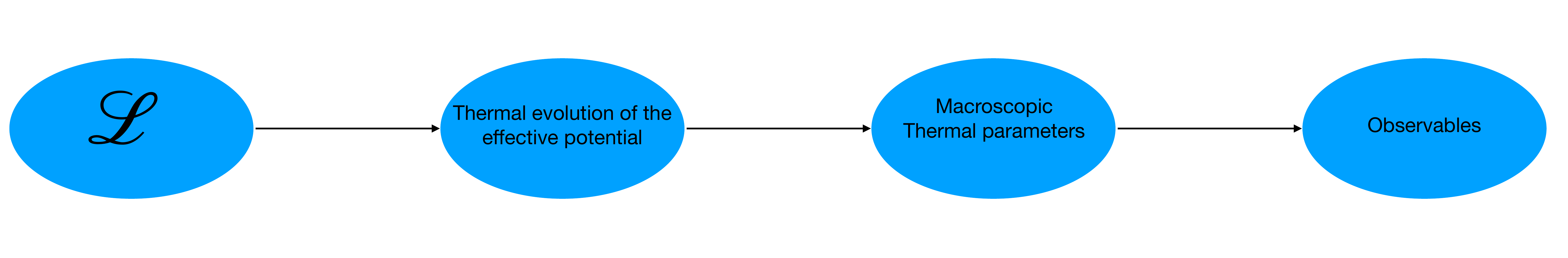}
    \caption{The uncertainty in linking a particular model with a set of observables is conceptually presented above. The break down of perturbation theory at finite temperature is the dominant error in the prediction of the evolution of the effective potential and ultimately non-perturbative methods might be required to predict macroscopic thermal parameters. The macroscopic thermal parameters of interest are often taken to be the latent heat, the time scale of the transition (usually approximated), the bubble wall velocity and the temperature of percolation, but if one desires to have an accurate prediction one needs the fluid velocity, the wall velocity, the mean bubble separation, the percolation temperature and the lifetime of the acoustic source (see also Fig. 1 of \cite{Schicho:2021gca}).}
    \label{fig:uncertainties}
\end{figure}

Given the enormous opportunity to shed light on both cosmology and particle physics, it is worth examining in detail the theoretical underpinnings of any given model in order to enumerate both theoretical uncertainties in basic methods and the degree of benefit in more accurate calculations or, equivalently the cost of various approximations. Approximations can arise in two steps in predicting an observable from a given model as shown in Fig. \ref{fig:uncertainties}. First the calculation of macroscopic thermal parameters, including the latent heat and the time scale of the transition, are often calculated using perturbative techniques which can introduce large errors \cite{Croon:2020cgk} in particular when long wavelength modes are not resummed carefully enough \cite{Kajantie:1995dw,Curtin:2016urg,Schicho:2021gca,Gould:2021dzl}.\footnote{Other important problems in common calculations are gauge dependence \cite{Patel:2011th} and the inhomogeneous background \cite{Garbrecht:2015yza,Ai:2020sru}.} The second step, which we focus on in this paper, converts macroscopic thermal parameters into a prediction for the spectrum - in particular the peak frequency and amplitude. Ultimately, both steps will likely require simulations to truly perform precision cosmology on a future hypothetical observation.\footnote{Infrared divergences of the dynamical mode for instance remain even after NLO resummation. As a result perturbation theory even at two loops disagrees substantially with montecarlo simulations very close to the critical temperature \cite{Niemi:2020hto}.} However, this is impractical for the analysis of large numbers of parameter sets for large numbers of models. We therefore examine several layers of improvement in the prediction of the peak amplitude that have recently arisen in several models involving physics beyond the standard model
\begin{itemize}
    \item The finite lifetime of the source first estimated in Ref. \cite{Ellis:2018mja} and derived in the sound shell model in an expanding background in Ref. \cite{Guo:2020grp}.
    \item Going beyond the bag model approximation in solving the hydrodynamic equations \cite{Giese:2020rtr,Giese:2020znk}.
    \item Calculating the mean bubble separation from the evolution of the bubble number density.
    \item Calculating the fraction of energy in the fluid from solving the hydrodynamic equations rather than using a fit \cite{Cutting:2019zws}.
    \item Including fits for the energy lost to vorticity modes \cite{Cutting:2019zws}.
\end{itemize}

\begin{figure}
    \centering
    \includegraphics[width=0.6\textwidth]{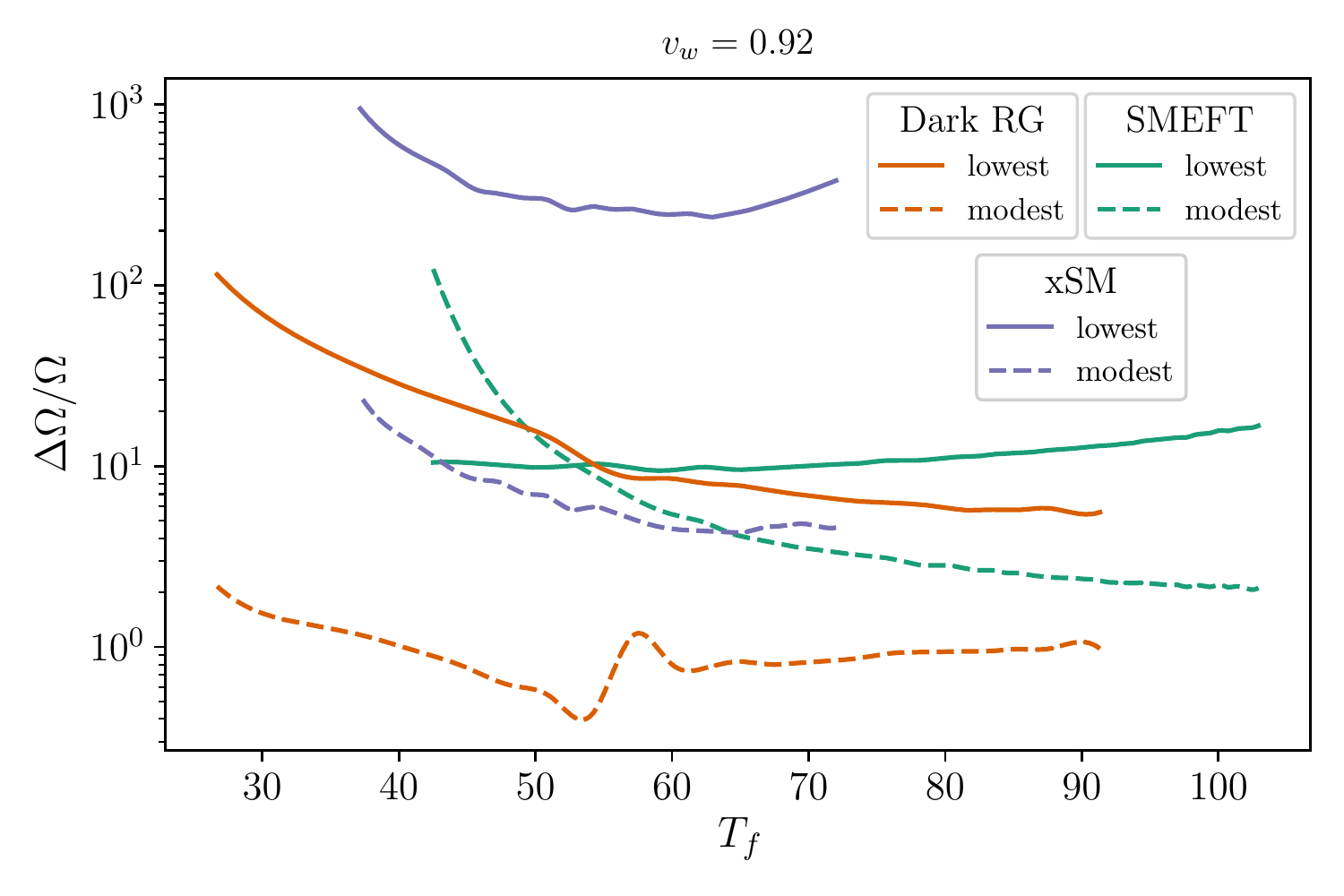}
    \caption{The relative error when using the  lowest and modest levels of diligence, compared to the highest level of diligence (for which $\Delta \Omega/\Omega = 0$). The vertical axis shows the peak (frequency-independent) gravitational wave energy density for detonation. The precise definition of $\Delta \Omega/\Omega$ is given in Eq.~\ref{eq:error}. The horizontal axis corresponds to the final temperature $T_f$ when the phase transition ends.  Three models are shown: SMEFT, a dark sector Higgs model  (Dark RG) and the singlet-extended Standard Model (xSM). The figures employ  calculations from  Eq.~[\ref{eq:omega1}, \ref{eq:omega-suppression-ellis}, \ref{eq:omega_suppressed}] and Eq.~\ref{eq:error}. The temperatures are set to $T_n$ (\ref{eq:Tn}), $T_p$ (\ref{eq:LISA_Tp}), and $T_f$ (\ref{eq:Tf}) for the lowest, modest, and highest diligence respectively. Both the modest and highest diligence contain suppression factors due to the lifetime of the source. The highest diligence contains the suppression factor due to vorticity effects in the plasma.}
    \label{fig:combined-error}
\end{figure}

In this paper we will enumerate the error in a number of models in order to get a broad understanding of the numerical importance of diligence. This avoids model-specific effects where accidental cancellations between different improvements could in principle occur. The models we consider include a toy model introduced for pedagogical purposes, the Standard Model Effective Field Theory (SMEFT), a dark sector Higgs and a real scalar singlet extension (xSM) of the Standard Model. For the benefit of the reader, a demonstration of the importance of diligence is provided at the outset in Fig. \ref{fig:combined-error}. Here, the relative error in the predicted peak amplitude is shown for SMEFT, the dark sector Higgs model (which we label throughout as ``Dark RG") and xSM. Our paper will be devoted to fully explaining Fig. \ref{fig:combined-error}; for now, we provide a feel for the comparative importance of these errors. For the Dark RG, for example, the relative error is far more manageable than what it is for SMEFT.  However, even for that model, the relative error is larger\footnote{up to the caveat that the gauge parameter is varied by an amount allowed by perturbativity considerations.} than the error from gauge dependence that is introduced in SMEFT in some commonly used methods. Thus, even this case, which may present an unrealistically optimistic picture, still motivates diligence in the calculation. 

The structure of this paper is as follows. In Section \ref{sec:PT} we outline three methods of various levels of diligence that we find used in the literature, including a level of diligence motivated by its use in the recent review \cite{Caprini:2019egz}. In Section \ref{sec:testmodels} we define the models we will use to demonstrate increasing levels of diligence, and in Section \ref{sec:results} we will present our results. We will end with our Conclusion.

\section{Phase Transition Dynamics}\label{sec:PT}
Gravitational waves produced from first order phase transitions is a finite temperature tunneling process, from some false vacuum to the true vacuum. When calculating this transition with perturbation theory, one needs to track the minima of the effective potential from the temperature at which the energy in each vacuum is degenerate - that is the critical temperature.  Below the critical temperature, bubbles of the true phase begin to form at some critical radius where the pressure is strong enough to cause expansion. The probability of such bubbles forming increases as the Universe cools, until the nucleation temperature at which there is an average of one bubble per Hubble volume. Slightly below this temperature is the percolation temperature at which bubble collisions are occurring and the final temperature when the phase transition ends. There are simple analytical expressions for these temperature scales which are the result of approximations used in the equations. However, the gravitational wave spectrum is sensitive to the level of diligence that goes into the computations and reducing the error is paramount to probing phase transitions at future gravitational wave detectors. We will now proceed to analyze the different level of diligence used in the literature. 

\subsection{Lowest diligence}
Here we will describe the level of lowest diligence in computing the gravitational wave spectrum. At this stage, we will only introduce the various parameter definitions and wait until the highest diligence section for a more in depth look at the numerical procedure. This level will involve computing all relevant parameters at the nucleation temperature. 

The tunneling rate per unit time per unit volume will have the general form 
\begin{equation}
	p(T) =  \bar{p}_0 T^4 e^{-S_3/T},
	\label{eq:tunneling_rate}
\end{equation}
where $\bar{p}_0$ is a dimensionless number that we will assume is $\mathcal{O}(1)$ and $S_3$  can be found by solving the bounce solutions that minimize the action given by 
\begin{equation}
	S_3 (\vec{\phi},T) = 4\pi \int dr r^2 \left[ \frac{1}{2} \left( \frac{d\vec{\phi}(t)}{dr}\right)^2 + V_{\text{eff}}(\vec{\phi},T)\right] \,\,.
\end{equation}
The nucleation temperature is defined as the temperature at which the probability of a single bubble being nucleated within a Hubble volume is $\mathcal{O}(1)$:
\begin{equation}
	\int_{0}^{t_n} p V_H (t) dt = \int_{T_n}^{\infty} \frac{dT}{T} \left( \frac{2 \xi M_{\text{pl}}}{T}\right)^4 \exp^{-S_3/T} \sim \mathcal{O}(1),
\end{equation}
where $M_{pl}$ is the Planck mass, $\xi \sim 3 \times 10^{-2}$, and $V_H(t)$ is the horizon volume. This equation will lead to the simple definition of the nucleation temperature \cite{Quiros:1999jp,Wainwright:2011kj,Ramsey-Musolf:2019lsf,Caprini:2019egz} 
\begin{equation}
	\frac{S_3(T_n)}{T_n} \approx 140.
	\label{eq:Tn}
\end{equation}
It is important to note that the above calculation assumes that the phase transition occurs in a radiation dominated era which is not guaranteed.

The strength of the gravitational wave spectrum will depend on  hydrodynamic parameters such as the amount of vacuum energy released during the phase transitions, the inverse time duration of the phase transition, and the fraction of latent heat that goes into the bulk motion of the plasma (referred to as the kinetic efficiency coefficient). We discuss each of these quantities in turn.

The strength of the phase transition is characterized as 
\begin{equation}
	\alpha(T_n) = \left. \frac{1}{\rho_{\text{rad}}} \left( \Delta V_{\text{eff}} - \frac{1}{4} \frac{d \Delta V_{\text{eff}}}{dT} \right) \right| _{T_n},
	\label{eq:alpha_Tn}
\end{equation}
where $V_{\text{eff}}$ is the finite temperature effective potential and the symbol $\Delta$ signifies the difference in the symmetric phase (false vacuum) and the broken phase (true vacuum). The energy density of radiation is given by $\rho_{\text{rad}} = \pi^2/30 \:g_* T_n^4$ where $g_*$ is the number of effective degrees of freedom at $T_n$. 

The inverse time duration of the phase transition evaluated at the nucleation temperature can be approximated as 
\begin{equation}
	\beta  = H_n T_n \frac{d (S_3/T)}{dT},
	\label{eq:betaHn}
\end{equation}
 where $H_n^2 = 8 \pi G \rho_{\rm rad}(T_n)/3$ is the Hubble parameter at the nucleation temperature. A smaller $\beta/H$ and larger $\alpha$ will result in stronger gravitational waves.  

The gravitational wave spectrum observed today has a simple broken power law fit \cite{Hindmarsh:2017gnf} in terms of the aforementioned parameters given by 
\begin{equation}
	h^2 \Omega_{\rm GW}(f) = 8.5 \times 10^{-6} \left( \frac{100}{g_n}\right)^{1/3} \left( \frac{\kappa \alpha}{1 + \alpha}\right)^2 \left( \frac{H_n}{\beta}\right) v_{w} S_{\rm SW}(f),
	\label{eq:omega1} 
\end{equation}
where $g_n$ is the number of degrees of freedom at the nucleation temperature and $\kappa$ is the efficiency coefficient that represents the fraction of the bulk kinetic energy in the plasma relative to the available vacuum energy. The numerical fits for the kinetic efficiency coefficient, $\kappa$, were derived in \cite{Espinosa:2010hh} for the different velocity profile types which we give in Appendix \ref{sec:kappa_fits}. The spectral shape, $S_{\rm SW}$, and the peak frequency, $f_{\rm SW}$ are given by 
\begin{eqnarray}
	S_{\rm SW} (f) = \left( \frac{f}{f_{\rm SW}}\right)^3 \left[ \frac{7}{4 + 3(f/f_{\rm SW})^2}\right]^{7/2} ,
\end{eqnarray}
\begin {eqnarray}
	f_{\rm SW} = 1.9 \times 10^{-5} \frac{1}{v_w} \left( \frac{\beta}{H_n}\right) \left( \frac{T_n}{100 \: \rm GeV}\right) \left( \frac{g_n}{100}\right)^{1/6} \rm Hz.
\end{eqnarray}
The gravitational wave spectrum may be rewritten in terms of the R.M.S velocity, $U_f^2 = \frac{3}{4} \kappa \alpha $, with the replacement  
\begin{equation}
	\left( \frac{\kappa \alpha}{ 1 + \alpha}\right) \rightarrow \Gamma U_f^2,
	\label{eq:kappa_to_Uf_bag}
\end{equation}
where $\Gamma \sim 3/4$ is the adiabatic index which is defined as the ratio of the enthalpy and energy density in the symmetric phase.  The term in the denominator on the left hand side, $(1 + \alpha)$, is the result of the energy density in the symmetric phase. 


\subsection{Moderate diligence}
The level of modest diligence is the approach most frequently used in the recent literature (including the recent LISA review \cite{Caprini:2019egz}). It closely resembles the lowest diligence with the exception that the thermal parameters are defined at the percolation temperature rather than the nucleation temperature and the finite lifetime of the source is taken into account with an ansatz correction to the peak amplitude.    
    The percolation temperature is here approximated by solving the equation
\begin{equation}
    \frac{S_3(T_p)}{T_p} = 131 - \log(A/T^4) - 4 \log\left( \frac{T}{100 \: \rm GeV}\right) - 4 \log\left( \frac{\beta(T)/H}{100}\right) + 3 \log(v_w),
    \label{eq:LISA_Tp}
\end{equation}
where $\log(A/T^4) \sim 14$ for an electroweak phase transition. Note that the derivative of the left hand side in Eq.~\ref{eq:LISA_Tp} appears on the right hand side, as can be seen from Eq.~\ref{eq:betaHn}. The percolation temperature is always below the nucleation temperature and hence closer to the final temperature when the phase transition ends. This makes using  the percolation temperature a better approximation to estimate the thermal parameters. However, if the percolation temperature is significantly far away from the nucleation temperature, one should check if the phase transition can even reach completion at all for cases of strong supercooling, since the universe may become vacuum dominated. The strength of the phase transition and the inverse time duration of the phase transition take on the same form as in Eq~\ref{eq:alpha_Tn}-\ref{eq:betaHn} but with the replacement $T_n \rightarrow T_p$ such that 
\begin{equation}
	\alpha(T_p) = \left. \frac{1}{\rho_{\text{rad}}} \left( \Delta V_{\text{eff}} - \frac{1}{4} \frac{d \Delta V_{\text{eff}}}{dT} \right) \right|_{T_p},
	\label{eq:alpha_Tp}
\end{equation}
and
\begin{equation}
	\beta  = H_p T_p \frac{d (S_3/T)}{dT},
	\label{eq:betaHp}
\end{equation}
where $H_p$ is now the value of the Hubble parameter at the percolation temperature.

The gravitational wave spectrum in Eq.~\ref{eq:omega1} assumed that the lifetime of the source is approximately one Hubble time, $H \tau_{sw} = 1$ 
\cite{Hindmarsh:2015qta}
. It was later pointed out in \cite{Ellis:2019oqb}, that a better approximation to the lifetime of the source is 
\begin{eqnarray}
	t_{\rm sw} = {\rm min} \left[ \frac{1}{H_p}, \frac{R_*}{U_f}\right],
	\label{eq:suppression-ellis}
\end{eqnarray}
where $R_*$ is the mean bubble separation and $U_f$ is the root mean squared velocity defined at $\alpha(T_p)$. The mean bubble separation is related to the inverse time duration using $R_\ast = (8\pi )^{1/3} v_w /\beta $.
We then take into account the finite lifetime of the source in the gravitational wave spectrum through    
\begin{equation}
    \Omega _{\rm sw} \to  \Omega _{\rm sw} t_{\rm sw} H,
    \label{eq:omega-suppression-ellis}
\end{equation}
and calculate all temperature dependent quantities at the percolation temperature $T_p$ defined in Eq.~\ref{eq:LISA_Tp}.

\subsection{High diligence}
The highest diligence with which one can calculate the gravitational wave spectrum involves a number of improvements to the predictions of the peak frequency and amplitude: 
\begin{itemize}
    \item[1] Improving on the bag model approximation for the fluid velocity and fraction of energy that is in gravitational waves;
    \item[2] Calculating the fluid velocity and efficiency from solving the hydrodynamic equations rather than using fits (related to the first);
    \item[3] Calculating the mean bubble separation from the number density of the bubbles;
    \item[4] Taking into account the finite lifetime of the soundwave source, derived in an expanding universe \cite{Guo:2020grp}; 
    \item[5] Calculating the suppression due to reheated droplets creating friction that slows collisions.
\end{itemize}
Note that Ref. \cite{Cutting:2018tjt} used the bag model in their simulations, so we assume that the suppression factor arising from kinetic energy lost in the fluid is independent of the change in the amplitude from improving on the bag model. Also in the last case, and only the last case, we use fits to estimate this degree of suppression as it relies on a full numerical simulation - methods to approximate this effect we leave to future work. In this section we outline in detail each of the other improvements.

\subsubsection{Calculation of the percolation temperature}
The false vacuum fraction at $t > t_c$ in an expanding universe is defined as 
\begin{eqnarray}
	g(t_c,t) = \exp \left[ -\frac{4 \pi}{3} \int_{t_c}^{t}  dt^\prime p(t^\prime) a^3(t^\prime) r(t^\prime,t)^3\right] \equiv \exp \left[ - I(t)\right],
	\label{eq:g(T)}
\end{eqnarray}
where $I(t)$ represents the volume of nucleated bubbles per comoving volume, double counting the overlapped space between bubbles and virtual bubbles within others \cite{Guth:1981uk}.  The comoving radius of a bubble nucleated at $t^\prime$ and measured at t is 
\begin{eqnarray}
	r(t^\prime,t) = \int_{t^\prime}^{t} dt^{\prime\prime} \frac{v_w}{a(t^{\prime\prime})} = v_w \left( \eta^\prime - \eta \right),
	\label{eq:r(t)}
\end{eqnarray}
for FLRW space where $\eta$ is the conformal time. This form of the comoving radius of the bubble assumes that the initial size of the bubble is negligible and the bubble wall velocity is constant.   

The above equations can be recast in terms of  integrals over temperature in order to render them more suitable for numerical computations. The scale factor of an expanding universe will have the form $a = c_a t^n$ where $c_a$ is a constant and $n $ is determined by the form of energy that is dominating the expansion, i.e. $n = 1/2$ and $n = 2/3$ for radiation and matter respectively. The expansion of the universe may be treated as an adiabatic process so that the entropy $s_R$ in the radiation sector is conserved per comoving volume:
\begin{eqnarray}
	s_R(T) a^3 = \rm const.
\end{eqnarray}   
The radiation sector will have $s_R \propto T^3$ which will typically give $T \propto 1/a \propto t^{-n}$. This is valid for both a radiation dominated universe and a matter dominated universe. However if non-relativistic matter can decay into radiation, the dependence is $T \propto a^{-3/8}$ \cite{Scherrer:1984fd}. In general, we may assume 
\begin{eqnarray}
	T \propto a^{- \gamma},
	\label{eq:T_to_a}
\end{eqnarray}
where $\gamma$ will depend on the amount of entropy injection.  The Hubble parameter is related to the rate of change of the scale factor, $H = \dot{a}/a$, and we then have 
\begin{eqnarray}
	\frac{1}{H} \frac{dT}{dt} = - \gamma T,
\end{eqnarray}
which allows for rates of change with respect to time to be exchanged with that of temperature. We can also relate the scale factor at the time of the critical temperature, $a_c$, to the scale factor at a later time by 
\begin{eqnarray}
	\frac{a}{a_c} = \left( \frac{T_c}{T}\right)^{1/\gamma}.
	\label{eq:scale_factor_ratio}
\end{eqnarray}  
For an expanding universe in a radiation dominated era, the total energy density is 
\begin{eqnarray}
	\rho_{\rm tot} = \rho_{\rm rad} + \Delta V,
	\label{eq:rho_tot}
\end{eqnarray}
which consists of the relativistic matter density $\rho_{\rm rad}$, and the vacuum energy released during the phase transition. 
Accordingly the Hubble parameter in an expanding universe in this notation yields
\begin{eqnarray}
	H(T)^2 = \frac{8 \pi G}{3} \rho_{\rm tot} \left( \frac{1}{y^4} + \frac{\Delta V}{\rho_{\rm tot}}\right) ,
\end{eqnarray}
with $y = a/a_c$. The above definitions can be used to convert any integral over time to that of temperature so that Eq.~(\ref{eq:g(T)}-\ref{eq:r(t)}) may be written as 
\begin{eqnarray}
	r(T^\prime,T) = \frac{v_w}{a_c} \int_{T}^{T^\prime} \frac{dT^{\prime\prime}}{T^{\prime\prime}} \frac{1}{\gamma H(T^{\prime\prime})} \left(\frac{T_c}{T^{\prime\prime}}\right)^{-1/\gamma} ,
\end{eqnarray}
\begin{eqnarray}
	I(T) = \frac{4 \pi}{3} \int_{T}^{T_c} \frac{dT^\prime}{T^\prime} \frac{1}{\gamma H(T^\prime)} \bar{p}_0 T^{\prime 4} \exp \left[-\frac{S_3(T^\prime)}{T^\prime}\right] \left(\frac{T_c}{T^{\prime}}\right)^{3/\gamma} \left[a_c  r\left(T^\prime,T\right)\right]^3,
\end{eqnarray}
with $g(T_c,T) = \exp \left[ -I(T) \right]$. 

With the false vacuum fraction now defined as a function of temperature, the percolation temperature occurs when the false vacuum fraction is $70\%$ of the total volume, i.e., when
\begin{eqnarray}
	g(T_p) = 0.7. 
	\label{eq:Tp}
\end{eqnarray} 
Similarly, the temperature when the phase transition ends occurs at the time when the volume of nucleated bubbles equals the comoving volume, i.e. $I(t_f) = 1$. This translates into 
\begin{eqnarray}
	g(T_f) = e^{-1}.
	\label{eq:Tf} 
\end{eqnarray} 
In most cases, the percolation temperature calculated from Eq.~\ref{eq:Tp} should roughly coincide with the final temperature calculated from Eq.~\ref{eq:Tf} and depend on $v_w$ and $\Delta V$. 

\subsubsection{Calculation of the nucleation temperature}
The evolution of the mean bubble density per proper volume is determined by
\begin{eqnarray}
	\frac{d \left[ n_b a^3(t) \right]}{dt} = p(t) g(t_c,t) a^3(t),
	\label{eq:mean_bubble_density}
\end{eqnarray}
which begins at $n_b(t_c) = 0$ and includes all the bubbles ever formed. We then transform the above result into an integral over temperature: 
\begin{eqnarray}
	n_b(T) = \left(\frac{T}{T_c}\right)^{3/\gamma} 	\int_T^{T_c} 
	\frac{d T^{\prime}}{T^{\prime}} \frac{1}{\gamma 	H(T^{\prime})}
	\bar{p}_0 T^{\prime 4} \text{exp}\left[- 	\frac{S_3(T^{\prime})}{T^{\prime}}\right] g(T_c, 	T^{\prime}) 
	\left(\frac{T_c}{T^{\prime}}\right)^{3/\gamma} .
\end{eqnarray}
The nucleation temperature can be numerically calculated as the temperature when there is one bubble per Hubble volume:
\begin{eqnarray}
	\frac{n_b(T_n)}{H(T_n)^3} = 1.
\end{eqnarray} 
The mean bubble density depends on the false vacuum fraction and therefore should change with $v_w$, $\kappa_m$, and $\Delta V$ as well.  

\subsubsection{Calculation of the mean bubble separation}
The mean bubble separation $R_*$ is related to the mean bubble number density and is given by 
\begin{eqnarray}
	R_* = \left( \frac{1}{n_b}\right)^{1/3}.
	\label{eq:Rstar_from_nb}
\end{eqnarray}
The gravitational wave spectrum is usually written in terms of the inverse time duration of the phase transition where $R_*$ and $\beta$ are interchangeable. This form of the mean bubble separation can similarly be found in an expanding universe by Taylor expanding Eq.~\ref{eq:tunneling_rate} about $\eta_0$ with respect to conformal time for exponential nucleation
\begin{eqnarray}
	p(\eta) = p_0(\eta_0) \exp \left[ -S_0 + \beta_c(\eta - \eta_0)\right],
\end{eqnarray}
where $\beta_c$ is the comoving inverse time duration of the phase transition:
\begin{eqnarray}
	\beta_c = \frac{d \ln p}{d \eta}\bigg|_{\eta=\eta_0}.
\end{eqnarray}
The above Taylor expansion is valid in the regime where the false vacuum fraction decreases faster than the rate $p$ increases which occurs when $I(t)$ becomes order 1 and most of the bubbles are nucleated. The exponent of the false vacuum can be transformed into an integration over conformal time, $dt = a d \eta$:
\begin{align}
	I(\eta) &= \frac{4 \pi}{3} \int_{\eta_c}^{\eta} d\eta^\prime a^4(\eta^\prime) p(\eta^\prime) r(\eta^\prime,\eta)^3 \nonumber\\
	&= 8 \pi \frac{v_w^3}{\beta_c^4} p_0(\eta_0) e^{-S_0 + \beta_c(\eta - \eta_0)},
\end{align}
which can then be used to calculate the false vacuum fraction for $\eta > \eta_c$. If we define the comoving bubble number density as $n_{b,c} = n_b a^3$, we can rewrite Eq.~\ref{eq:mean_bubble_density} in terms of conformal time:
\begin{eqnarray}
	\frac{d \left( n_{b,c}\right)}{d \eta} = p(\eta) g(\eta_c,\eta) a^4(\eta).
\end{eqnarray} 
We now integrate over $\eta$ to have 
\begin{eqnarray}
	n_{b,c} = \frac{1}{\beta _c} p_0(\eta_0) e^{-S_0 + \beta _c(\eta _f - \eta _0)} = \frac{\beta_c^3(v_w)}{8 \pi v_w^3},
\end{eqnarray}
where we used the relation that $I(\eta_f) = 1$ for $\eta_0 \sim \eta_f$. The relation between the comoving mean bubble density and the mean bubble separation is now 
\begin{eqnarray}
	R_{*,c} = (8 \pi)^{1/3} \frac{v_w}{\beta_c(v_w)},
	\label{eq:beta_Tf}
\end{eqnarray} 
where $\beta_c$ and $R_{*,c}$ are both functions of the bubble wall velocity. The mean bubble separation at conformal time $\eta$ is now related to the inverse time duration when the phase transition ends:
\begin{eqnarray}
	R_*(\eta) = \frac{a(\eta)}{a(\eta_f)} (8 \pi)^{1/3} \frac{v_w}{\beta(v_w)},
	\label{eq:Rstar}
\end{eqnarray} 
where $R_{*,c} = R_*/a(\eta)$ and $\beta_c = a(\eta_f) \beta$. The above result can be transformed into a function of temperature using a  relation similar to Eq.~\ref{eq:scale_factor_ratio} to give 
\begin{eqnarray}
	\frac{a(\eta)}{a(\eta_f)} = \left( \frac{T_f}{T}\right)^{1/\gamma}.
	\label{eq:scale_factor_ratio_Tf}
\end{eqnarray}
The procedure to calculate the inverse time duration of the phase transition at $T_f$ for a fixed $v_w$ is 
\begin{enumerate}
	\item Find $T_f$ from the false vacuum fraction $g(T_c,T)$ for when $I(T_f) = 1$ ;
	\item Calculate $n_b(T_f)$ ;
	\item Use $R_*(T_f) = \left( \frac{1}{n_b(T_f)}\right)^{1/3}$ ;
	\item Solve Eq.~\ref{eq:Rstar} using Eq.~\ref{eq:scale_factor_ratio_Tf} to get $\beta(v_w)$.
\end{enumerate}

\subsubsection{Going beyond the Bag model}
The free energy density $f$ of a model with $g_*$ degrees of freedom consists of a zero temperature scalar potential and a thermal potential that involves the summation over all relativistic species that interact with the scalar $\phi$. In the Standard Model, this involves the standard electroweak Higgs field and degrees of freedom $g_{\rm SM} = 106.75$. The free energy density gives 
\begin{eqnarray}
    f = V_0(\phi) + T^4 \left[ \sum_B J_B \left( \frac{M_B}{T} \right) + \sum_F J_F \left( \frac{M_F}{T} \right)\right],
    \label{eq:free_energy_density_sum}
\end{eqnarray}
where the summations run over all relativistic bosons $B$ and fermions $F$ at temperature $T$. In the high temperature expansion, the free energy density can be written as 
\begin{eqnarray}
    f = - \frac{1}{3} \frac{\pi^2}{30} g_* T^4 + V_T(\phi),
    \label{eq:free_energy_density_highT}
\end{eqnarray}
where $V_T(\phi)$ is the thermal effective Higgs potential and we explicitly separate out the scalar independent terms that go as $T^4$. The hydrodynamics of the plasma can be described as a perfect fluid in terms of the energy density $e$, pressure $p$, and enthalpy $\omega$. These thermodynamic quantities can be explicitly calculated from the free energy density through the relation $p = -f(\phi,T)$ for the pressure. The energy density and enthalpy are then related to the pressure through 
\begin{eqnarray}
    e = T \frac{\partial p}{\partial T} - p, \quad \omega = e + p = T \frac{\partial p}{\partial T}.
\end{eqnarray}
These quantities together, along with the velocity of the fluid $v$, will give the energy momentum tensor of the plasma 
\begin{equation}
    T^{\mu\nu} = u^\mu u^\nu \omega + g^{\mu \nu} p,
\end{equation}
where $g^{\mu \nu}$ is the inverse Minkowski spacetime metric and $u^\mu(v)$ is the four velocity of the fluid.

During the period of gravitational wave production, bubbles of the new phase will collide and generate sound waves in the perturbed plasma. The velocity profile around a single bubble is determined by the hydrodynamic quantities and the junction conditions across the bubble wall.  We will denote the broken phase inside the bubbles by subscripts "$b,-$" and the symmetric phase outside the bubbles as "$s,+$". The signs "$-,+$" are used to describe quantities behind and in front of the bubble wall. The continuity equations $\partial_\mu T^{\mu\nu} = 0$ can be integrated in the wall frame across the bubble wall to give the junction conditions 
\begin{eqnarray}
    \frac{v_+}{v_-} &= \frac{e_b(T_-) + p_s(T_+)}{e_s(T_+) + p_b(T_-)}, \\
    v_+ v_- &= \frac{p_s(T_+) - p_b(T_-)}{e_s(T_+) - e_b(T_-)},
\end{eqnarray}
where $v_\pm$ and $T_\pm$ corresponds to the velocity and temperature of the fluid behind and in front of the bubble wall. We may assume that $T_+ \approx T_- $ and expand the thermodynamic quantities about the symmetric phase so that the ratio of velocities about the junction give 
\begin{equation}
    \frac{v_+}{v_-} \simeq \frac{\left(v_+ v_-/ c_{s,b}^2 -1\right) + 3 \alpha_{\bar{\theta}_+}}{\left(v_+ v_-/ c_{s,b}^2 -1\right) + 3 v_+ v_- \alpha_{\bar{\theta}_+}},
\end{equation}
where $c_{s,b}$ is the speed of sound in the broken phase and $\alpha_{\bar{\theta}_+}$ is the phase transition strength \cite{Giese:2020znk}. The speed of sound can be calculated in both the symmetric and broken phase and is defined in terms of the pressure as   
\begin{eqnarray}
    c_s^2 = \frac{dp/dT}{de/dT} = \frac{p^\prime(T)}{T p^{\prime\prime}(T)},
    \label{eq:cs2}
\end{eqnarray}
where $(...)^\prime$ denotes derivatives with respect to temperature. In terms of the free energy density, the pressure in the symmetric phase is $\phi$ independent with $p_s = -f(0,T)$ and the pressure in the broken phase is $p_b = -f(\phi_{\rm min},T)$ which is evaluated at true vacuum $\phi_{\rm min}$. It is important to keep every term in the free energy density when calculating the speed of sound in order to properly account for the full temperature dependence of the model. This includes keeping all light degrees of freedom that do not acquire field dependent masses that affect the Higgs effective potential. The new phase transition strength parameter $\alpha_{\bar{\theta}_+}$ is dependent on the speed of sound in the broken phase and has a similar form to the bag parameter built up from the trace anomaly, 
\begin{eqnarray}
    \alpha_{\bar{\theta}_+} = \frac{\bar{\theta}_s(T_+) - \bar{\theta}_b(T_+)}{3 \omega_s(T_+)}, \:  \rm{with}  \quad \bar{\theta} = e - p/c_{s}^2
\end{eqnarray}
\cite{Giese:2020znk}.
Going back to the free energy density, we can define the phase transition strength as a function of temperature
\begin{equation}
    \alpha_{\bar{\theta}}(T) = \frac{1}{3 \omega_s} \left( (1 + c_s^{-2}) \Delta V_{T} - T \frac{d \Delta V_{T}}{dT}\right),
    \label{eq:alpha_psuedotrace}
\end{equation}
where $V_T$ is the Higgs effective potential defined in Eq.~\ref{eq:free_energy_density_sum}-\ref{eq:free_energy_density_highT} and $\Delta$ denotes the difference between the symmetric and broken phase. The phase transition strength has the same form as the bag model $\alpha_{\theta}$ defined in Eq.~\ref{eq:alpha_Tn} when $c_s^2 = 1/3$ and $\omega_s = 4/3 \rho_{\rm rad}$. The junction conditions to obtain $v_{\pm}$ will serve as boundary conditions for solving the velocity profile obtained from projecting the continuity equations into the parallel and perpendicular motions of the fluid flow. The hydrodynamic equations become
\begin{eqnarray}
    (\xi - v) \partial_{\xi} e &=& \omega \left[ 2 \frac{v}{\xi} + \gamma^2 ( 1 - \xi v) \partial_{\xi} v\right], \\ 
    (1 - v \xi) \partial_\xi p &=& \omega \gamma^2 (\xi - v) \hfill \partial_\xi v,
\end{eqnarray}
where $\xi = R/t$ is a self similar coordinate defined as the ratio between the distance to the bubble center and the time since nucleation. These hydrodynamic equations may be combined to give the enthalpy profile and the velocity profile: 
\begin{eqnarray}
    2 \frac{v}{\xi} &= \gamma^2(1 - v \xi) \left[\frac{\mu^2}{c_s^2} - 1 \right] \partial_\xi v, \\
    \partial_\xi \omega &= \omega \left( 1 + \frac{1}{c_s^2}\right) \gamma^2 \mu(\xi,v) \partial_\xi v,
\end{eqnarray}
where $\partial_\xi = \partial_\xi e + \partial_\xi p$ and $c_s^2 = dp/de$ are used to connect the equations. The Lorentz boost transformation used in the equations is defined in terms of the self similar coordinate $\mu(\xi,v) = \left(\xi - v \right)/\left(1 - \xi v \right)$. In detonations, the fluid velocity ahead of the bubble wall is always zero so that the hydrodynamic profiles are independent of the speed of sound in the symmetric phase. Deflagrations have a non-zero bubble wall velocity ahead of the bubble wall and the equations will then depend on the speed of sound in the symmetric phase. Both profile types will always depend on the speed of sound in the broken phase through the junction conditions. In detonation, it is sufficient to use $\alpha_{\bar{\theta}+} = \alpha_{\bar{\theta}}(T)$ with $T$ usually taken as the nucleation temperature or the percolation temperature. Deflagrations and hybrid modes take $\alpha_{\bar{\theta}}$ as input and require a shooting method by varying $\alpha_{\bar{\theta}}$ until $\alpha_{\bar{\theta}}$ is reached far away from the bubble. 

The quantity of interest for the peak gravitational wave energy density is the kinetic energy fraction $K$ which can be solved by the hydrodynamic equations:
\begin{equation}
    K = \frac{\rho_{fl}}{e_s},
\end{equation}
where $\rho_{\rm fl} = 3/v_w^3 \int d \xi \xi^2 v^2 \gamma^2 \omega$ is the fluid's kinetic energy. We use the publicly available code in \cite{Giese:2020znk} to numerically compute the kinetic energy efficiency $\kappa_{\bar{\theta}}$ for a given set of $c_s^2$ and $\alpha_{\bar{\theta}}$ when comparing to calculations in the bag model.  The kinetic energy fraction is related to the efficiency parameter through  
\begin{equation}
    K = \left( \frac{\bar{\theta}_s - \bar{\theta}_b}{4 e_s} \right) \kappa\left(\alpha_{\bar{\theta}},c_{s,s}^2,c_{s,b}^2\right),
    \label{eq:K_beyond_bag}
\end{equation}
where 
\begin{eqnarray}
    \kappa = \frac{4 \rho_{\rm fl}}{3 \alpha_{\bar{\theta}} \omega_s}
\end{eqnarray}
\cite{Giese:2020znk}.
The quantities $c_{s,s}$, $c_{s,b}$, $e_s$,$\omega_s$, $\alpha_{\bar{\theta}}$, and $\kappa$ in determining the kinetic energy fraction $K$ are all calculated at the final temperature $T_f$ when the phase transition ends.
The enthalpy-weight root-mean-square fluid velocity around a single bubble may be found from the kinetic energy fraction,
\begin{eqnarray}
    \bar{U}_f^2 = \frac{e_s}{\omega_s} K,
    \label{eq:Uf_to_K_beyond_bag}
\end{eqnarray}
which is evaluated at $T_f$.

\subsubsection{Gravitational Wave Spectrum}
\begin{figure}
    \centering
    \includegraphics[width=0.6\textwidth]{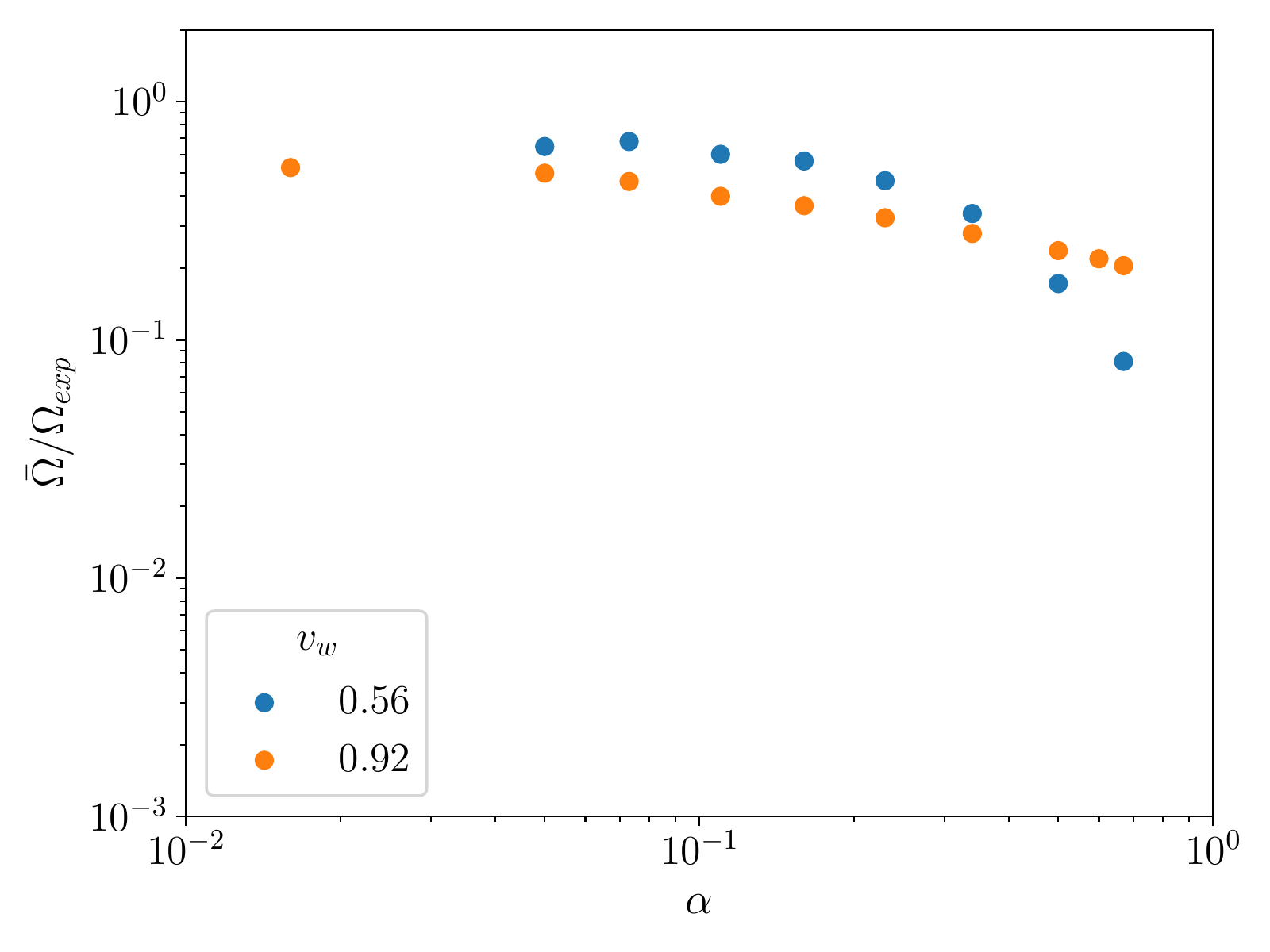}
    \caption{Suppression factor with respect to the strength of the phase transition due to vorticity and reheating effects in the plasma. $\bar{\Omega}$ is the reduced peak gravitational wave energy density and $\Omega_{\rm exp}$ is the expected peak gravitational wave energy density. The data is taken from \cite{Cutting:2019zws}. }
    \label{fig:vort-suppression}
\end{figure}
The first numerical simulations of strong first order phase transitions for $\alpha \gtrsim 0.1$ were undertaken in \cite{Cutting:2019zws} which showed that previous results overestimated the gravitational wave energy density. The rotational component of the fluid velocity, particularly for deflagrations, increases in relative size as the transition strength grows. This reduces the amount of available kinetic energy that can be transferred to the fluid. The rotational component for detonations, however, remains small and constant.  The probable explanation for this effect is the formation of reheated droplets of the metastable phase that are produced during the collisions of the bubble walls. These droplets can then slow down the bubble walls and reheat the surrounding regions. The simulations considered a simple bag equation of state where the results only depend on input parameters of $v_w$ and $\alpha$. For a given $v_w$, we use an interpolation of their results to estimate the corresponding suppression factor to the $\Omega_{\rm GW} h^2$ as a function of $\alpha$. We show the suppression factor in Fig.~\ref{fig:vort-suppression} for two representative bubble wall velocities. We utilize extrapolation for when $\alpha$ is out of range.   Although the results were performed using a bag equation of state, we numerically compute $\alpha$ in the beyond the bag model and assume that the suppression from vorticity and reheating derived with a bag model applies without modification. We will test this assumption in future work. Furthermore, the simulations suggest that the RMS fluid velocity $\bar{U}_f$ reaches a maximum that is under-approximated by calculating the expected $\bar{U}_f$ around a single bubble. We use the results of $\bar{U}_{f,\rm max}/\bar{U}_{f,\rm exp}$ to estimate the maximum fluid velocity in the highest diligence after calculating $\bar{U}_f$ in the beyond the bag model. 

A careful calculation of the gravitational wave production in an expanding universe will result in a suppression factor of the form 
\begin{equation}
    \Upsilon_{RD} = 1 - \frac{1}{y},
\end{equation}
for a radiation dominated era where $y = a/a_s$ is a dimensionless scale factor ratio normalized by when the source of production becomes active. For a radiation dominated era, the time elapsed since some reference time, $t_s$, is
\begin{equation}
    (t - t_s) H_s = \frac{y^2 - 1}{2}.
\end{equation}
Due to the presence of shocks and turbulence in the plasma, the time elapsed is unlikely to last an arbitrarily long time.  The effective lifetime of the source is then given by the timescale of turbulence, $\tau_{\rm sw} = R_*/\bar{U}_f$. This was the basis of the suppression factor used in Eq.~\ref{eq:suppression-ellis}-\ref{eq:omega-suppression-ellis}. We use the estimated maximum of the fluid velocity in the beyond the bag model when calculating the lifetime of the source. The gravitational wave energy density will then be suppressed by 
\begin{eqnarray}
    \Upsilon = 1 - \frac{1}{\sqrt{1 - 2 \tau_{\rm sw} H_s}},
    \label{eq:suppression-upsilon}
\end{eqnarray}
which approaches the asymptotic value $\Upsilon = 1$, the lowest diligence, when $\tau_{\rm sw} H_s \rightarrow \infty$. When $\tau_{\rm sw}H_s \ll 1$, the suppression factor is approximately $\Upsilon = \tau_{\rm sw} H_s$, the modest diligence.  

The peak gravitational wave energy density after taking into consideration the suppressions arising from vorticity and reheating effects in the plasma as well as the lifetime of source becomes 
\begin{equation}
	h^2 \Omega_{\rm GW} = 8.5 \times 10^{-6} \left( \frac{100}{g_*}\right)^{1/3} K^2 \left( \frac{H_*}{\beta}\right) v_{w} \Upsilon \left(\bar{U}_{f,\rm max},R_* \right) \left( \frac{\bar{\Omega}(v_w,\alpha)}{\bar{\Omega}_{\rm exp}(v_w,\alpha)}\right),
	\label{eq:omega_suppressed} 
\end{equation}
where  $K$ is calculated in the beyond the bag, $\beta/H_*$ is calculated from the mean bubble separation, and the last factor arises from the vorticity and reheating effects in the plasma. 

\section{Test models}\label{sec:testmodels}
In this section we examine the numerical difference in predictions arising from different levels of diligence in several models
\begin{itemize}
    \item[1] The Standard Model Effective Field Theory (SMEFT), itself close to a toy model when it comes to cosmological phase transitions \cite{Postma:2020toi}, but which allows for a comparison to the uncertainties arising from gauge dependence and the breakdown of perturbation theory as outlined in \cite{Croon:2020cgk}.
    \item[2] Dark Higgs models \cite{Croon:2018erz}, the simplest phase transition that can occur in a dark sector and has only three free parameters.
    \item[3] A real scalar singlet extension to the standard model (xSM) \cite{OConnell:2006rsp}. A model that allows a tree level barrier, like SMEFT, but is on firmer footing as a physical theory.
\end{itemize}
Using a spectrum of models gives a realistic account of the size of the relative errors for different level of diligence without being overly sensitive to model specific effects. We will also present a toy model in the appendix \ref{sec:toy}, that has the convenient property that much of the analysis can be done analytically.

\subsection{SMEFT}
SMEFT is a model independent method of examining many extensions of the Standard Model by augmenting it with a tower of high dimensional operators, each suppressed by higher and higher powers of the cutoff scale corresponding to the scale of new physics. Unfortunately, the Standard Model requires such a large change to its potential that the scale of new physics needs to be quite low to augment a strong first order phase transition \cite{Postma:2020toi}. In such a case the SMEFT only provides a qualitative description of the UV complete scalar sector, and then only in special circumstances \cite{Postma:2020toi}. In the SMEFT the tree level potential is augmented by a single higher dimensional operator
\begin{equation}
	V_{\rm tree} = \mu_h^2 \phi^\dagger \phi + \lambda \left( \phi^\dagger \phi\right)^2 + \frac{1}{M^2} (\phi^\dagger \phi)^3 + \Delta V,
\end{equation}
where $M$ characterizes the cut off scale, $\phi$ is the SM Higgs doublet and $\Delta V_h$ is chosen such that the zero-temperature minimum is shifted to the origin. 
We will consider the full free energy density at  one-loop given by 
\begin{eqnarray}
	f(\phi,T) = V_{\rm tree} + V_{\rm CW} + V_{\rm CT} + V_T ,
\end{eqnarray}
where $V_{\rm CW}$ is the Coleman-Weinberg contribution and $V_T$ is the finite-temperature correction.  These are given by 
\begin{eqnarray}
	V_{\rm CW}(h,\mu) = \frac{1}{64 \pi^2} \sum_{\alpha} N_{\alpha}  M_{\alpha}^4(h) \left[ \log \frac{M_\alpha^2(h)}{\mu^2} - C_\alpha \right],
	\label{eq:CW-term}
\end{eqnarray}
and 
\begin{eqnarray}
	V_{T}(h,T) = \frac{T^4}{2 \pi^2}\sum_{\alpha}N_{\alpha} \int_0^\infty dx \: x^2 \log \left[1 \pm e^{-\sqrt{x^2 + M_\alpha^2(h)/T^2}} \right] \\
	+ \frac{T}{12\pi} \sum_{\textrm{bosons} \alpha} N_\alpha \left[ M_\alpha^3(h) - M_{T,\alpha}^3(h,T) \right] ,
\label{eq:VT-term}	
\end{eqnarray}
where $N_\alpha$ counts the number of degrees of freedom of each particle and $C_\alpha$ is a constant that is $5/6$ for gauge bosons and $3/2$ for all others. We note that the daisy terms in Eq.~\ref{eq:VT-term} are the result of a high temperature expansion which may cause an IR-divergence in the speed of sound for low temperatures \cite{Giese:2020znk}. We explicitly check this by including a Boltzmann suppression term when $M_{\alpha} \lesssim 2.2 T$ . The sums run over the top quark, $W$ and $Z$ bosons, and the Higgs boson $h$.  The total degrees of freedom in SMEFT is the Standard Model value $g_{\rm SM} = 106.75$. The calculation of the speed of sound requires including all the relativistic particle species in the free energy density. We will account for the remaining light particles that were neglected in $V_T$ by including the term: 
\begin{equation}
	\delta V_T(h) = - \frac{\pi^2}{90} g^\prime_{\ast} T^4,
\end{equation}
to the free energy density where $g_\ast^\prime = 345/4$. However, in the bag model, the speed of sound is taken to be $c_s^2 = 1/3$ and the light species can be ignored as they do not affect the phase transition dynamics. The last term in $V_T$ corresponds to the resummation of the daisy terms of the scalar bosons. To calculate the effective potential and the counter-terms at zero-temperature, we fix the zero-temperature $\overline{\rm MS}$-parameters by matching the physical observables at the $Z$ boson pole mass $m_Z$ using the full self energies. 
To go beyond the bag model, we need the absolute pressure in each phase, and not just the relative pressure. We therefore add an overall constant in the potential such that the pressure in the broken phase at zero temperature vanishes at one loop.
The scale of the Coleman-Weinberg potential is taken to be at $\mu \sim T$ for the dynamics of the phase transition and we run the parameters to this scale. 
  
\subsection{Dark Renormalizable Models}
Here we will consider a dark Higgs model \cite{Schwaller:2015tja,Huang:2020mso,Croon:2018erz,Huang:2021rrk} of the type $SU(N)/SU(N-1)$ with renormalizable operators following the conventions in \cite{Croon:2018erz}.  The overall scale $\Lambda$ and the zero temperature vacuum expectation value $v$ are the only inputs of the model. We can then define the zero temperature parameters such as the tachyonic mass and self coupling as 
\begin{eqnarray}
	-\mu^2(0) = -\frac{\Lambda^4}{v^2}, \\
	\lambda(0) = \frac{\Lambda^4}{v^4},
\end{eqnarray}
where factors of $v/\Lambda$ will control the thermal parameters. The tree level potential is then defined as 
\begin{equation}
	V(H) = \Lambda^4 \left[ - \frac{1}{2} \left( \frac{h_D}{v}\right)^2 + \frac{1}{4} \left( \frac{h_D}{v}\right)^4 \right] + \Delta V,
\end{equation}
where $h_D$ is the dark Higgs of the doublet $H$ and $\Delta V$ shifts that potential at the minimum to zero. The degrees of freedom of the full dark sector in consideration are 
\begin{equation}
	n_H = 1, \: n_G = 2 N-1, \: n_{GB} = 3 \times (2N -1), \: n_f = 2 \times N \times N_f,
\end{equation}
where $n_G$ is the number of Goldstone bosons, $n_{GB}$ are the gauge bosons, $N_f$ is the number of fermions,and $N$ is the rank of the group.   

The free energy density in consideration is 
\begin{equation}
	f(h_D,T) = V_{\rm tree} + V_{\rm CW} + V_{\rm CT} + V_{T} + \delta V_T, 
\end{equation}
where $V_{\rm CW}$ and $V_{T}$ are defined in Eq.~(\ref{eq:CW-term}-\ref{eq:VT-term}). Here the summations now only run over the dark sector particles. This requires us to add on the additional relativistic particles not included in the sum which now include the full degrees of freedom of the Standard Model:
\begin{equation}
	\delta V_T = -\frac{\pi^2}{90} g_{\ast} T^4,
\end{equation}
where $g_{\ast} = 106.75$. We add the term $\Delta V$ so that the minimum of the tree-level potential is shifted to zero. We choose the scale of the one-loop potential to be at $\Lambda$.  

The inclusion of the CW-term will shift the zero-temperature vacuum expectation value away from $v$. We add the counter-term potential 
\begin{equation}
	V_{\rm CT}(h_D) = -\frac{\delta \mu^2}{2} h_D^2 + \frac{\delta \lambda}{4} h_D^4 + \delta \Delta V_h,
\end{equation}
where $\delta \mu^2$, $\delta \lambda$, and $\delta \Delta V_h$ are chosen such that 
\begin{eqnarray}
	\frac{\partial f(h_D,0)}{\partial h_D} \Big|_{v} = 0, \\
	\frac{\partial^2 f(h_D,0)}{\partial h_D^2} \Big |_{v} = m^2(0) \equiv 2 \frac{\Lambda ^4}{v^2},\\
	f(v,0) = 0,
\end{eqnarray}
to maintain the tree-level structure of the potential. We work in the Landau gauge where the Goldstone bosons have zero mass at the $h_D = v$ which causes an IR-divergence in the one-loop potential. This causes an issue in the evaluation of the counter-terms.  One prescription is to remove the Goldstone bosons from the sum in the CW potential. For the purpose of this work, we will follow the procedure in \cite{Elias_Mir__2014} to evaluate the counterterms. This shifts the Goldstone mass by its mass at the one-loop level, i.e 
\begin{equation}
	m_{G}(h_D) \rightarrow m_{G}(h_D) + \frac{1}{h_D} \frac{\partial V^{(1)}}{\partial h_D},
\end{equation}  
where 
\begin{equation}
	m_{G}(h_D) = \Lambda^4 \left( \frac{h_D^2}{v^4} - \frac{1}{v^2}\right),
\end{equation}
is the field dependent mass of the Goldstone bosons.  

\subsection{xSM}
The singlet extended SM, known as "xSM", consists of the standard SM Higgs doublet $H^T = \left( G^+, \left( v_{\rm EW} + h + i G^0 \right)/\sqrt{2} \right)$ and a real gauge singlet $S = v_s + s$ where the electroweak vacuum is $\left( v_{\rm EW}, v_s \right)$ \cite{Barger:2007im,Profumo:2007wc,Alves:2020bpi,Alves:2019igs,Alves:2018jsw,Alves:2018oct,Liu:2021jyc,Chiang:2018gsn,Fuyuto:2014yia}. The tree level potential in this setup is defined as 
\begin{align}
	V(H,S) = &-\mu^2 H^\dagger H + \lambda  \left(H^\dagger H\right)^2 + \frac{a_1}{2} H^\dagger H S  \nonumber \\ 
	  &+ \frac{a_2}{2} H^\dagger H S^2 + \frac{b_2}{2} S^2 + \frac{b_3}{3} S^3 + \frac{b_4}{4} S^4 + \Delta V,
\end{align}
where $\Delta V$ shifts the minimum of the potential to zero. The mass parameters $\mu^2$ and $b_2$ are related to the other model parameters through the minimization conditions around the electroweak vacuum, 
\begin{align}
	\mu^2 &= \lambda v_{\rm EW} ^2 + \frac{1}{2} v_s \left(a_1 + a_2 v_s \right) \nonumber,  \\
	b_2       &= \frac{1}{4 v_s} \left[ v_{\rm EW}^2 \left( a_1 + 2 a_2 v_s\right) + 4 v_s^2 \left( b_3 + b_4 v_s \right) \right].
\end{align}
The  parameters $\lambda$, $a_1$, and $a_2$ are related to the physical parameters $\theta$, $m_{h_1}$, and $m_{h_2}$  through the mass matrix diagonalization as 
\begin{align}
	\lambda &= \frac{m_{h_1}^2 c_\theta^2 + m_{h_2} s_\theta^2}{2 v_{\rm EW}^2} ,\nonumber \\
	a_1 &= \frac{2 v_s}{v_{\rm EW}^2} \left[ 2 v_s^2 \left( 2 b_4 + \tilde{b}_3\right) - m_{h_1}^2 - m_{h_2} + c_{2 \theta} \left( m_{h_1}^2 - m_{h_2}^2\right)\right] , \nonumber \\
	a_2 &= -\frac{1}{2 v_{\rm EW}^2 v_s}  \left[-  2 v_s \left( m_{h_1}^2 + m_{h_2}^2 - 4 b_4 v_s^2 \right)  + \left( m_{h_1}^2 - m_{h_2}^2\right)\left( 2 c_{2\theta} v_s - v_{\rm EW} s_{2 \theta} + 4 \tilde{b}_3 v_s^3 \right)\right],
\end{align}
where $\tilde{b}_3 \equiv b_3/v_s$ and $\theta$ is the mixing angle of the physical fields $h_1$ and $h_2$ defined as 
\begin{eqnarray}
	h_1 = c_{\theta} h + s_\theta s,\quad h_2 = -s_\theta h + c_\theta s,
\end{eqnarray}
with $s_\theta \equiv \sin(\theta)$ and $ c_\theta \equiv \cos(\theta)$. Here we associate $h_1$ as the SM Higgs and $h_2$ is some heavier scalar.

The free energy density we consider in the xSM presented here contains only the high temperature expansion approximation of the full finite temperature one loop effective potential since the phase transition is primarily driven by the cubic terms.  The free energy is then 
\begin{align}
	f(h,s,T) = &-\frac{1}{2} \left[ \mu^2 - \Pi_h(T)\right]h^2 - \frac{1}{2} \left[ -b_2 - \Pi_s(T) \right] s^2 \nonumber \\
	& + \frac{1}{4} \lambda h^4 + \frac{1}{4} a_1 h^2 s + \frac{1}{4}a_2 h^2 s^2 + \frac{b_3}{3}s^3 + \frac{b_4}{4}s^4 \nonumber \\
	& - \frac{1}{3} \frac{\pi^2}{30} g_* T^4,
\end{align}
where we take $g_* = 106.75$. The field dependent thermal mass $\Pi_h(T)$ and $\Pi_s(T)$ are  
\begin{align}
	\Pi_h(T) &= \left( \frac{2 m_w^2 + m_z^2 + 2 m_t^2}{4v^2}  + \frac{\lambda}{2} + \frac{a_2}{24} \right) T^2, \nonumber \\
	\Pi_s(T) &= \left(\frac{a_2}{6} + \frac{b_4}{4} \right)	T^2,
\end{align}
where the physical masses of the $W$, $Z$, and $t$-quark are used to define the gauge and Yukawa couplings to $h$. 

\section{Results}\label{sec:results}
The resulting gravitational wave spectrum is dependent on the level of precision of the thermal parameters. Until recently, the bag model was assumed to compute the phase transition strength and the kinetic energy of the fluid. Going beyond the bag model will require the calculation of the speed of sound in the plasma for both the symmetric and broken phase which should result in a more accurate treatment of the thermal parameters. These quantities are temperature dependent and will change depending on the temperature at which they are computed.  Furthermore, the temperature scales of the phase transition such as the nucleation and percolation temperature are also sensitive to level of diligence in the calculations. We use the publicly available codes CosmoTransitions \cite{Wainwright:2011kj} and BubbleProfiler \cite{Athron:2019nbd} to compute the actions in order to find the relevant transition temperatures.  

The lowest diligence level will compute the thermal parameters  at the estimated nucleation temperature defined in Eq.~\ref{eq:Tn}.  The strength of the phase transition is calculated using Eq.~\ref{eq:alpha_Tn},  the inverse time duration of the phase transition is calculated using Eq.\ref{eq:betaHn}, and the peak gravitational wave energy density is calculated using Eq.~\ref{eq:omega1}.  
\begin{figure}
	\centering
        \includegraphics[width=0.49\textwidth]{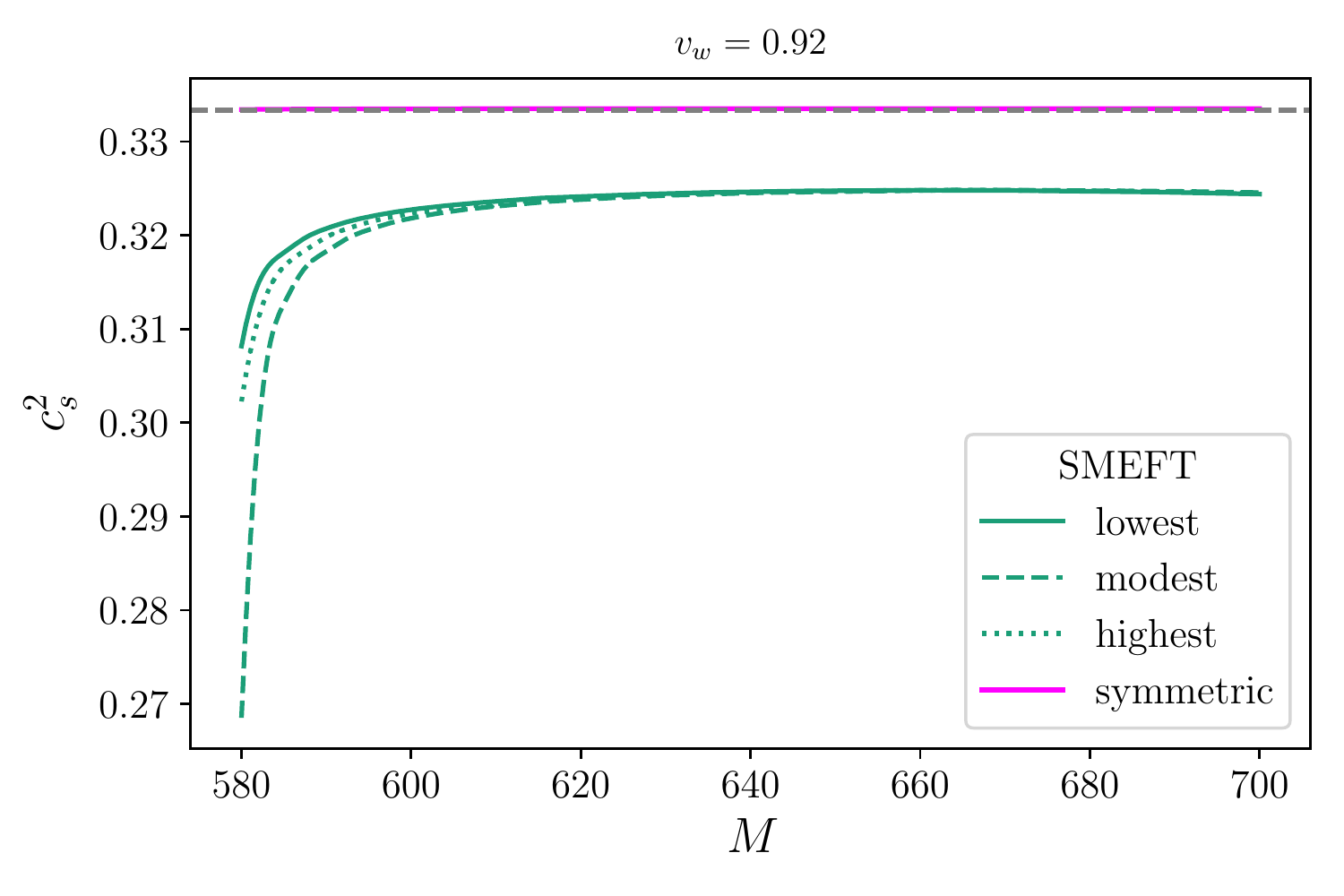}
        \includegraphics[width=0.49\textwidth]{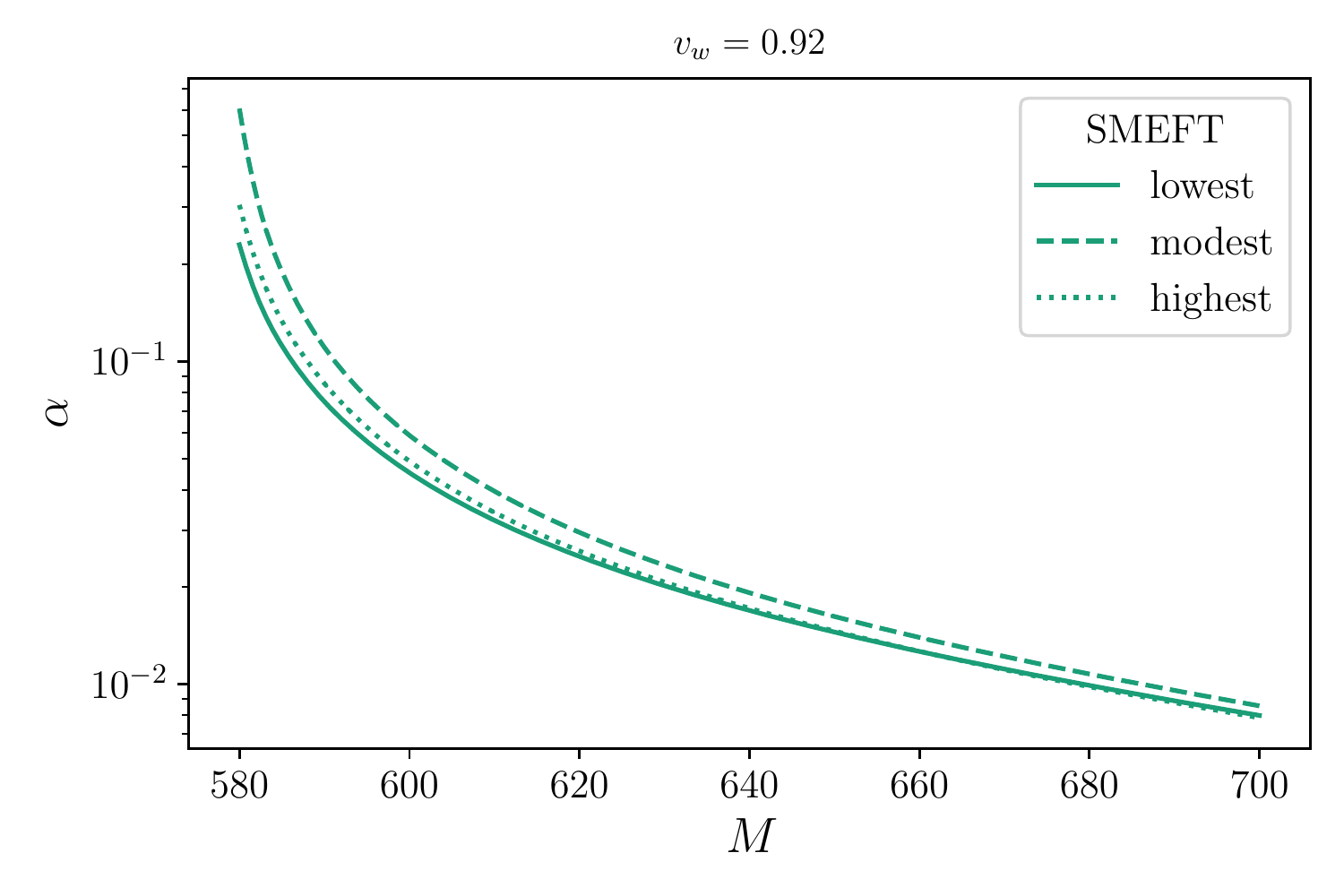}
        \includegraphics[width=0.49\textwidth]{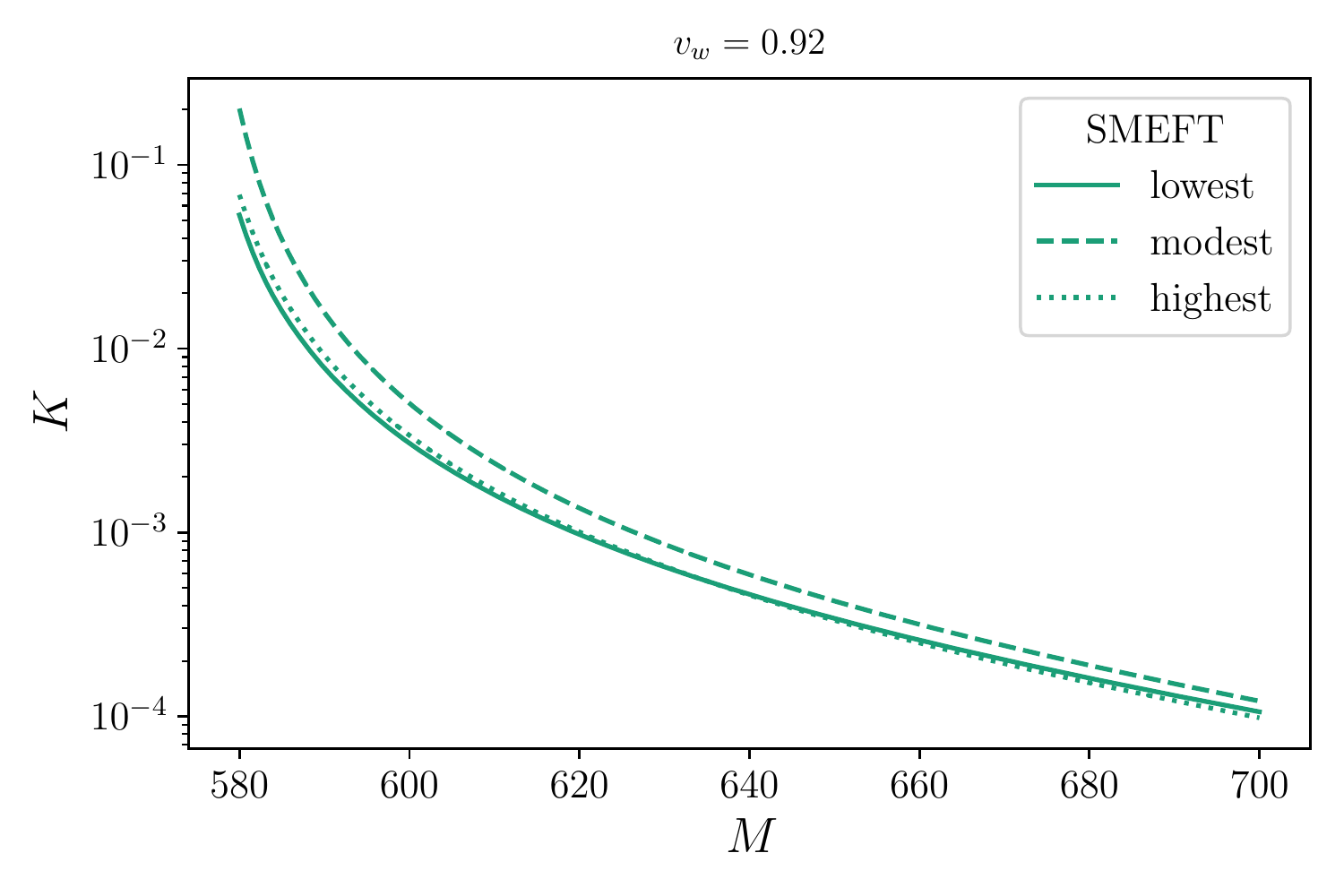}
 	\caption{\textbf{SMEFT}: The top left panel shows the speed of sound calculated in the symmetric and broken phase using Eq.~\ref{eq:cs2} at the different levels of diligence. The gray dashed line corresponds to the bag model with $c_s^2 = 1/3$. The symmetric phase (solid magenta) is only shown at highest diligence. The top right panel shows the strength of the phase transition at the different levels of diligence using Eq.~[\ref{eq:alpha_Tn}, \ref{eq:alpha_Tp}, \ref{eq:alpha_psuedotrace}]. The bottom  panel shows the kinetic energy fraction at the different levels of diligence where the lowest and modest diligence use fits for $\kappa$ to get $K$ and the highest diligence uses Eq.~\ref{eq:K_beyond_bag}. The temperatures are set to $T_n$ (\ref{eq:Tn}), $T_p$ (\ref{eq:LISA_Tp}), and $T_f$ (\ref{eq:Tf}) for the lowest, modest, and highest diligence respectively. The numerical calculation of the speed of sound only enters in the highest diligence of $\alpha$. }
	\label{fig:smeft-cs2-alpha-K}
\end{figure}

The modest diligence level will compute the thermal parameters  at the estimated percolation temperature in Eq.~\ref{eq:LISA_Tp}.  We will use Eq.~\ref{eq:alpha_Tp}, Eq.~\ref{eq:betaHp}, and Eq.~\ref{eq:omega-suppression-ellis} to estimate the strength of the phase transition, inverse time duration, and peak gravitational wave spectrum respectively.  The lifetime of the source is estimated using Eq.~\ref{eq:suppression-ellis}. There is an  ambiguity as to which temperature to employ in the calculation of the thermal parameters:  the nucleation temperature or the percolation temperature.  The true percolation temperature $T_p$, Eq.~\ref{eq:Tp}, should lie close to the temperature at which the phase transition ends $T_f$. 
The inverse time duration should be computed from the mean bubble separation, Eq.~\ref{eq:Rstar_from_nb} and Eq.~\ref{eq:Rstar_from_nb}, which is evaluated at $T_f$. The highest diligence will evaluate all thermal parameters at $T_f$ to ensure that all quantities are evaluated at the same temperature as the inverse time duration.
From here on out,  we will associate $T_p$ with Eq.~\ref{eq:LISA_Tp} when referring to modest diligence.  The highest diligence will also utilize the beyond the bag model to calculate the strength of the phase transition Eq.~\ref{eq:alpha_psuedotrace} and the kinetic energy fraction Eq.~\ref{eq:K_beyond_bag} which requires the numerically calculated speed of sound in Eq.~\ref{eq:cs2}.  The lowest and modest diligence calculations will assume $c_s^2 = 1/3$ as is done in the bag model. The peak gravitational wave spectrum is found from Eq.~\ref{eq:omega_suppressed} which accounts from the lifetime of the source, Eq.~\ref{eq:suppression-upsilon}, as well as vorticity and reheating effects in the plasma. We note that the suppression factor due to the finite lifetime used in Eq.~\ref{eq:suppression-ellis} is a valid approximation of Eq.~\ref{eq:suppression-upsilon} only when $\tau_{\rm SW} H \ll 1$.

In the following sections, we will compare the different levels of diligence in SMEFT, the dark renormalizable model, and xSM. The error in the peak gravitational wave energy density,
\begin{equation}
	\frac{\Delta \Omega}{\Omega} = \frac{|\Omega^{j} h^2 - \Omega^{{\rm high}}h^2|}{{\rm min} \left[ \Omega^{j}h^2, \Omega^{{\rm high}}h^2\right]},
	\label{eq:error}
\end{equation}
where $j=({\rm low, mod})$ refers to lowest, modest, and highest diligence respectively, will be used to compare the different levels of diligence. 
\subsection{SMEFT}
\begin{figure}
	\centering
        \includegraphics[width=0.49\textwidth]{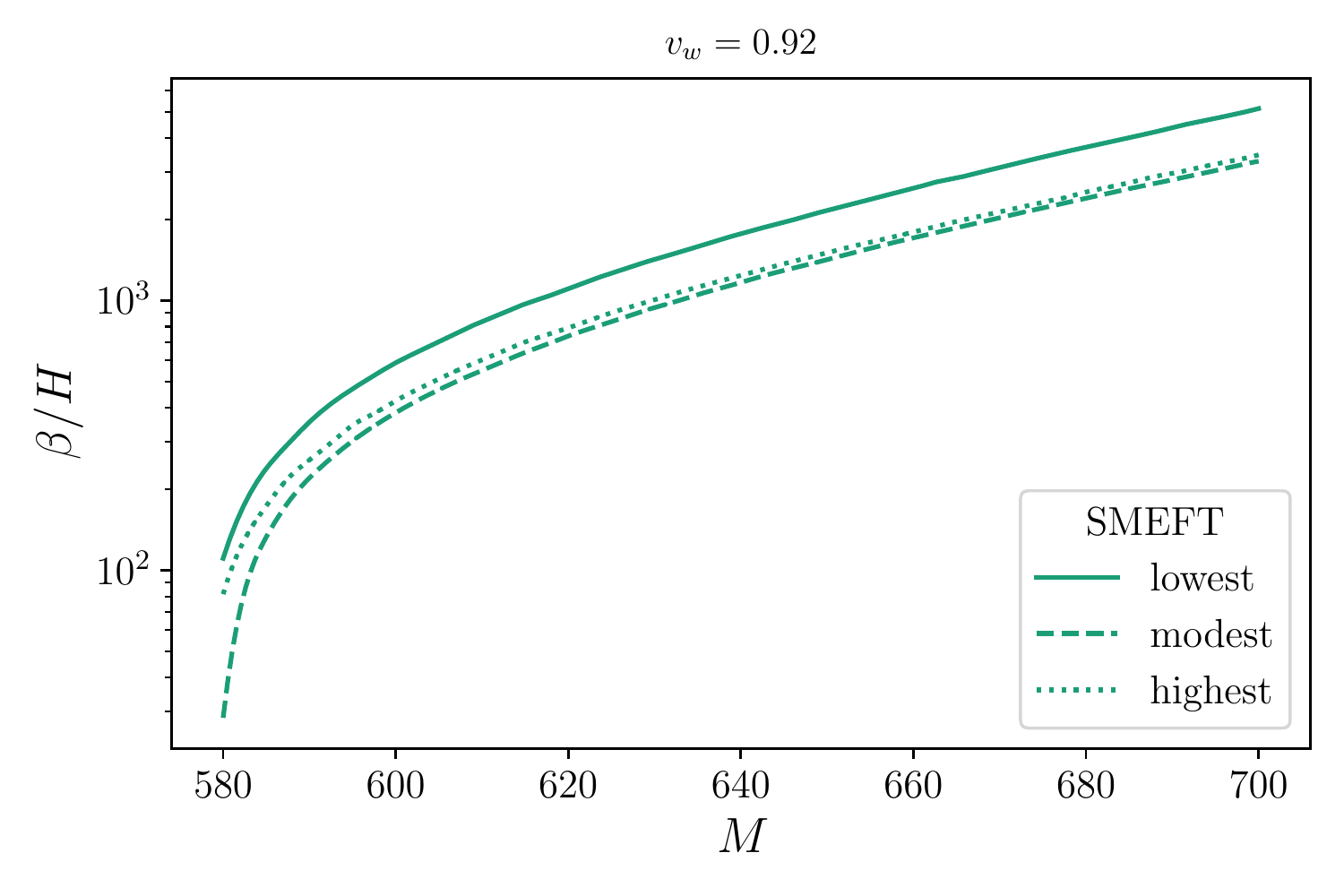}
        \includegraphics[width=0.49\textwidth]{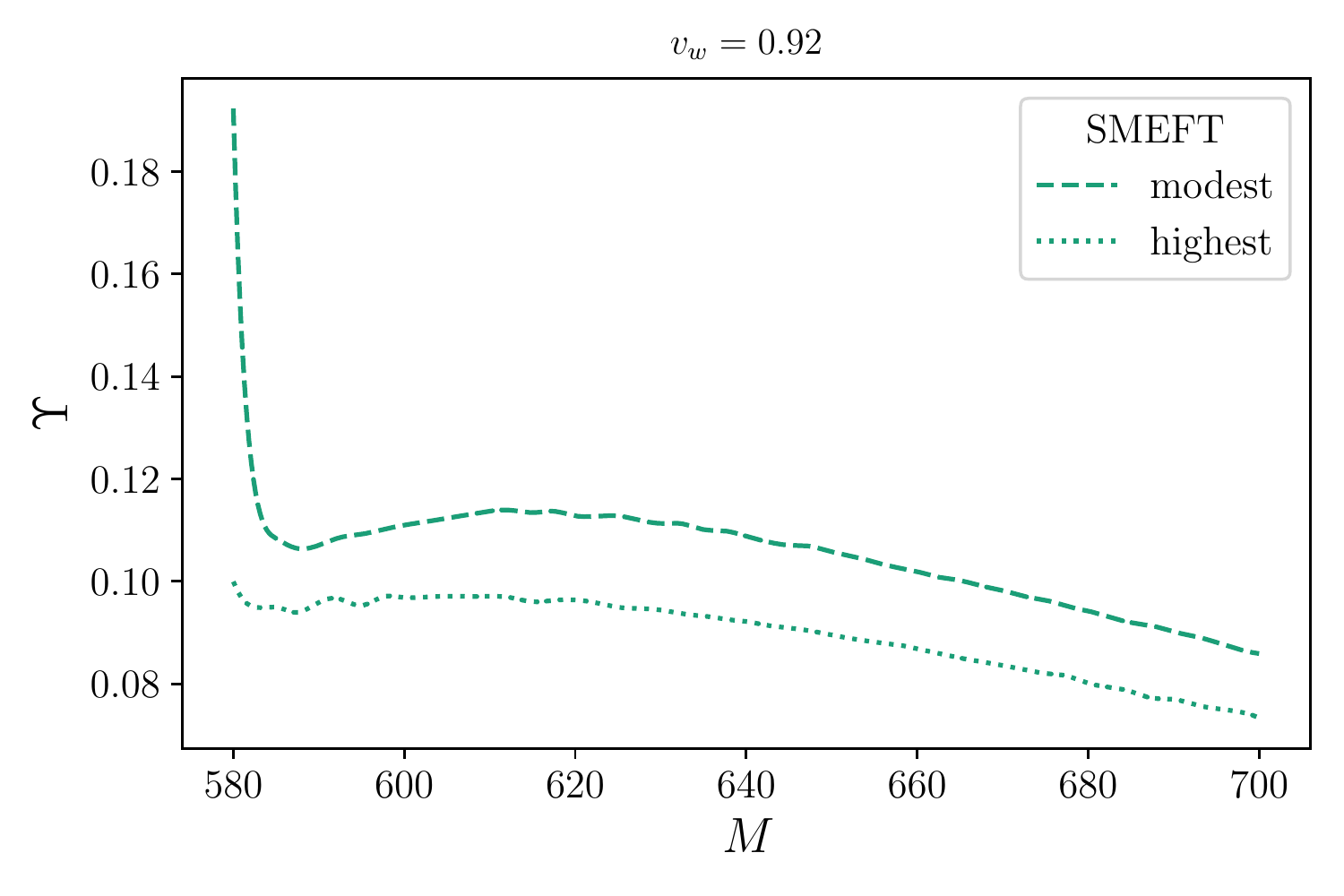}
	\caption{\textbf{SMEFT}: The left panel shows the inverse time duration of the phase transition at the different levels of diligence using Eq.~[\ref{eq:betaHn}, \ref{eq:betaHp}, \ref{eq:beta_Tf}]. The lowest and modest diligences are estimated using the first derivative of the action $dS/dT$ and the highest diligence is computed directly from the mean bubble separation, Eq.~\ref{eq:Rstar_from_nb} and Eq.~\ref{eq:Rstar}. The right panel shows the suppression factor due to the lifetime of the source using Eq.~\ref{eq:suppression-ellis} and Eq.~\ref{eq:suppression-upsilon} for modest and highest diligence respectively.  The lowest diligence corresponds to $\Upsilon \rightarrow 1$. The temperatures are set to $T_n$ (\ref{eq:Tn}), $T_p$ (\ref{eq:LISA_Tp}), and $T_f$ (\ref{eq:Tf}) for the lowest, modest, and highest diligence respectively.}
	\label{fig:smeft-beta-upsilon}
\end{figure}

The SMEFT model we consider has the scale of the zero-temperature one loop potential set to $\mu = T$ as well as temperature dependence in the running of the couplings. This will contribute additional temperature dependence to the speed of sound calculations in the broken phase.  We note that Ref.~\cite{Wang:2020nzm} also considered the beyond the bag model in SMEFT using a high temperature expansion of the effective potential. The speed of sound  will never reach the bag model with $c_s^2 = 1/3$ as seen in the top figure left of Fig.~\ref{fig:smeft-cs2-alpha-K}. The green curves show the different levels of diligence for the speed of sound in the broken phase and the dashed gray curve represents $c_s^2=1/3$. The magenta curve is the speed of sound calculated in symmetric phase which is approximately the same in each level and does not deviate far from the bag model. We do not consider any additional relativistic degrees of freedom and thus expect little deviations between the speed of sound in the symmetric phase. As the scale $M$ grows large, the speed of sound in the broken phase approaches a constant value of $c_s^2 \sim 0.32$. There is noticeable disagreement between the different levels below $M=600$ where there is mild supercooling. For a given $M$, the speed of sound is only a function of temperature.  The differences in $c_s^2$ in the broken phase is the result of these different temperatures at which the speed of sound is set to when calculating the strength of the phase transition $\alpha_{\bar{\theta}}(c_s^2)$. The large difference in $T_p$ and $T_f$ is due to the approximations of $T_p$ in Eq.~\ref{eq:LISA_Tp} which is less accurate when $S_3/T$ acquires a minimum for smaller $M$. 

On the top right panel of Fig.~\ref{fig:smeft-cs2-alpha-K} we show the strength of the phase transition computed at the different levels of diligence. Both the lowest and modest diligence curves have $c_s^2 =1/3$ whereas the highest diligence curve corresponds to the beyond the bag calculation with $c_s^2$ shown in the top left panel. Although each level is computed at different temperatures, the lowest diligence is a better approximation of the strength of the phase transition compared to level 2 which over approximates $\alpha$. This is a result of $T_p$ computed in the modest diligence placing far below $T_f$ which results in a higher estimated $\alpha$. The difference between the different levels on the strength of the phase transition is negligible for large $M$ as a result of the asymptotic behavior observed in $c_s^2$ and the better approximation of $T_p$ when there is no barrier present. There is also a dependence on the bubble wall velocity in both $c_s^2$ and $\alpha$ for the modest and highest diligence curves in computing $T_p$ and $T_f$ but we only show detonation with $v_w = 0.92$ because the difference is minor. The kinetic energy fraction is shown in the bottom panel of Fig.~\ref{fig:smeft-cs2-alpha-K} and should depend on the speed of sound and phase transition strength.  Similar to what was seen in $\alpha$, the lowest and highest diligence curves are closer together for large $M$ while the modest diligence curve is the least accurate. For each of the levels, the largest error in both $\alpha$ and $K$ occurs for smaller $M$ where the speed of sound is significantly lower than $c_s^2 = 1/3$ and $T_p$ is far from $T_f$. 

\begin{figure}
	\centering
        \includegraphics[width=0.49\textwidth]{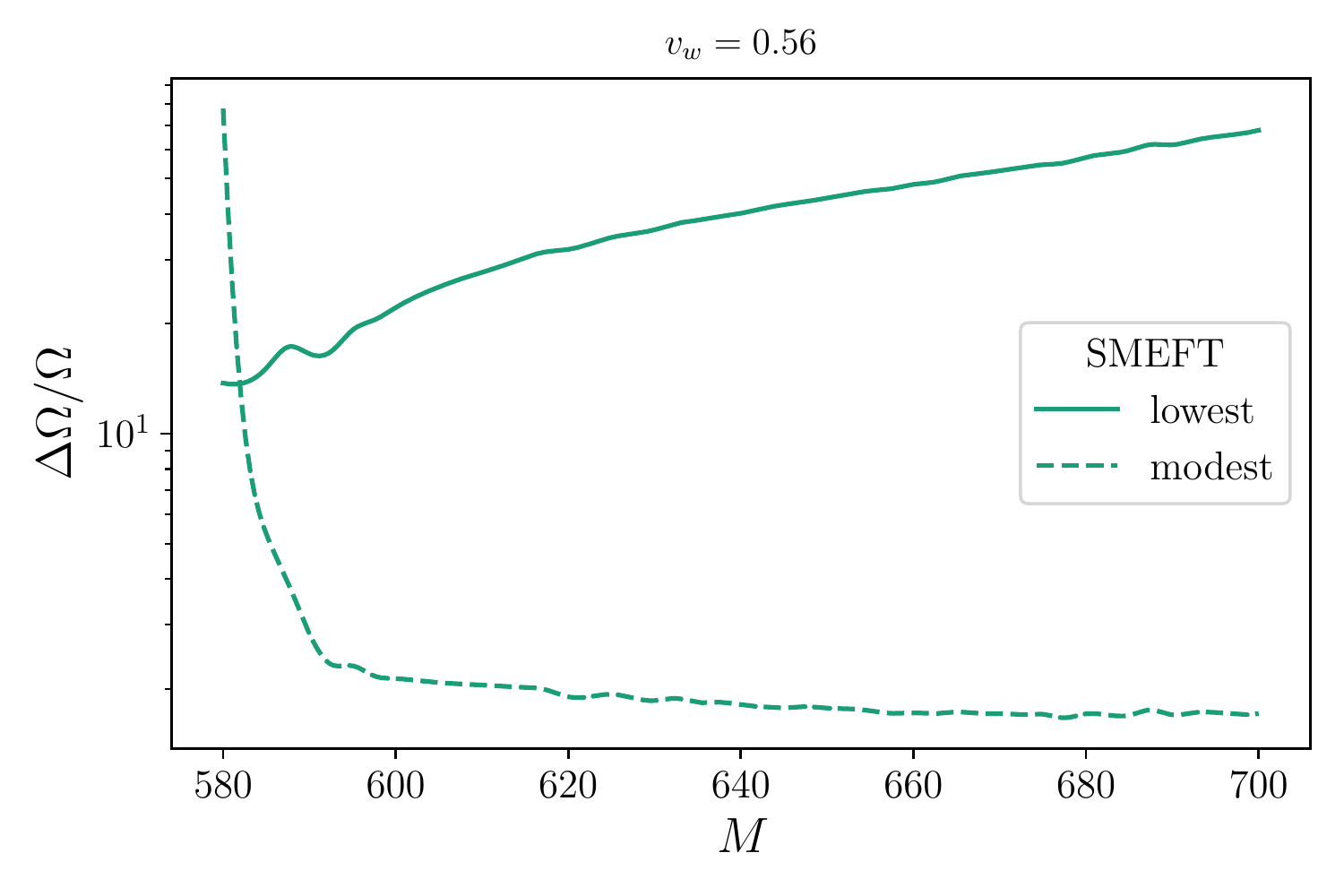}
        \includegraphics[width=0.49\textwidth]{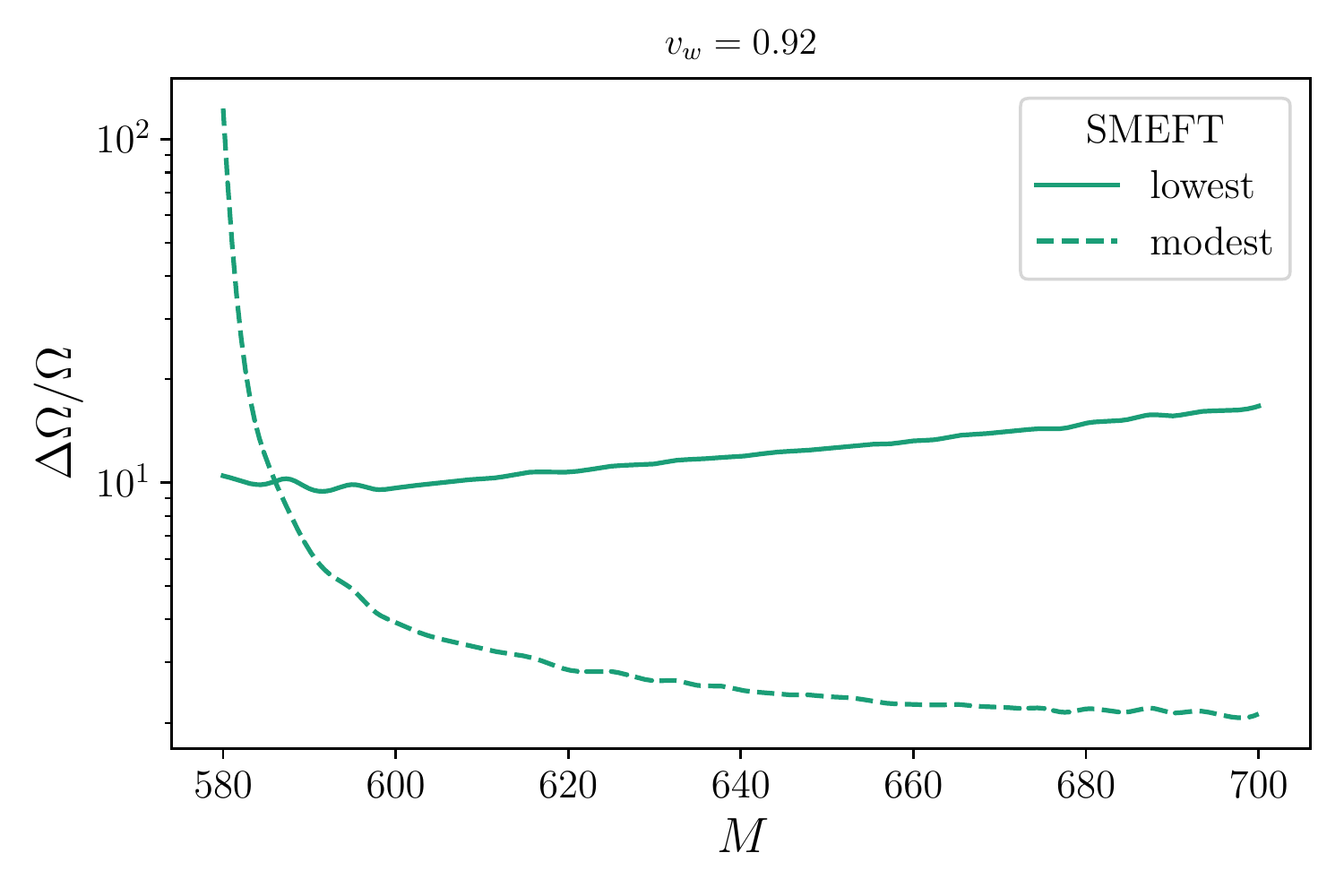}
	 \caption{\textbf{SMEFT}: The relative error when using the  lowest and modest levels of diligence, compared to the highest level of diligence (for which $\Delta \Omega/\Omega = 0$). The vertical axis shows the peak (frequency-independent) gravitational wave energy density for detonation. The precise definition of $\Delta \Omega/\Omega$ is given in Eq.~\ref{eq:error}. The horizontal axis corresponds to the cutoff scale $M$. 
	 $\Delta \Omega/\Omega$ is displayed for deflagration and detonation at  different levels of diligence using Eq.~[\ref{eq:omega1}, \ref{eq:omega-suppression-ellis}, \ref{eq:omega_suppressed}] and Eq.~\ref{eq:error}. The temperatures are set to $T_n$ (\ref{eq:Tn}), $T_p$ (\ref{eq:LISA_Tp}), and $T_f$ (\ref{eq:Tf}) for the lowest, modest, and highest diligence respectively. Both the modest and highest diligence contains suppression factors due to the lifetime of the source. The highest diligence contains the suppression factor due to vorticity effects in the plasma.}
	\label{fig:smeft-spectrum}
\end{figure}
In the left panel of Fig.~\ref{fig:smeft-beta-upsilon}, we show the inverse time duration $\beta/H$ of the phase transition for detonation.  The largest difference between modest diligence and highest diligence occurs for small $M$ and is due to the following reason:  the minimum formed in the action where $T_p$ calculated in Eq.~\ref{eq:LISA_Tp} along with $\beta/H$ in Eq.~\ref{eq:betaHp} are inaccurate when there is a minimum present. The lowest diligence is a better approximation than the modest diligence in this regime. The modest and the highest diligence become indistinguishable for large $M$ when there is no minimum in the action. For small $M$, the lowest diligence curve appears to be a good approximation for modest diligence. Although $\beta/H$ estimated from the action is not accurate when there is a minimum, the error using $T_n$ appears to do better than using the approximation of $T_p$. Contrary to the $\alpha$ and $K$, $\beta$ in the lowest diligence never approaches the highest diligence for large $M$ where the error appears to get worse. This is due to the inaccuracy in using the approximate $T_n$. In this regime, $T_p$ is a better approximation of the inverse time duration as there is no minimum present in the action.  The right panel of Fig.~\ref{fig:smeft-beta-upsilon} shows the suppression factor due to the lifetime of the source in the modest and highest diligence.  The error between the two levels gets worse for small $M$ which is the result of the error in $T_p$.  The error approaches a constant as $M$ gets large. 

In Fig.~\ref{fig:smeft-spectrum} we show the relative error in the peak gravitational wave spectrum for both deflagration and detonation. The error with respect to the highest diligence is estimated using Eq.~\ref{eq:error}. For both deflagration and detonation, the modest diligence spikes to an error of $\Delta \Omega/\Omega \sim 10^2$ for small M. This is correlated with the large error observed in $\alpha$, $K$, $\beta$, and $\Upsilon$ seen in Fig.~\ref{fig:smeft-cs2-alpha-K} and Fig.~\ref{fig:smeft-beta-upsilon}.  The modest diligence error for both profile types slowly approach a constant value as M grows large which is the result of the minimal error in $\alpha$, $\beta/H$, $K$.  The error in $\Upsilon$  appears to become a constant for large $M$. The suppression factors due to the lifetime of the source grow to zero as $M$ grows large which results in the increasing behavior of the peak error in the lowest diligence which does not include any suppression factors. Overall we notice an error in the peak gravitational wave energy density of $10^1 - 10^2$ for lowest diligence and $10^0 - 10^2$ for modest diligence. 

\subsection{Dark Renormalizable Models}
\begin{figure}
	\centering
        \includegraphics[width=0.49\textwidth]{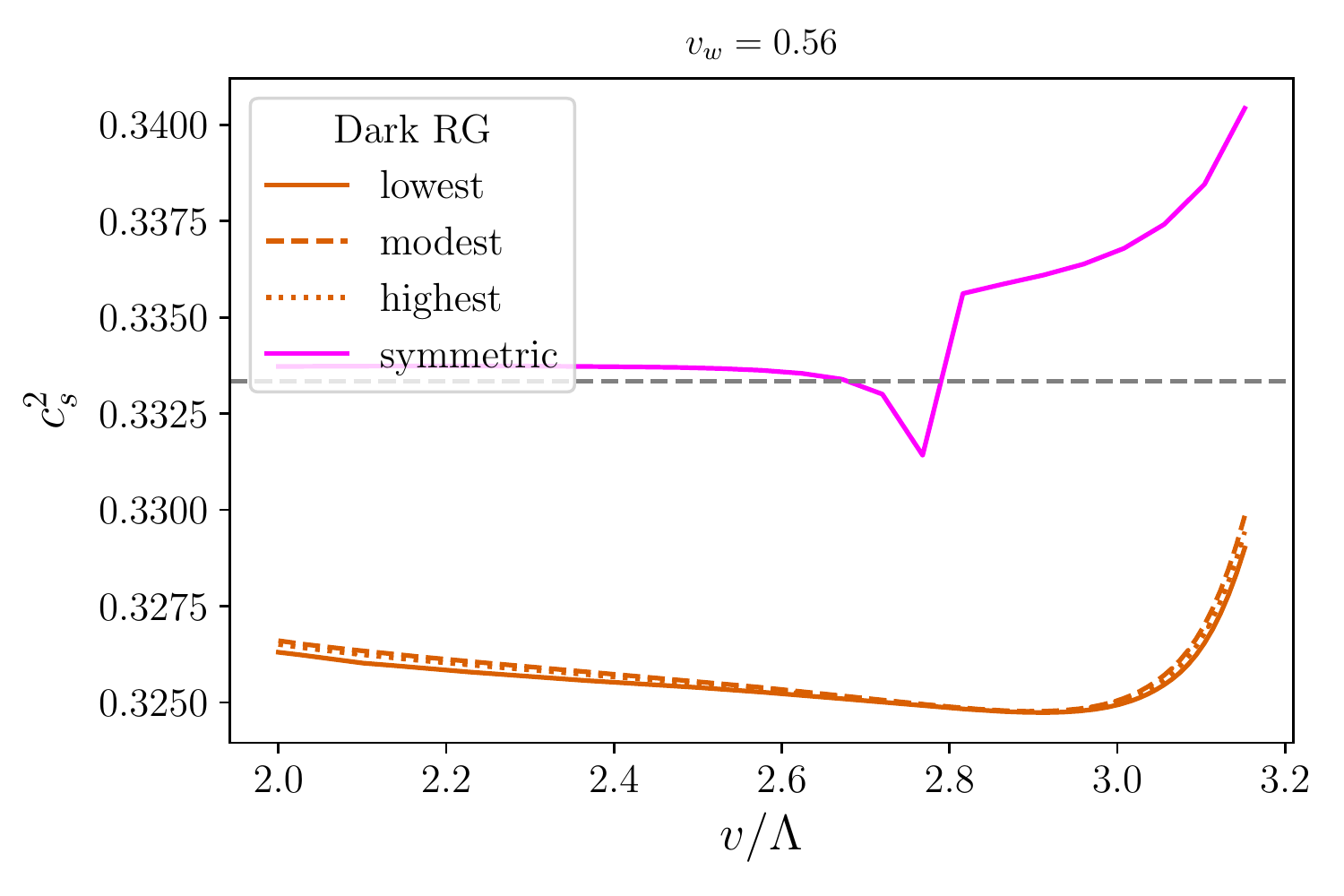}
        \includegraphics[width=0.49\textwidth]{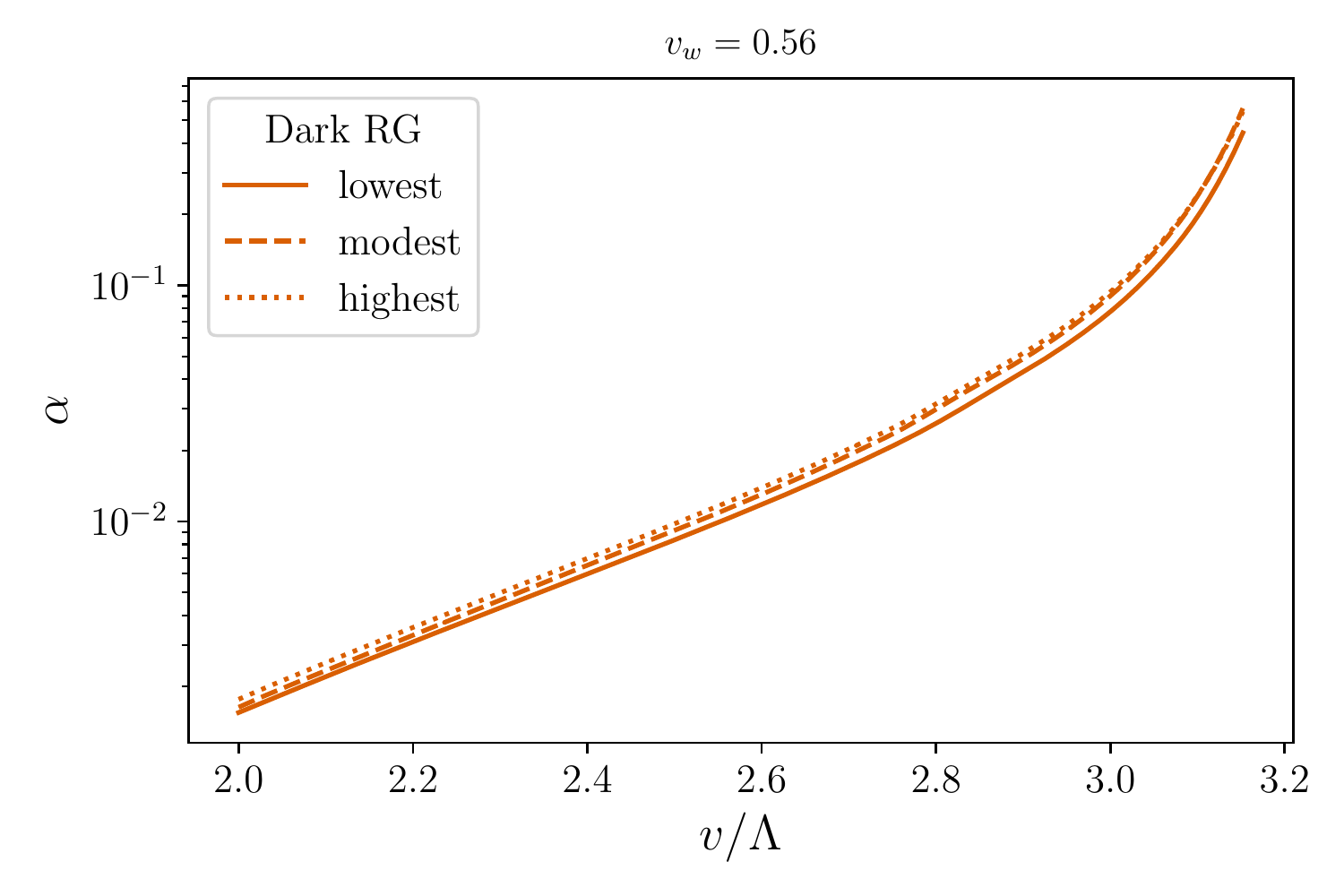}
 	\caption{\textbf{Dark RG}: The left panel shows the speed of sound calculated in the symmetric and broken phase using Eq.~\ref{eq:cs2} at the different levels of diligence. The gray dashed line corresponds to the bag model with $c_s^2 = 1/3$. The symmetric phase (solid magenta) is only shown at highest diligence. The right panel shows the strength of the phase transition at the different levels of diligence using Eq.~[\ref{eq:alpha_Tn}, \ref{eq:alpha_Tp}, \ref{eq:alpha_psuedotrace}]. The temperatures are set to $T_n$ (\ref{eq:Tn}), $T_p$ (\ref{eq:LISA_Tp}), and $T_f$ (\ref{eq:Tf}) for the lowest, modest, and highest diligence respectively. The numerical calculation of the speed of sound only enters in the highest diligence of $\alpha$. }
	\label{fig:darkRG-cs2-alpha}
\end{figure}

The dark renormalizable model considered in the analysis  does not couple to the Standard Model and will consist of a $N = 10$ group, and $2 N - 1$ gauge bosons with charge $g_D = 0.8$. The scale of the one-loop potential is also $T$ independent.  These will result in a speed of sound in the symmetric phase that differs from the one seen in SMEFT.  

We show the speed of sound calculated using Eq.~\ref{eq:cs2} on the left panel of  Fig.~\ref{fig:darkRG-cs2-alpha} for the different levels of diligence. The differences between the levels of diligence in the speed of sound are only minor.  We show only the highest diligence curve for the speed of sound in the symmetric phase. For small $v/\Lambda$, the speed of sound in the symmetric phase remains constant with a value slightly above the one given in the bag model. This is attributed to the additional degrees of freedom arising from the dark sector. The speed of sound above $v/\Lambda = 2.6$ begins to decrease until it reaches a discontinuity near $v/\Lambda = 2.8$. It then jumps to $c_s^2 = 0.336$ where it begins to monotonically increase. This discontinuity is a result of the daisy terms in the effective potential. With out the daisy terms, the speed of sound in the symmetric phase would be smoothly connected and monotonically increasing. 

The strength of the phase transition is plotted on the right panel Fig.~\ref{fig:darkRG-cs2-alpha}.    The different levels appear to agree very well with each other with the lowest diligence becoming slightly worse at high $v/\Lambda$. For most of the parameter space, the highest diligence has the greatest $\alpha$ because it is  computed at the numerically calculated values for $c_s^2$ in the broken phase which results in a amplification compared to the other two levels. This is due to the factor $(1 + c_s^{-2})$ in $\alpha_{\bar{\theta}}$. The error between the modest and highest diligence begins to decrease as $v/\Lambda$ increases which is related to the speed of sound approaching $c_s^2 = 1/3$. Despite the differences between the different levels, the speed of sound in the broken phase lies between $c_s^2 \sim 0.325 - 0.330$ and does not contribute a significant source of error to the strength of the phase transition.

\begin{figure}
	\centering
        \includegraphics[width=0.49\textwidth]{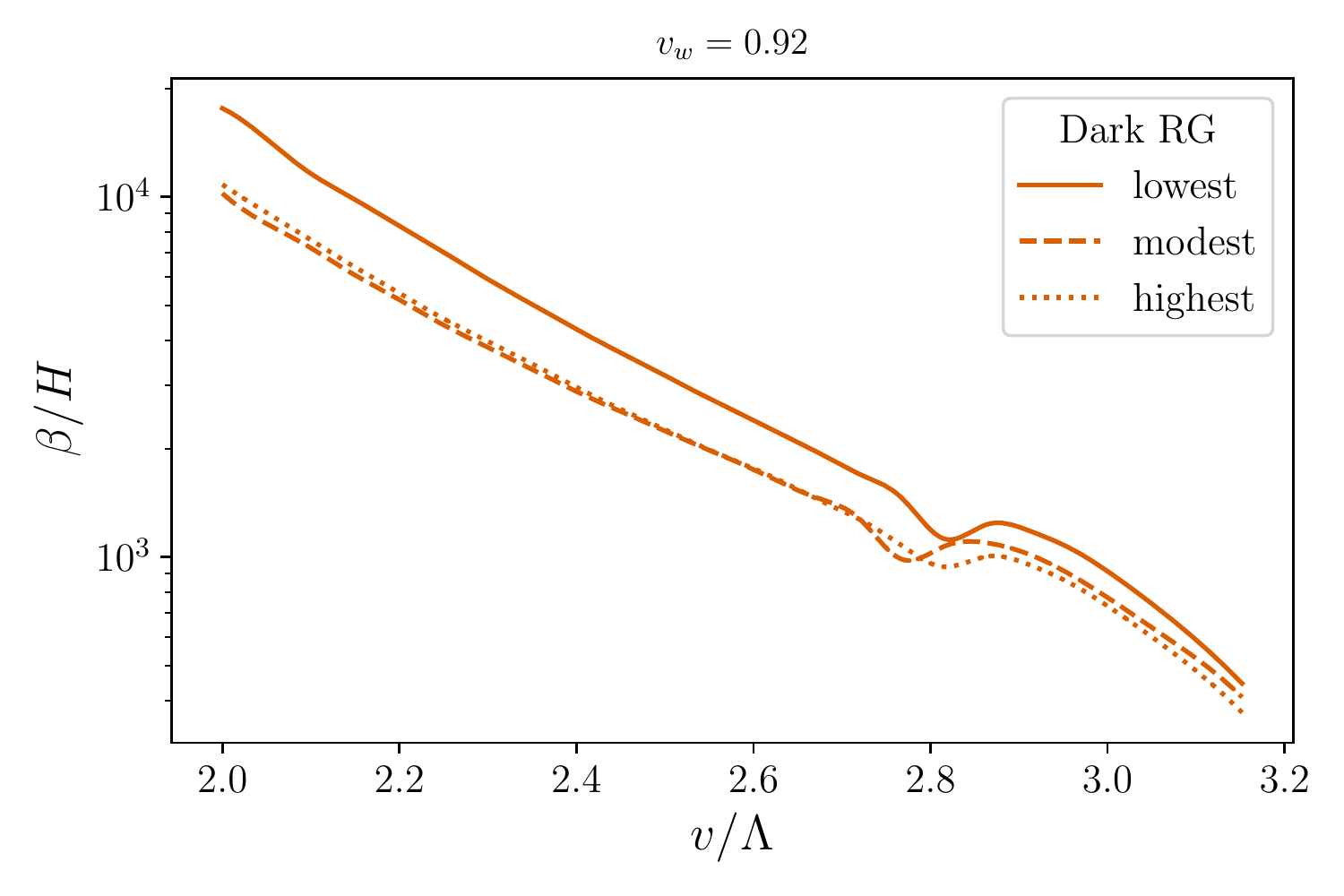}
        \includegraphics[width=0.49\textwidth]{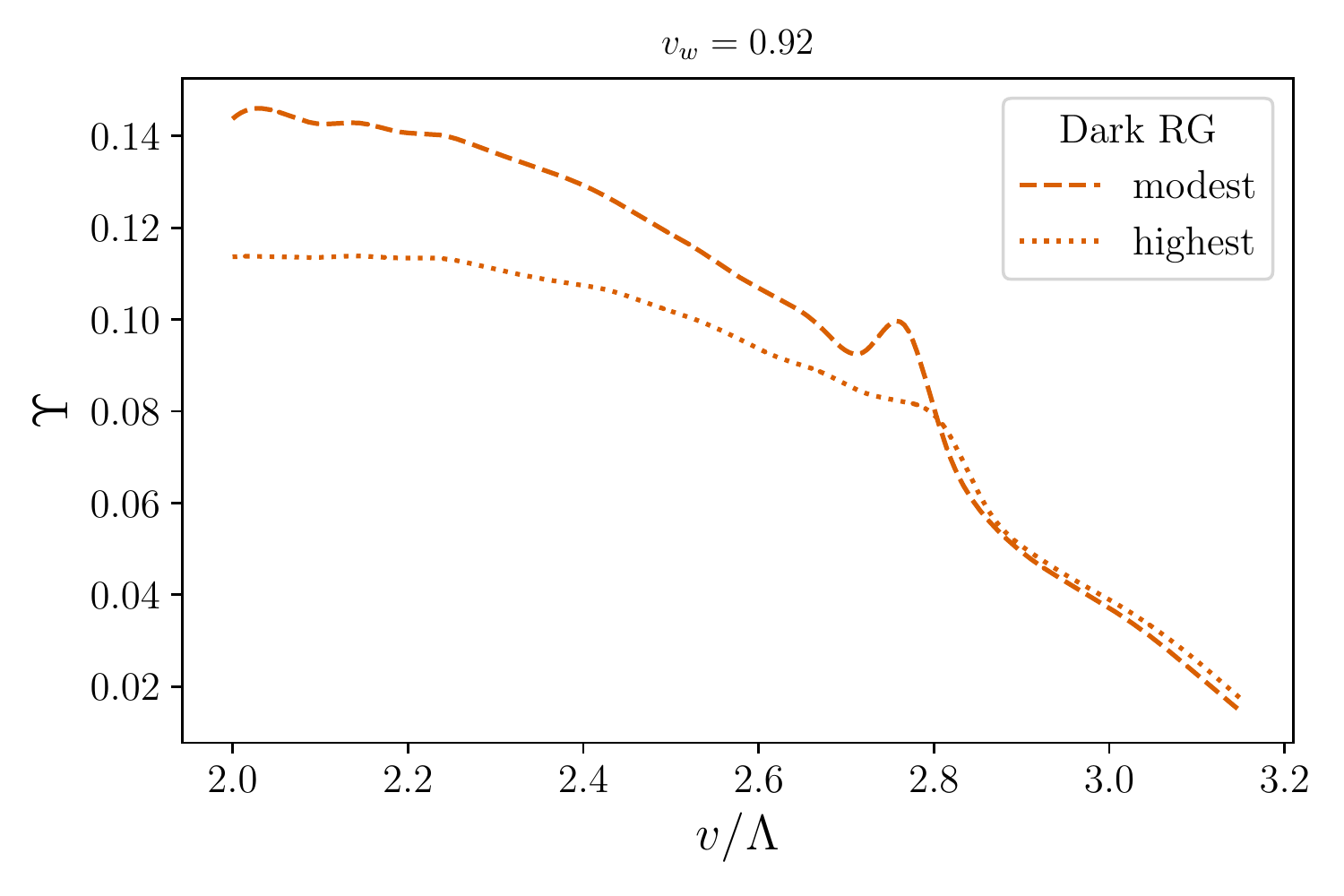}
 	\caption{\textbf{Dark RG}: The inverse time duration of the phase transition at the different levels of diligence using Eq.~[\ref{eq:betaHn}, \ref{eq:betaHp}, \ref{eq:beta_Tf}]. The lowest and modest diligences are estimated using the first derivative of the action $dS/dT$ and the highest diligence is computed directly from the mean bubble separation.  The right panel shows the suppression factor due to the lifetime of the source using Eq.~\ref{eq:suppression-ellis} and Eq.~\ref{eq:suppression-upsilon} for modest and highest diligence respectively.  The lowest diligence corresponds to $\Upsilon \rightarrow 1$.  The temperatures are set to $T_n$ (\ref{eq:Tn}), $T_p$ (\ref{eq:LISA_Tp}), and $T_f$ (\ref{eq:Tf}) for the lowest, modest, and highest diligence respectively.}
	\label{fig:darkRG-beta-upsilon}
\end{figure}
The inverse duration of the phase transition is plotted in the left panel of Fig.~\ref{fig:darkRG-beta-upsilon} for detonation. The lowest diligence calculated using Eq.~\ref{eq:betaHn} consistently over approximates $\beta/H$ while modest diligence calculated using Eq.~\ref{eq:betaHp} agrees well with the highest diligence found from the mean bubble separation.  There were  no minima found in the action for any of the parameters in consideration so the difference between $T_p$ calculated using Eq.~\ref{eq:LISA_Tp} in the modest diligence and $T_f$ calculated using Eq.~\ref{eq:Tf} in the highest diligence is only minor. The dips near $v/\Lambda > 2.8$ in $\beta/H$ are the result of the shape of $S(T)/T$ which causes the highest error between the modest and highest diligence. This dip also effects the suppression factor due to the lifetime of the source as seen in the right panel of Fig.~\ref{fig:darkRG-beta-upsilon}. The modest diligence over-approximates the suppression factor up until $v/\Lambda \sim 2.8$ where they eventually become approximately equal in magnitude.   The large $v/\Lambda$ regime has a small $\beta/H$ and large $\alpha$ which results in a small lifetime of the source $\tau_{\rm sw} $. The modest diligence suppression factor is then an appropriate approximation in this regime. 

\begin{figure}
	\centering
        \includegraphics[width=0.49\textwidth]{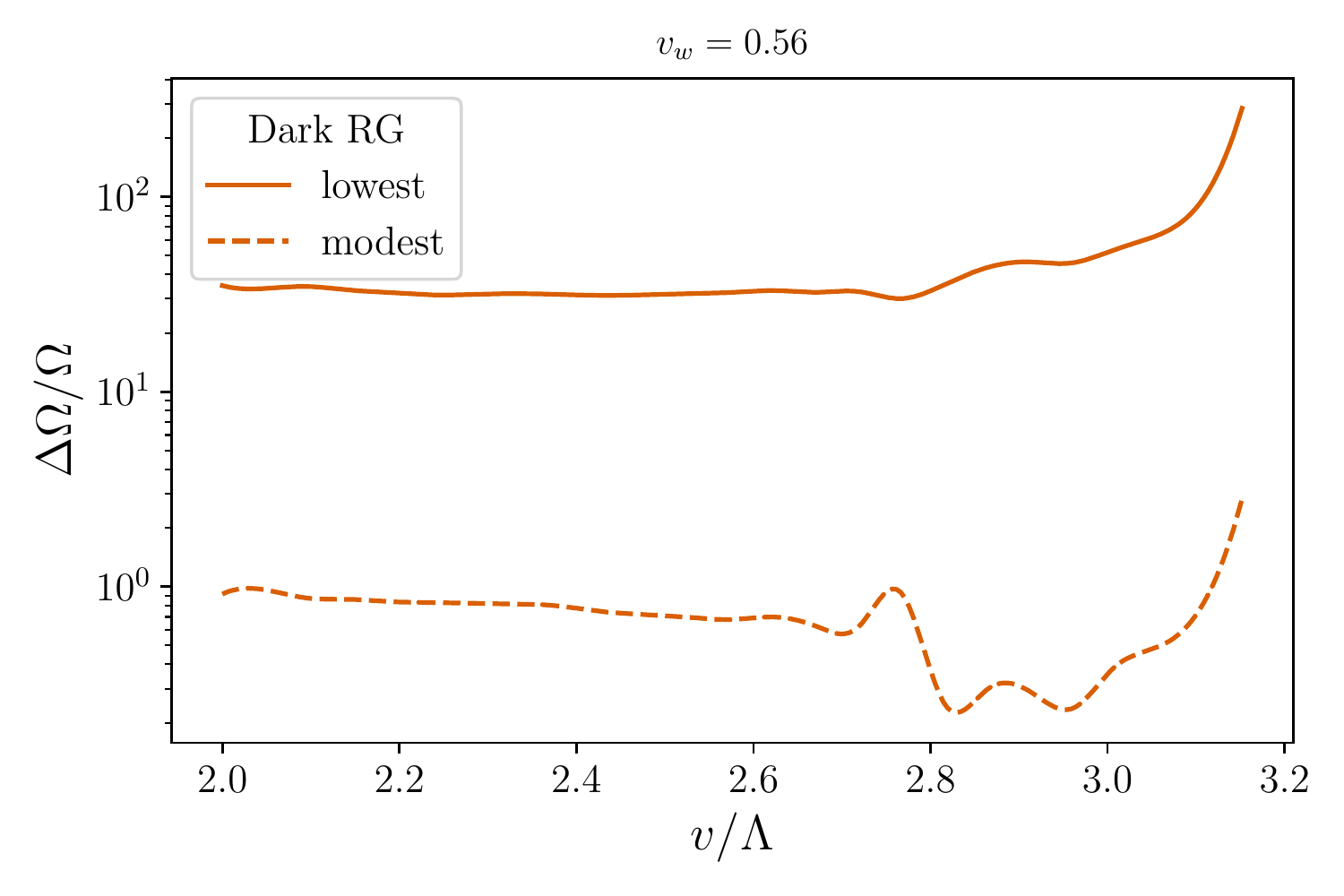}
        \includegraphics[width=0.49\textwidth]{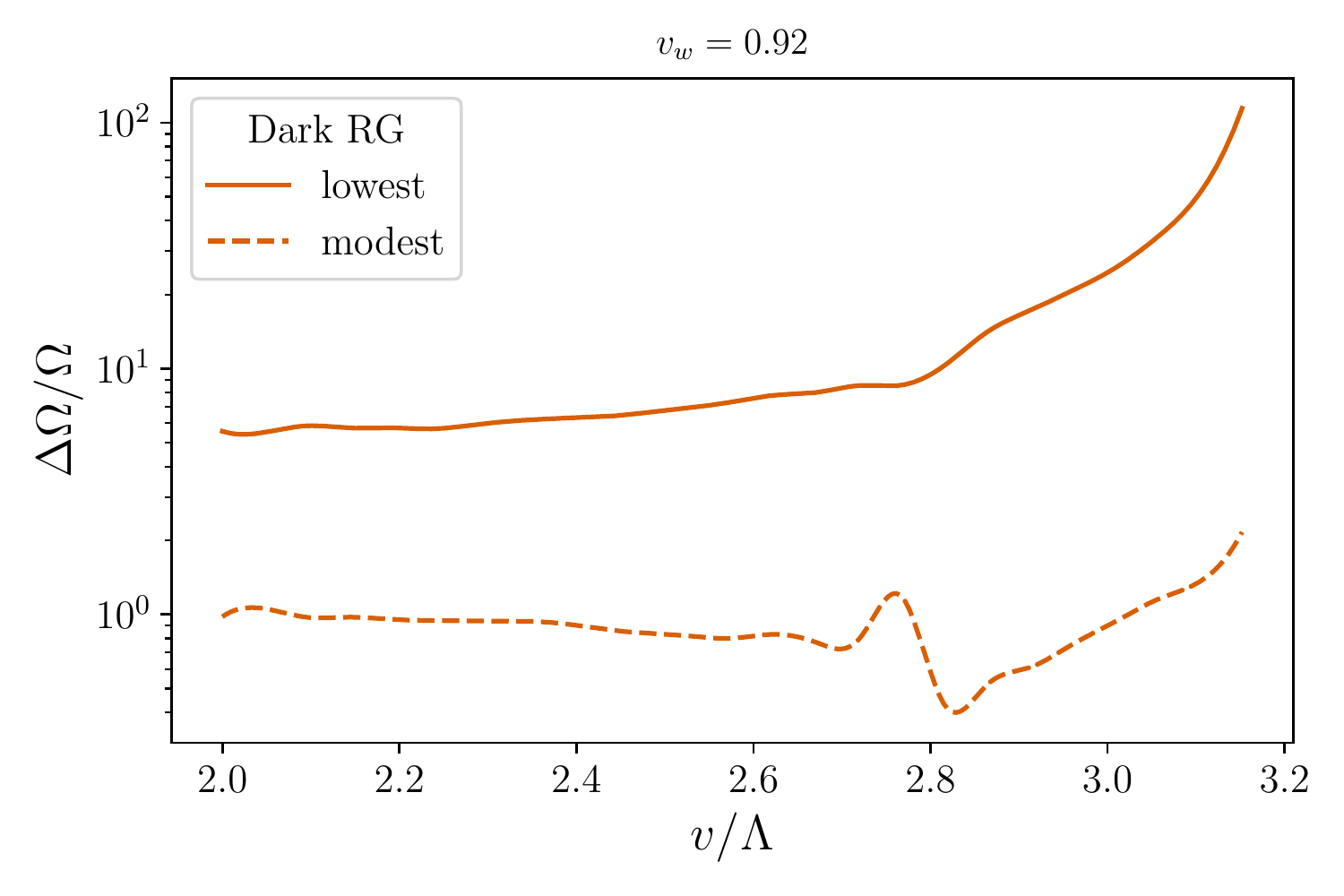}
	\caption{\textbf{Dark RG}: The relative error when using the  lowest and modest levels of diligence, compared to the highest level of diligence (for which $\Delta \Omega/\Omega = 0$). The vertical axis shows the peak (frequency-independent) gravitational wave energy density for detonation. The precise definition of $\Delta \Omega/\Omega$ is given in Eq.~\ref{eq:error}. The horizontal axis corresponds to the ratio of the tree level v.e.v to the cut off scale $v/\Lambda$. 
	$\Delta \Omega/\Omega$ is displayed for deflagration and detonation at  different levels of diligence using Eq.~[\ref{eq:omega1}, \ref{eq:omega-suppression-ellis}, \ref{eq:omega_suppressed}] and Eq.~\ref{eq:error}. The temperatures are set to $T_n$ (\ref{eq:Tn}), $T_p$ (\ref{eq:LISA_Tp}), and $T_f$ (\ref{eq:Tf}) for the lowest, modest, and highest diligence respectively. Both the modest and highest diligence contains suppression factors due to the lifetime of the source. The highest diligence contains the suppression factor due to vorticity effects in the plasma.}
	\label{fig:darkRG-deltaomega}
\end{figure}
The error in the gravitational wave spectrum is shown in Fig.~\ref{fig:darkRG-deltaomega} for deflagration and detonation.  For both the lowest and modest diligence, the error remains roughly constant until $v/\Lambda \sim 2.8$ where it exhibits some oscillations.  This behavior is related to the dips in Fig.~\ref{fig:darkRG-beta-upsilon}.  Past $v/\Lambda \sim 2.8$, both the lowest and modest diligence begin to increase. The error in the lowest diligence past this point is dominated by the lack of suppression factor due to the lifetime of the source. The suppression factor remains roughly constant until $v/\Lambda \sim 2.8$ where it begins to approach zero and as a result increases the error. The increasing behavior in the modest diligence is likely due to the separation in $\beta/H$ between the modest and highest diligence in Fig.~\ref{fig:darkRG-beta-upsilon} and the suppression factor from vorticity and reheating effects in the plasma which are stronger for larger $\alpha$.  The values of the speed of sound in the symmetric and broken phase calculated at $T_f$ are not far from the bag model of $c_s^2  = 1/3$ and we do not consider it a strong source of error in the peak gravitational wave energy density spectrum. Overall we notice an error in the peak gravitational wave energy density of $10^1 - 10^3$ for lowest diligence and $10^{-1} - 10^1$ for modest diligence. 

\subsection{xSM}
\begin{figure}
    \centering
    \includegraphics[width=0.49\textwidth]{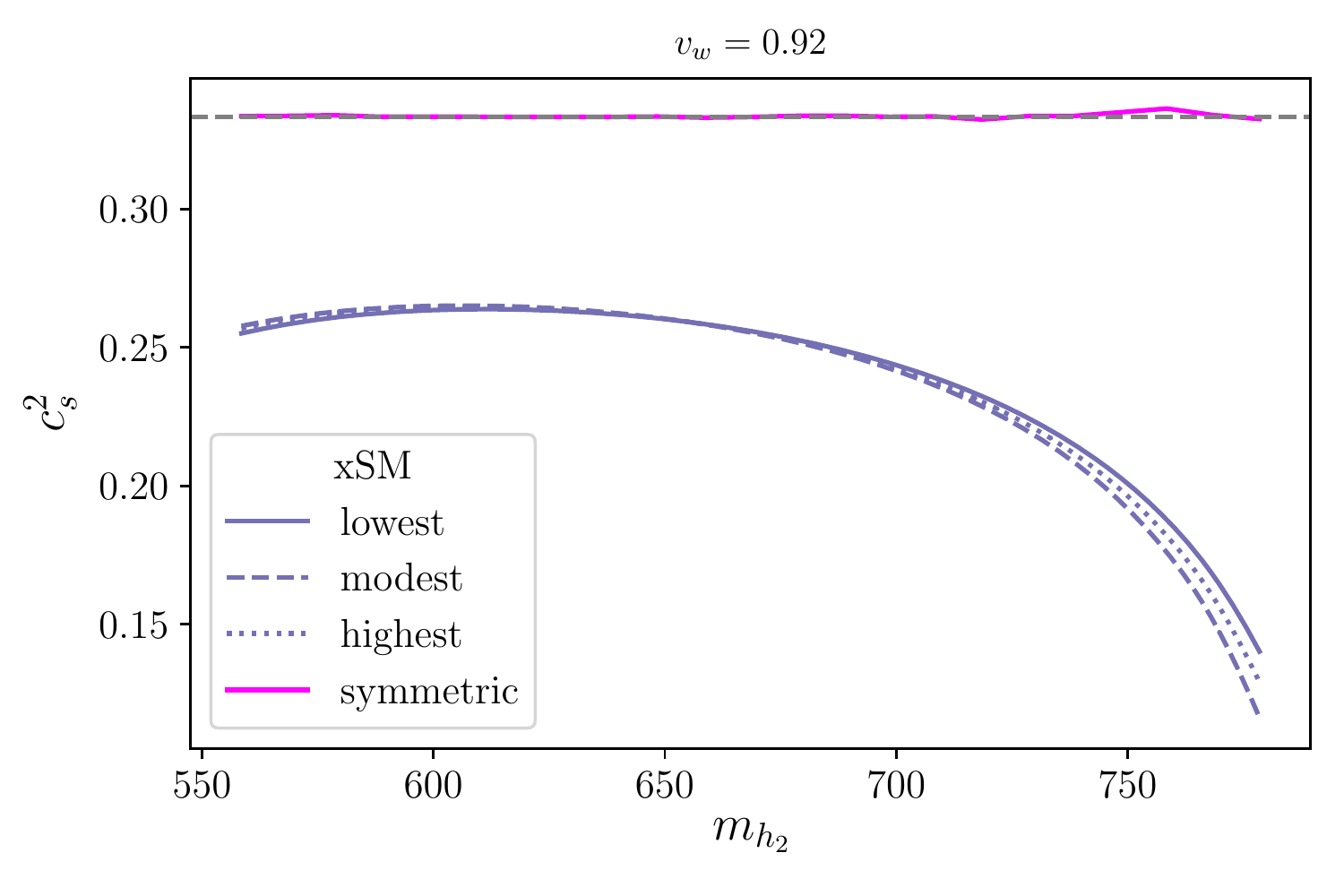}
    \includegraphics[width=0.49\textwidth]{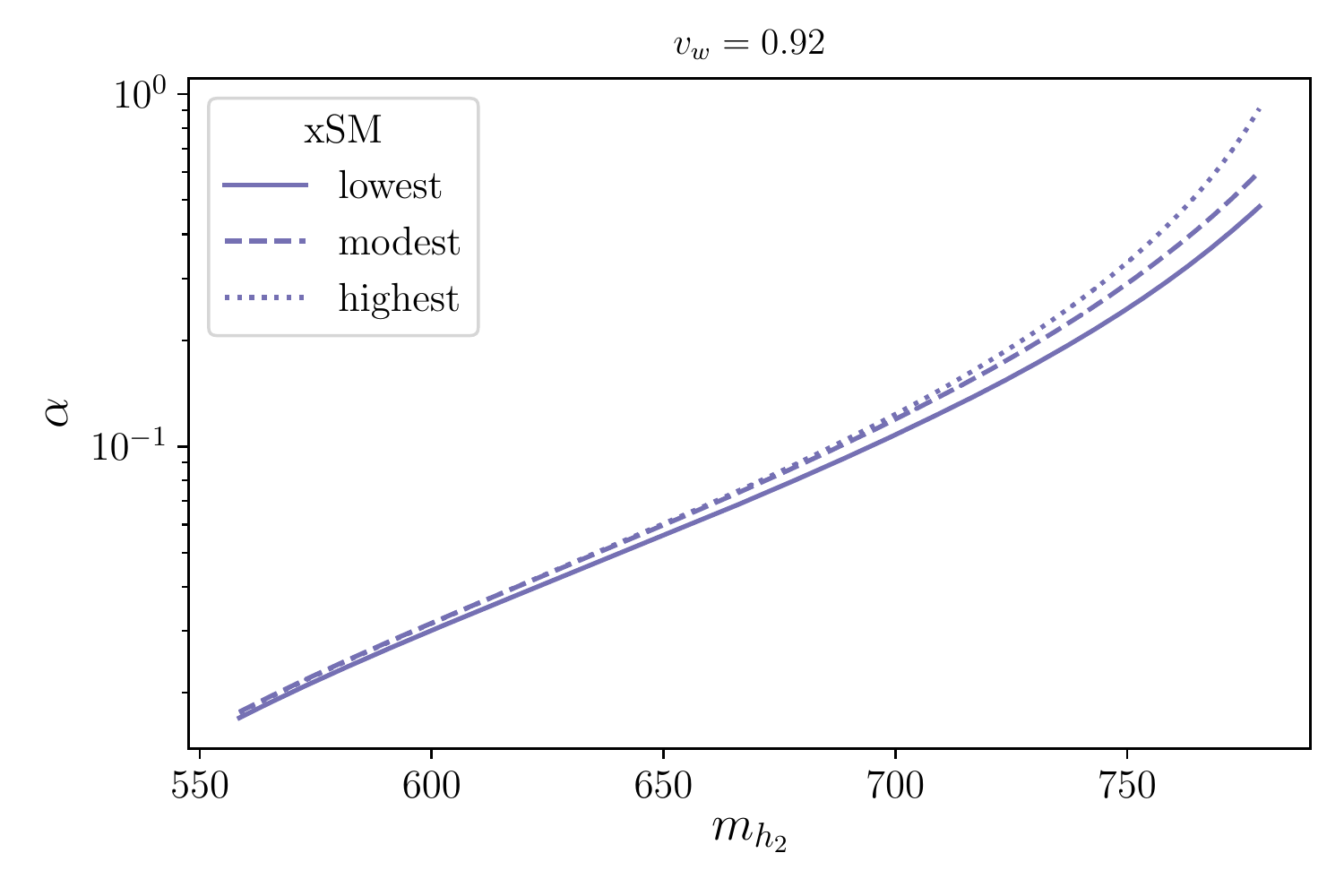}
    \includegraphics[width=0.49\textwidth]{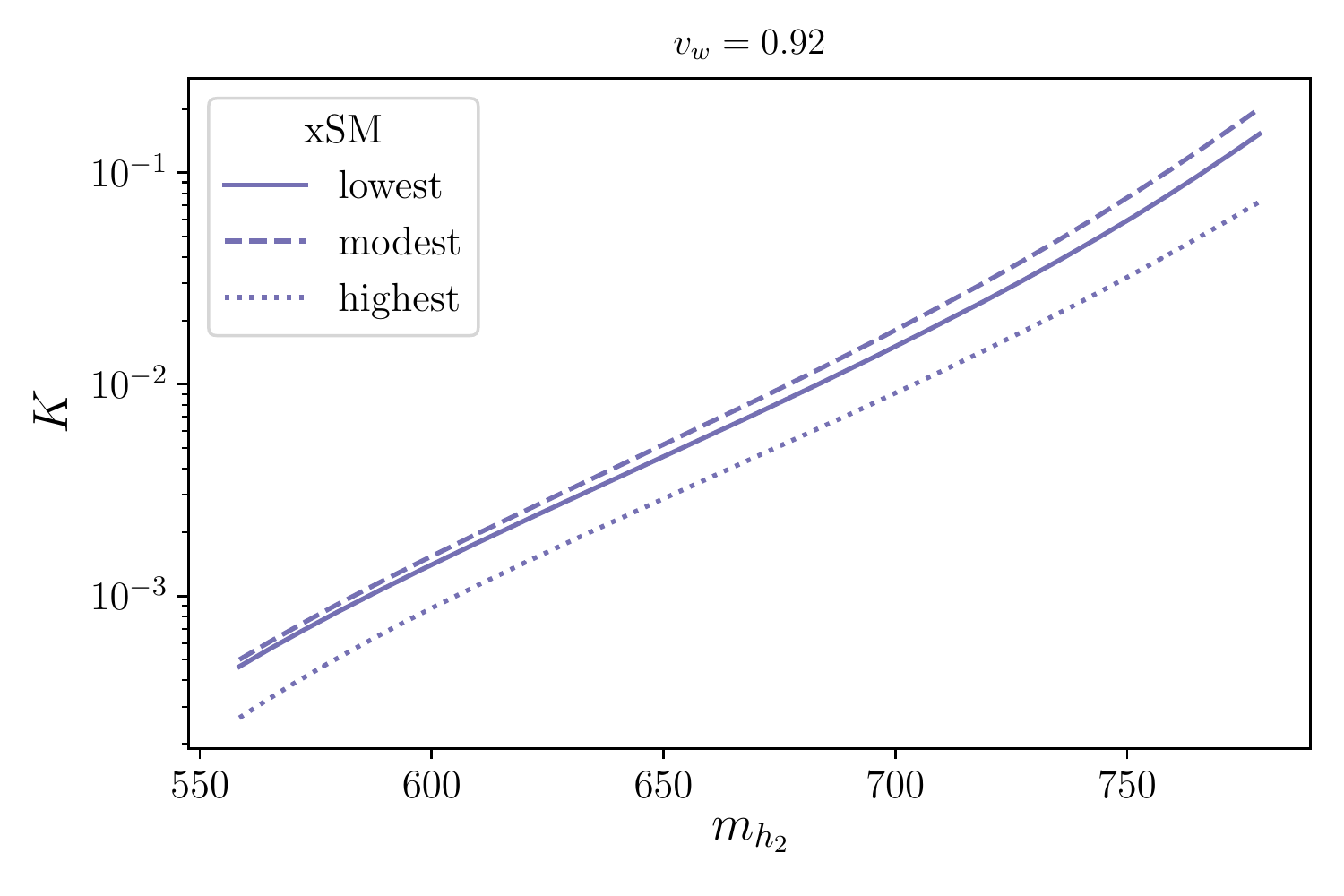}
    \caption{\textbf{xSM}: The top left panel shows the speed of sound calculated in the symmetric and broken phase using Eq.~\ref{eq:cs2} at the different levels of diligence. The gray dashed line corresponds to the bag model with $c_s^2 = 1/3$. The symmetric phase (solid magenta) is only shown at highest diligence. The top right panel shows the strength of the phase transition at the different levels of diligence using Eq.~[\ref{eq:alpha_Tn}, \ref{eq:alpha_Tp}, \ref{eq:alpha_psuedotrace}]. The bottom panel shows the kinetic energy fraction at the different levels of diligence where the lowest and modest diligence use fits for $\kappa$ to get $K$ and the highest diligence uses Eq.~\ref{eq:K_beyond_bag}. The temperatures are set to $T_n$ (\ref{eq:Tn}), $T_p$ (\ref{eq:LISA_Tp}), and $T_f$ (\ref{eq:Tf}) for the lowest, modest, and highest diligence respectively. The numerical calculation of the speed of sound only enters in the highest diligence of $\alpha$. }
    \label{fig:xsm-cs2-alpha-K}
\end{figure}

We show in the top left panel of Fig.~\ref{fig:xsm-cs2-alpha-K} the speed of sound in the symmetric and broken phase for a scan over the heavy singlet mass in the xSM model while holding all other parameters constant. The speed of sound in the symmetric phase is approximately $c_s^2 = 1/3$ as in the bag model.  The speed of sound in broken phase deviates far from the bag model where it approaches zero as $m_{h_2} \rightarrow 0$.  The speed of is strongly correlated with the cubic term that arises from the extra scalar who also acquires a tree level vacuum expectation value. The speed of sound can then be suppressed by increasing the $b_3$ parameter. 
This strong suppression in the broken phase speed of sound will lead to an amplification in the strength of the phase transition as seen in the top right panel of Fig.~\ref{fig:xsm-cs2-alpha-K}. The strength of the phase transition in the highest diligence grows larger compared to the other levels as the singlet gets heavier. This is directly related to the suppression in the speed of sound in the broken phase. There is a minor difference in the lower singlet mass range.  The kinetic energy fraction is shown in the bottom panel of Fig.~\ref{fig:xsm-cs2-alpha-K}. The lowest and modest diligence both overestimate $K$ for the entire range of the parameter space which is not observed in $\alpha$.  This can be attributed to the approximations used in the kinetic energy fraction pre-factor $ \alpha/(1 + \alpha) $ used in the peak gravitational wave energy density in  Eq.~\ref{eq:omega1} and the speed of sound dependence in solving the beyond the bag model hydrodynamic equations. 
\begin{figure}
    \centering
    \includegraphics[width=0.6\textwidth]{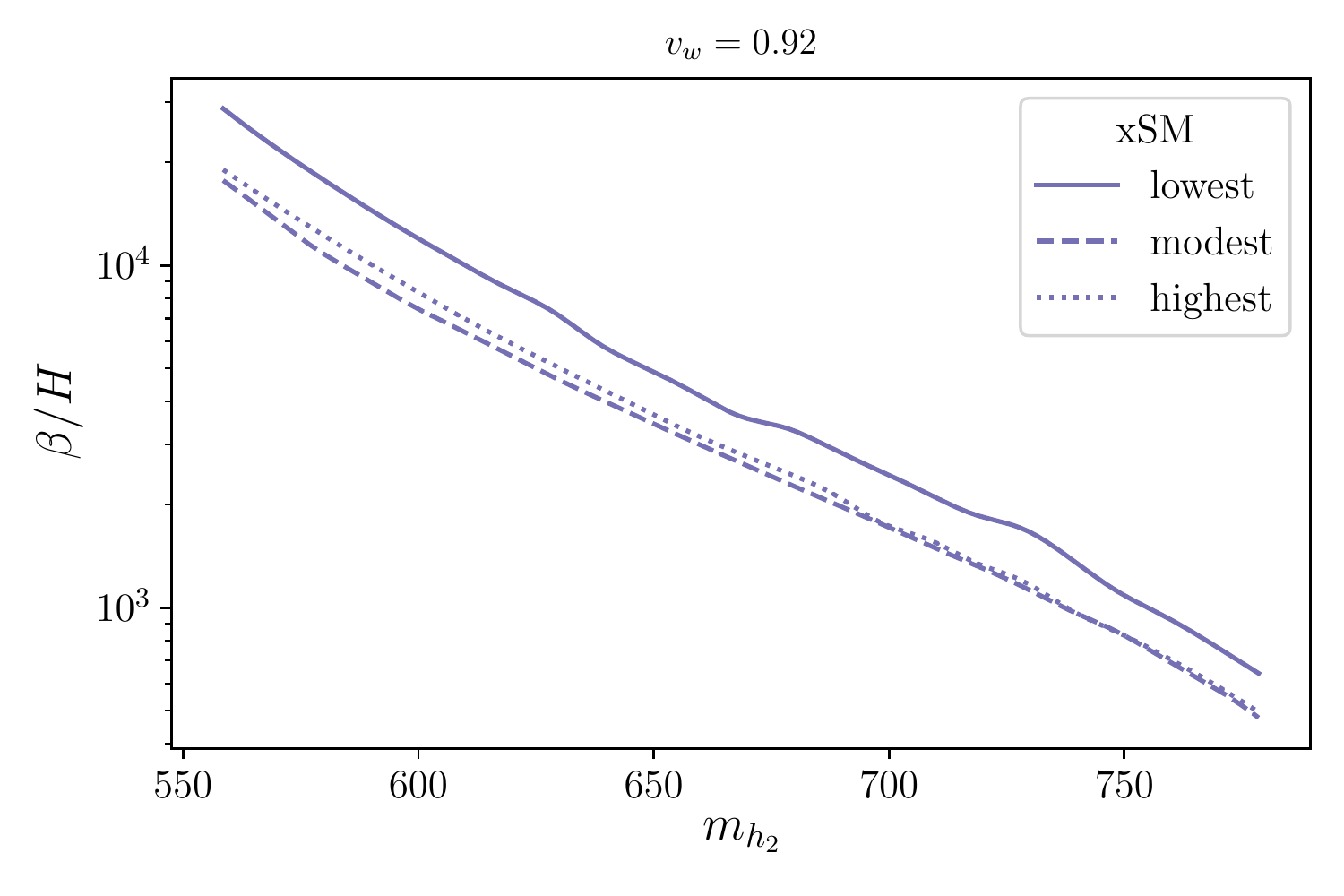}

    \caption{\textbf{xSM}: The inverse time duration of the phase transition at the different levels of diligence using Eq.~[\ref{eq:betaHn}, \ref{eq:betaHp}, \ref{eq:beta_Tf}]. The lowest and modest diligences are estimated using the first derivative of the action $dS/dT$ and the highest diligence is computed directly from the mean bubble separation.  The temperatures are set to $T_n$ (\ref{eq:Tn}), $T_p$ (\ref{eq:LISA_Tp}), and $T_f$ (\ref{eq:Tf}) for the lowest, modest, and highest diligence respectively.}
    \label{fig:xsm-beta}
\end{figure}

The inverse time duration of the phase transition is plotted in  Fig.~\ref{fig:xsm-beta} for the different levels of diligence.  The modest diligence is a better approximation than that of the first level for $\beta/H$ but slightly under-approximates the spectrum for lower mass ranges. The lowest diligence is a poor approximation for $\beta/H$ for the entire parameter space.

\begin{figure}
    \centering
    \includegraphics[width=0.49\textwidth]{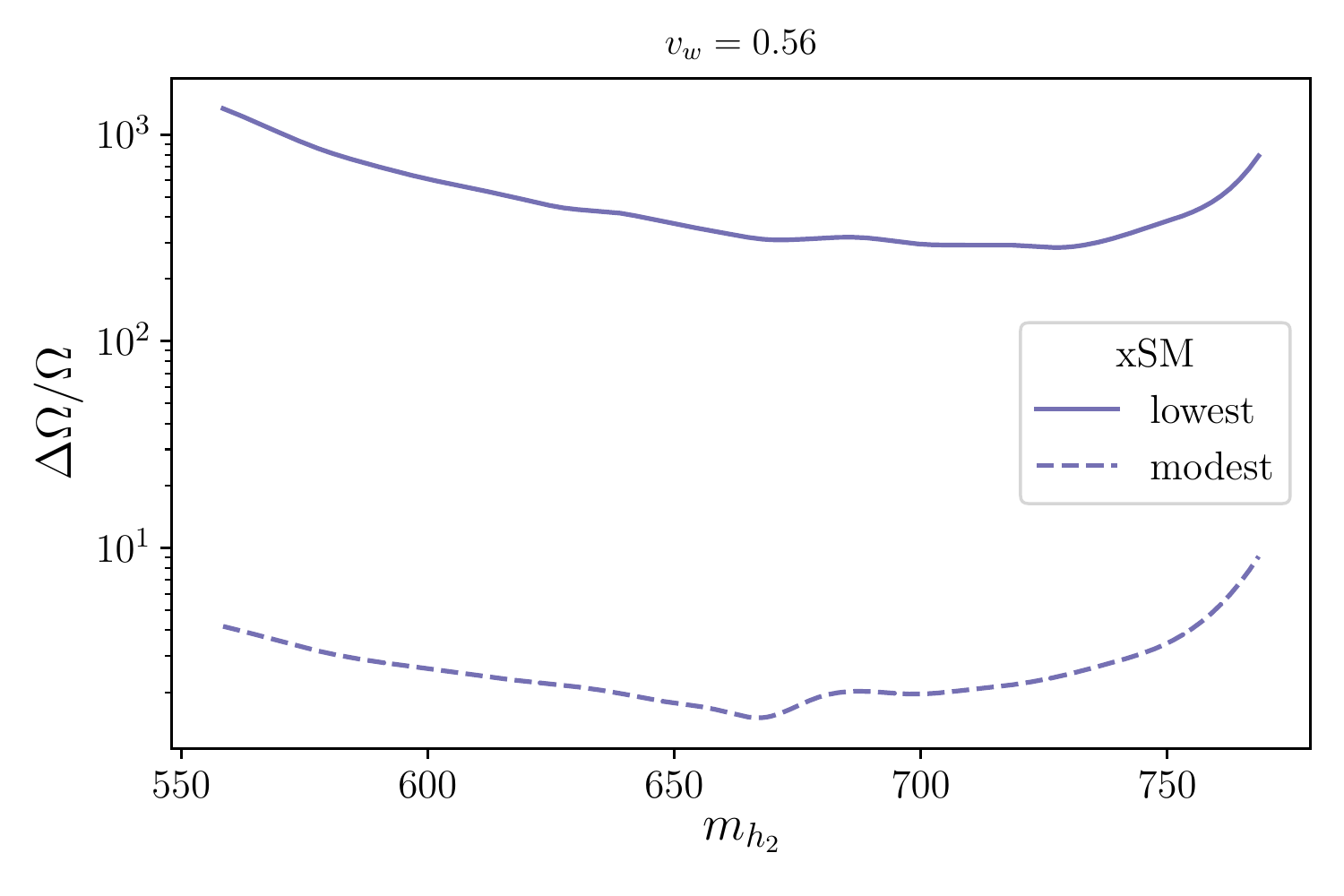}
    \includegraphics[width=0.49\textwidth]{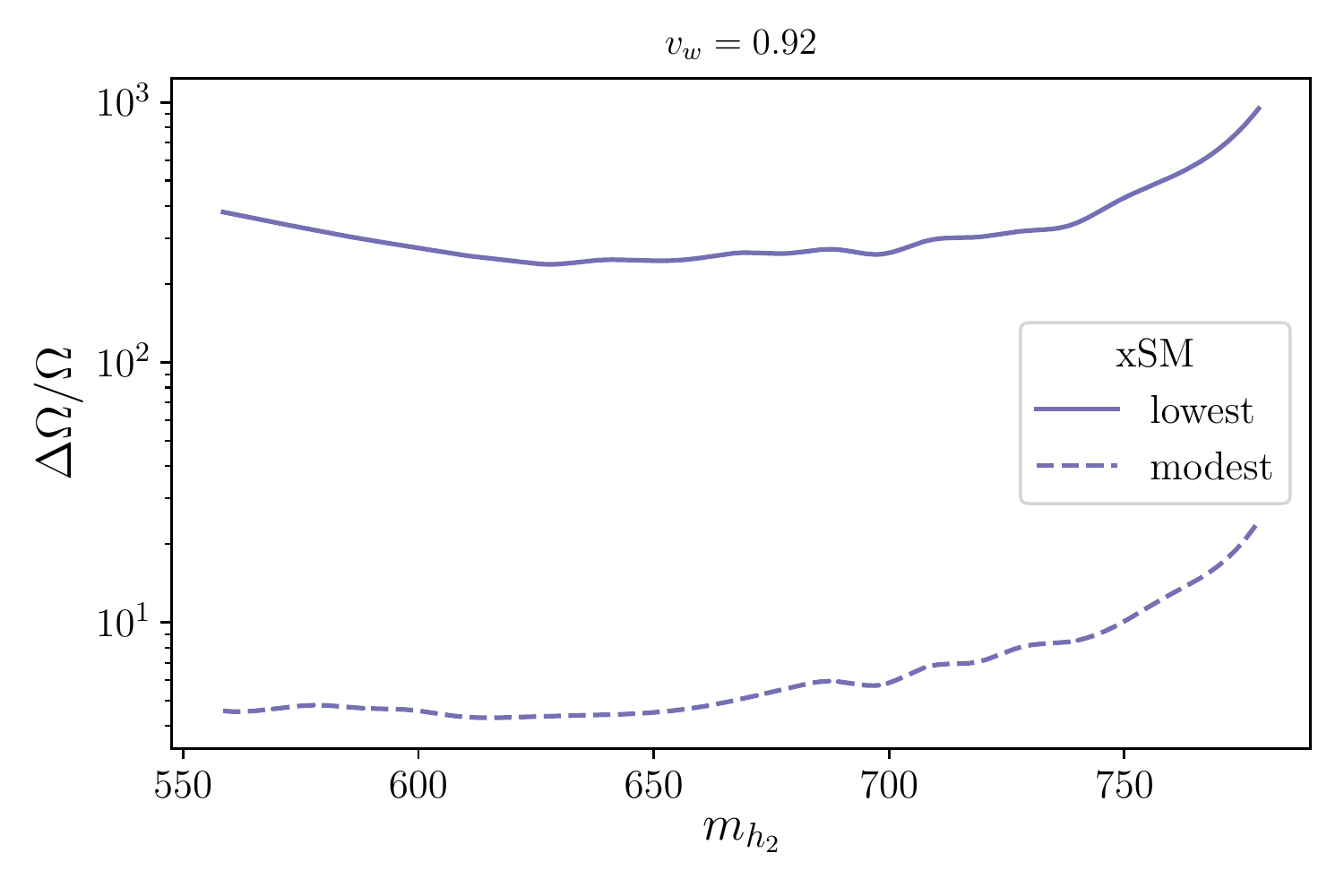}
	\caption{\textbf{Dark RG}: The relative error when using the  lowest and modest levels of diligence, compared to the highest level of diligence (for which $\Delta \Omega/\Omega = 0$). The vertical axis shows the peak (frequency-independent) gravitational wave energy density for detonation. The precise definition of $\Delta \Omega/\Omega$ is given in Eq.~\ref{eq:error}. The horizontal axis corresponds to the heavy singlet mass $m_{h_2}$. 
	$\Delta \Omega/\Omega$ is displayed for deflagration and detonation at  different levels of diligence using Eq.~[\ref{eq:omega1}, \ref{eq:omega-suppression-ellis}, \ref{eq:omega_suppressed}] and Eq.~\ref{eq:error}. The temperatures are set to $T_n$ (\ref{eq:Tn}), $T_p$ (\ref{eq:LISA_Tp}), and $T_f$ (\ref{eq:Tf}) for the lowest, modest, and highest diligence respectively. Both the modest and highest diligence contains suppression factors due to the lifetime of the source. The highest diligence contains the suppression factor due to vorticity effects in the plasma.}
    \label{fig:xsm-deltaomega}
\end{figure}

The error in the gravitational wave spectrum compared to the highest diligence for deflagration and detonation is given in Fig.~\ref{fig:xsm-deltaomega}. The largest error in the spectrum occurs for the lowest diligence and this is due to the lack of suppression factor for the finite lifetime of the source and the larger uncertainty in $\beta/H$.  The  suppression factor for the modest diligence case is an under-approximation to the finite lifetime of the source particularly in the higher singlet mass regions.   Both the lowest and modest diligence receive significant errors from neglecting the beyond the bag contributions to the kinetic energy which over estimates the peak spectrum which also  gets worse for higher singlet masses. Overall the  range of error in the peak gravitational wave energy density is between $[10^{2} \sim 10^{3}]$  and $[10^{0} \sim 10^{2}]$ for the different levels of diligence. All of the  points above the range in $m_{h_2}$ shown are certainly viable points and may even reach higher levels of error. However, this range in $m_{h_2}$ is chosen such that all the points remain in either deflagration or detonation for both consistency and the lack of numerical simulations for hybrids.

\subsection{Mean Bubble Separation vs Inverse Time Duration}
The gravitational wave energy density is dependent on determining the mean bubble separation when the phase transition ends at temperature $T_f$.  An approximation to the mean bubble separation can be determined by calculating the inverse time duration, $\beta/H$, from the first derivative of the action.  This calculation is typically only valid when there is a negligible barrier at zero temperature. However, if there is a barrier at tree level, a minimum in the action will develop near $T_f$ and the second derivative, $\beta_2$ will become relevant while the first derivative will vanish.   The bubble nucleation rate can then take on the form
\begin{equation}
	p = p_0 \exp \left[ -S_*- \frac{1}{2} \beta_2^2 (t - t_*)^2 \right],
\end{equation}
where $t_*$  is the time when the temperature  is near $T_f$ and $S_* = S_3(T_*)/T_*$. The above result will ultimately lead to a new relation between the mean bubble separation $R_*$ and the inverse time duration of the phase transition $\beta$.   

This subtlety is not usually taken into account and the relation between $R_*$ and $\beta$ that is useful for gravitational wave calculations is simply given by the  approximate formula
\begin{equation}
	 H R_* = (8 \pi)^{1/3} v_w \left( \frac{H_*}{\beta} \right), \: {\rm with} \quad \beta = HT dS/dT ,
	 \label{eq:HRvsBeta_H}
\end{equation}  
where $\beta$ is related to the first derivative of the action regardless of the presence of a barrier. Out of the models we consider, SMEFT and xSM can acquire tree level barriers that result in a minimum in the action. The lowest and modest diligence results presented here assume Eq.~\ref{eq:HRvsBeta_H} always hold which can result in significant errors for these two models. Furthermore, the percolation temperature at which $\beta/H$ is estimated is a function of $\beta/H$, Eq.~\ref{eq:betaHp}, which can also acquire error if the barrier is not sufficiently taken care of.  The highest diligence results can side-step these issues by numerically calculating the mean bubble separation from the number bubble density which is independent of any assumptions about the curvature of the action, i.e 
\begin{equation}
	H R_*  = \left( \frac{n_b}{H^3}\right)^{-1/3},
	\label{eq:HRvsnb}
\end{equation}    
evaluated at the final temperature $T_f$. The final temperature as well does not depend on any underlying assumptions about the curvature of the action because it is numerically calculated from the false vacuum fraction.  For comparisons between the inverse time durations with respect to the highest diligence,  we first calculate $H R_*$ and use Eq.~\ref{eq:HRvsBeta_H} to determine an effective $\beta/H$.    
\begin{figure}
    \centering
    \includegraphics[width=0.49\textwidth]{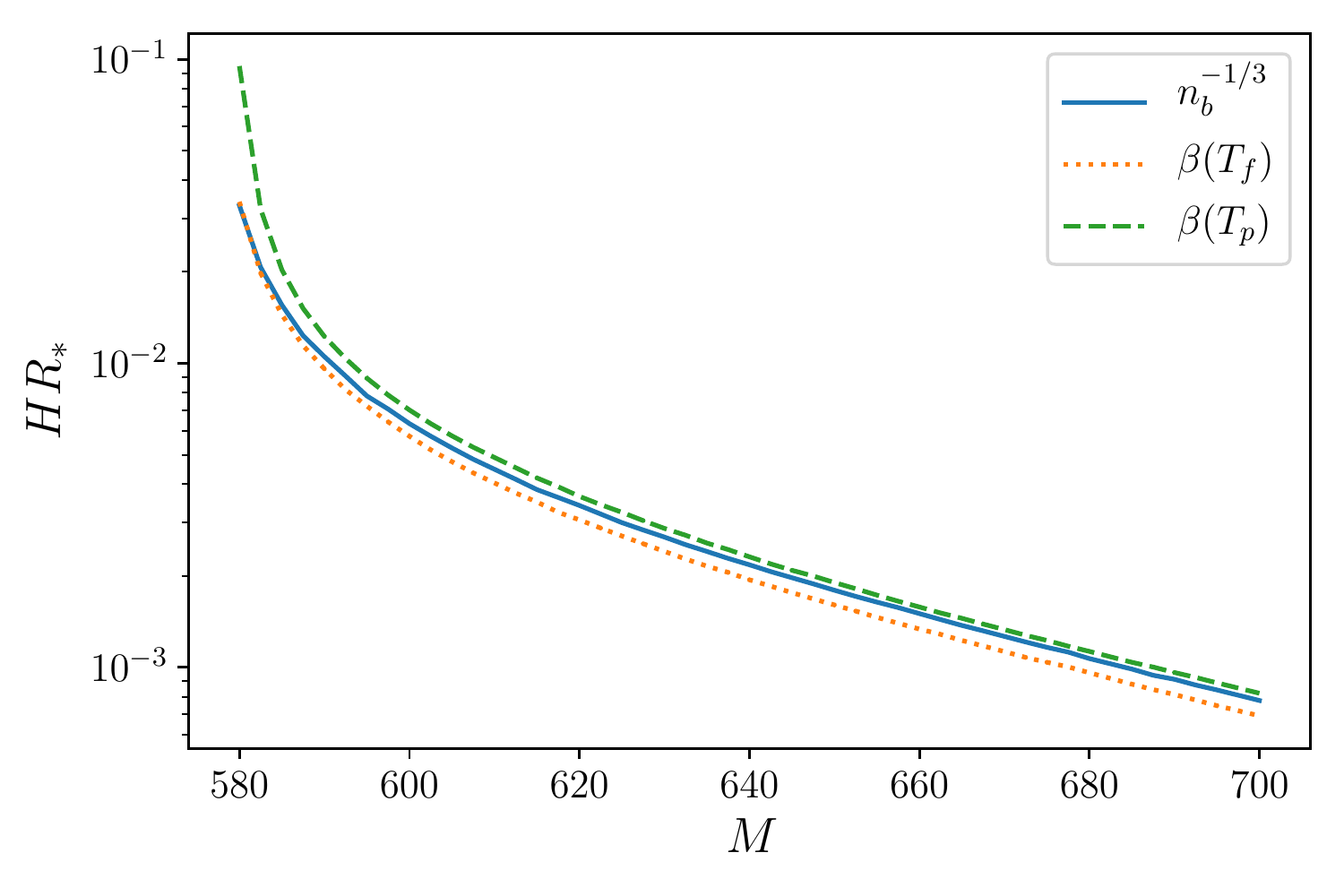}
    \includegraphics[width=0.49\textwidth]{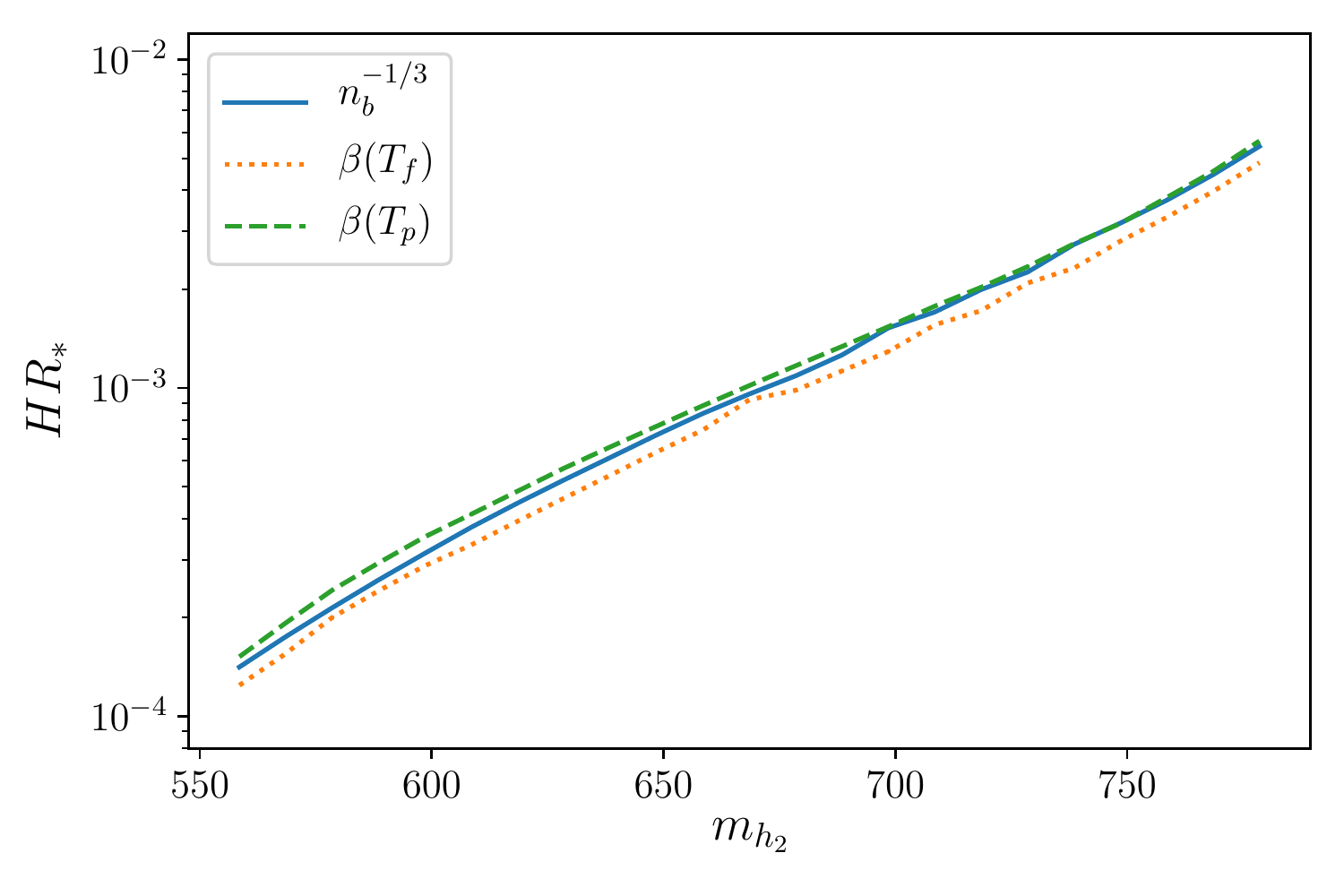}
    \caption{The mean bubble separation times the Hubble parameter for SMEFT (\textbf{left} ) and xSM (\textbf{right}). The solid line corresponds to the numerically calculated value defined in Eq.~\ref{eq:HRvsnb} evaluated at $T_f$ (\ref{eq:Tf}). The dashed and dotted lines correspond to the estimated value using Eq.~\ref{eq:HRvsBeta_H} evaluated at $T_f$ (\ref{eq:Tf}) and $T_p$(\ref{eq:LISA_Tp}) respectively.}
    \label{fig:SMEFT-xSM-HR}
\end{figure}

The comparison between $H R_*$ using Eq.~\ref{eq:HRvsBeta_H} and Eq.~\ref{eq:HRvsnb} is shown in Fig.~\ref{fig:SMEFT-xSM-HR} where the left figure corresponds to SMEFT and right figure corresponds to xSM. The solid lines represent the proper mean bubble separation calculated at $T_f$. The dotted and dashed lines correspond to the mean bubble separation calculated first from $\beta$ at $T_f$ and $T_p$ respectively.  We denote $T_p$ to refer to the estimation given in Eq.~\ref{eq:LISA_Tp}. Below $M = 600$ GeV in SMEFT, the action acquires a minimum as a result of the tree level barrier at zero temperature which causes $\beta(T_p)$ to significantly over-approximate $HR_*$. As mentioned previously, this is a result of the underlying assumptions in approximating both $T_p$ and $\beta$ which ignore the barrier.  The mean bubble separation calculated from $\beta(T_f)$ performs better than $\beta(T_p)$ in this regime with nearly identical $HR_*$ predictions compared to $n_b^{-1/3}$. This is largely due to $T_f$ being independent of any assumptions on the action. 

The xSM model consists of a second scalar and several parameters which when varied may induce either first step or second step phase transitions.  The bench marks chosen involve scanning of the heavy singlet mass while holding the other model parameters fixed. All of the points resulted in a one step phase transition along with no minimum in the action.  On the right of Fig.~\ref{fig:SMEFT-xSM-HR}, we see that all three methods resulted in a roughly consistent approximation of $HR_*$ with slightly better performance from $\beta(T_p)$ for large $m_{h_2}$. This can be attributed to the lack of minimum in the action observed in the parameter space.  We reserve a further analysis of the mean bubble separation in xSM for future work.  

\section{Summary of Results}
The previous results involved fixing certain characteristics associated with each of the outlined levels of diligence.  In this section, we will fix all of the quantities as high diligence while varying the level of a single quantity to determine its impact on the error of $\Delta \Omega/\Omega$. Table \ref{tab:Error} shows the range of error we observe associated with varying the level of diligence in the calculation of the transition temperature, mean bubble separation, fluid velocity, finite lifetime of the source, and vorticity effects.  The base level of comparison will use $\Omega_{\rm GW}$ calculated using Eq.~\ref{eq:omega_suppressed} which assumes the transition temperature is at $T_f$ and includes beyond the bag effects and the suppression factors due to the finite lifetime of the source and reheating effects in the plasma. We will now proceed to describe how the range of errors are calculated. 
\begin{table}[h]
    \centering
    \begin{tabular}{l|c|c|c}
    Effect     & Range of error (medium) &Range of error (low) & Type of error  \\ \hline 
    Transition temperature     & ${\cal O}(10^{-4} - 10^{1})$ & ${\cal O}(10^{-1} - 10^0)$ & Random \\
    Mean bubble separation & ${\cal O} (0 - 10^{-1})$ & ${\cal O}(10^{-1} - 10^0)$& Suppression \\
    Fluid velocity & ${\cal O}(10^{-2} - 10^{0})$ & ${\cal O}(10^{-2} - 10^{0})$ & Random \\
    Finite lifetime & ${\cal O}(10^{-3} - 10^{-1})$ & ${\cal O}(10^1 - 10^3)$ & Enhancement \\
    Vorticity effects & ${\cal O}(10^{-1}-10^{0})$ & $-$ & Random
    \end{tabular}
    \caption{Full range of error of $\Delta \Omega /\Omega $ for each individual effect comparing the medium diligence and low diligence approaches to the high diligence approach.}
    \label{tab:Error}
\end{table}

The transition temperatures used for the different levels were $T_n$ (\ref{eq:Tn}) and $T_p$ (\ref{eq:LISA_Tp}. The frequency independent $\Omega_{\rm GW}$ is now calculated at high diligence using Eq.~\ref{eq:omega_suppressed} for both the lowest and modest diligence. This is to show how $\Omega_{\rm GW}$ can change purely by the temperature at which the transition is assumed to take place.  The lowest diligence will use $T_n$ to calculate $\Delta \Omega/\Omega$ while the modest diligence will use $T_p$.  The base level comparison is $\Omega_{\rm GW}$ at $T_f$ which corresponds to the previously defined high diligence. Varying the transition temperature leads to an error of $(10^{-1} - 10^{0})$ and $(10^{-4} - 10^{1})$ for lowest and modest diligence respectively.  The modest diligence can experience a larger error than the lowest diligence and this is due to the result of the strong super-cooling observed in SMEFT when $M \simeq 600$. The approximations used in calculating $T_p$ break down when a minimum develops in the action and results in the $10^1$ peak in the error for modest diligence. The error in the lowest diligence results in an enhancement in the spectrum which is attributed to $T_n > T_f$.  The modest diligence experienced both enhancement and suppression which is due to $T_p$ being much closer to $T_f$. The lowest diligence primarily had $\Omega_{\rm GW}^{\rm low} < \Omega_{\rm GW}^{\rm high}$ and modest diligence had $\Omega_{\rm GW}^{\rm med} > \Omega_{\rm GW}^{\rm high}$. For these reasons, we conclude that the type of error due to the transition temperature is random and dependent on the underlying model. 

The estimation of the error due to the mean bubble separation will involve calculating $R_* H$ from the $\beta/H$ at $T_n$ for the lowest diligence and $R_* H$ from $n_b$ at $T_f$ for modest diligence.  We use $T_f$ for the modest diligence in determining the relevant quantities in $\Omega_{\rm GW}$ in this case to minimize error which may arise from using the $T_p$ approximation. All quantities in $\Omega_{\rm GW}$ are calculated in high diligence at $T_n$ and $T_f$ for lowest and modest diligence respectively. The  lowest diligence exhibits the largest error with a range of $(10^{-1} - 10^{0})$ while modest diligence has the range $(10^{-3} - 10^{-1})$.  The error in modest diligence observed in the table is only due to the approximation of the mean bubble separation from the inverse time duration but it is expected to be higher if $T_p$ is used as opposed to $T_f$ which helps to correct the error. Both the lowest and modest diligence had mostly $\Omega_{\rm GW}^{\rm low,med} < \Omega_{\rm GW}^{\rm high}$ with modest diligence having a couple benchmarks with  $\Omega_{\rm GW}^{\rm med} > \Omega_{\rm GW}^{\rm high}$. We denote this type of error as predominately suppression. 

The error estimate from the fluid velocity involves comparing the fits for kappa given in Appendix \ref{sec:kappa_fits} to solving the hydrodynamic profiles numerically. The fluid velocity is related to the kappa through the kinetic energy fraction $K$ in Eq.~(\ref{eq:kappa_to_Uf_bag},\ref{eq:K_beyond_bag},\ref{eq:Uf_to_K_beyond_bag}). The lowest diligence calculates $\Omega_{\rm GW}$ at $T_f$ in the highest diligence with the replacement that $K$ and $U_f$ are now calculated using the fits to $\kappa$ and the bag calculation for $\alpha$. The modest diligence is the same as the lowest diligence except that $\kappa$ is calculated using the hydrodynamic profiles with $c_s^2 = 1/3$ in the bag model. The error associated with the different treatments of the fluid velocity is $(10^{-2} - 10^{0})$ for lowest diligence and $(10^{-3} - 10^{0})$ for modest diligence. This represents the amount of error that one might expect in these models when performing precise calculations of $\Omega_{\rm GW}$ but without taking into consideration the beyond the bag treatment of the speed of sound in the plasma. The type of error we observe for the fluid velocity is random. 

To determine the impact of the suppression factor due to the finite lifetime of the source has on the error, we compare $\Omega_{\rm GW}$ calculated in Eq.~\ref{eq:omega_suppressed} with out $\Upsilon$ for the lowest diligence and with the replacement $\Upsilon \rightarrow \tau_{\rm SW} H$ corresponding to Eq.~\ref{eq:suppression-ellis} for the modest diligence. All quantities are evaluated at $T_f$. Modest diligence will also contain the suppression to $U_f$ due to vorticity and reheating effects in the plasma. Note that this suppression is less dramatic than what one might naively expect from ref \cite{Cutting:2019zws}, as the suppression in the fluid velocity results in a longer lifetime for the soundwaves. For the range of models we consider, the error for modest diligence is in the range $(10^{-3} - 10^{-1})$ and represents the validity of the approximation used in $\Upsilon$.   The error in the lowest diligence experiences the highest error with a range of $(10^{1} - 10^{3})$. For all of the models, $\Omega_{\rm GW}^{\rm low,med} > \Omega_{\rm GW}^{\rm high}$. This type of error is an enhancement. 

The last row in Table \ref{tab:Error} corresponds to the error in $\Omega_{\rm GW}$ calculated using Eq.~\ref{eq:omega_suppressed} without suppression factors arising from vorticity and reheating effects in the plasma. This is compared to the full suppression in highest diligence which uses $U_{f, {\rm max}}$ in the lifetime of the source as well. The range of error we observe is in the range of $10^{-1} - 10^{1}$. Neglecting $U_{f,{\rm max}}$ in the suppression factor will contribute at most an error of $0.62$. The lowest diligence experienced $\Omega_{\rm GW}^{\rm low} < \Omega_{\rm GW}^{\rm high}$ for the all of the models.  The modest diligence experienced mostly random error. The primary focus should be on modest diligence so we denote this type of error as random. 







\section{Conclusion}

In this work we have examined the cost of various short-cuts and approximations used in the literature when predicting the gravitational wave spectrum generated by a cosmological first order phase transition. Even in the case where some modest diligence has been used in the calculation, we found the cost to often be comparable to problems in finite temperature QFT such as the scale  dependence that arises from the break down of perturbation theory as well as the gauge dependence. Assuming detonations, the dominant cost in cases where there is a fair amount of super-cooling is from poor estimates of the percolation temperature in Eq. \ref{eq:LISA_Tp}. The poor estimate of the percolation temperature has a knock on effect in enhancing the errors that arise from using the bag model and an estimate for the suppression factor. In the case where there is no tree-level barrier delaying the nucleation of the phase transition, the dominant error is due to the bag model approximation. 
\par 
Although the errors are often as large as finite temperature QFT errors, they are arguably easier to reduce. At present, all of these errors can be handled except for the reheating and vorticity effects where we had to rely on interpolations. High diligence calculations for multiple models were considerably more tractable than the two loop calculations required to bring scale dependence at finite temperature under control \cite{Croon:2020cgk}. We recommend future phenomenological calculations of gravitational wave signals from primordial phase transitions at the very least take the level of theoretical uncertainties into consideration. 

\section{Acknowledgements}

HG and KS are supported by DOE Grant desc0009956. GW was supported by World Premier International Research Center Initiative (WPI), MEXT, Japan.

\appendix{}
\section{Kinetic Energy Efficiency Coefficient}\label{sec:kappa_fits}
The kinetic energy efficiency coefficient may be solved by integrating over the enthalpy and velocity profiles around a single bubble,
\begin{equation}
    \kappa = \frac{3}{\epsilon v_w^3} \int_0^{\xi_{max}} d\xi \xi^2 \omega \gamma^2 v^2, 
\end{equation}
where $\epsilon$ is the bag constant and $\xi = r/t$ is a self similar coordinate in terms of the distance form the bubble center $r$ and the time since nucleation $t$. The fits to $\kappa$ are provided in \cite{Espinosa:2010hh} and are valid in the range $10^{-3} < \alpha < 10$ to a precision of $10 \%$. The fits are found by splitting the parameter space of $v_w$ into three regions and four boundary conditions. The boundary conditions are 
\begin{align}
    \kappa_A &\simeq v_w^{6/5} \frac{6.9 \alpha}{1.36 - 0.037 \sqrt{\alpha} + \alpha}, \: {\rm for} \: v_w \ll c_s, \\
    \kappa_B &\simeq \frac{\alpha^{2/5}}{0.017 + (0.997 + \alpha)^{2/5}}, \: {\rm for} \: v_w = c_s, \\
    \kappa_C &\simeq \frac{\sqrt{\alpha}}{0.135 + \sqrt{0.98} + \alpha}, \: {\rm for} \: v_w = v_J = \frac{\sqrt{\frac{2}{3} \alpha + \alpha^2} + \sqrt{1/3}}{1 + \alpha}, \\
    \kappa_D &\simeq \frac{\alpha}{0.73 + 0.083 \sqrt{\alpha} + \alpha}, \: {\rm for} \: v_w \rightarrow 1,
\end{align}
where $v_J$ is the Jouguet velocity and $c_s$ is the speed of sound. 
Subsonic deflagrations in the region $v_w \lesssim c_s$ have a kinetic energy cofficient approximated by 
\begin{equation}
    \kappa \simeq \frac{c_s^{11/5} \kappa_A \kappa_B}{\left( c_s^{11/5} - v_w^{11/5}\right) \kappa_B + v_w c_s^{6/5} \kappa_A},
\end{equation}
and detonations in the region $v_w > v_J$ by 
\begin{equation}
    \kappa \simeq \frac{\left(v_J - 1\right)^3 v_J^{5/2} v_w^{-5/2} \kappa_C \kappa_D}{\left[\left( v_J - 1\right)^3 - \left(v_w - 1\right)^3 \right]v_J^{5/2} \kappa_C + \left( v_w - 1 \right)^3 \kappa_D},
\end{equation}
Supersonic deflagrations, hybrid, in the region $c_s \lesssim v_w \lesssim v_J$ can be approximated by 
\begin{eqnarray}
    \kappa \simeq \kappa_B + \left( v_w - c_s \right) \delta \kappa + \frac{\left(v_w - c_s \right)^3}{\left(v_J - c_s\right)} \left[ \kappa_C - \kappa_B - \left(v_J - c_s \right)\delta \kappa\right],
\end{eqnarray}
where 
\begin{equation}
    \delta \kappa \simeq -0.9 \log \frac{\sqrt{\alpha}}{1 + \sqrt{\alpha}}.    
\end{equation}

\section{Toy Model}\label{sec:toy}
\subsection{Toy Effective Potential}
A general free energy density of a single scalar field, $\phi$, under a high temperature expansion can be written in the form 
\begin{align}
	f(\phi,T) = D(T^2 - T_0^2)\phi^2 - E T \phi^3 + \frac{1}{4}\lambda \phi^4  + \Delta V -\frac{1}{3}a T^4 ,
 \label{eq:toy_model}
\end{align}
where $\Delta V$ is added to the potential to cancel out the zero temperature minimum such that $f(\phi_{\text{min}},0) = 0$. The Standard Model effective potential was considered in Ref. \cite{Dine:1992wr}. We require $D>0$, $E>0$, $\lambda > 0$ to ensure symmetry is broken at low temperature and generate a barrier between the symmetric and broken phase. The vacuum terms are not necessary for determining the phase transition structure of the model, however, they are necessary for determining the temperature dependence of the speed of sound. 

The structure of the potential along with the constraints on the parameters allows for simple analytical forms for the minima as a function of temperature. The minimum is found by minimizing Eq.~\ref{eq:toy_model} with respect to the scalar field which results in 
\begin{equation}
	\phi_{\text{min}} = \frac{3 E T \pm \sqrt{9E^2T^2 - 8 D \lambda(T_0^2 - T^2)}}{2 \lambda},
\end{equation}
where the '$+$' sign gives the local minimum. When $T$ is large, the global minimum will sit at the origin with $\phi_{\text{min}} = 0$.  As $T$ decreases, a second minimum will develop at 
\begin{equation}
	T = \sqrt{\frac{T_0^2}{1 - \frac{9}{8}\frac{E^2}{\lambda D}}}.
\end{equation}
This minimum will eventually become degenerate with minimum at the origin at the critical temperature when 
\begin{equation}
	T_C = \sqrt{\frac{T_0^2}{1 - \frac{E^2}{\lambda D}}}.
	\label{eq:toy-TC}
\end{equation}
The Euclidean action of a bounce configuration, $S_3$, will start from infinity at $T = T_c$ and decrease with temperature. There is an analytical form for the action given by 
\begin{eqnarray}
	\frac{S_3(T)}{T} = \frac{M^3}{4 E^2 T^3} \tilde{S}_3(\sigma), \\
	\tilde{S}_3(\sigma) = 4 \times 4.85 \times \left[ 1 + \frac{\sigma}{4} \left( 1 + \frac{2.4}{1 - \sigma} + \frac{0.26}{(1 - \sigma)^2}\right) \right],
\end{eqnarray}
where $\sigma = \lambda M^2/(2E^2T^2)$ controls the overall shape of the potential \cite{Dine:1992wr}. The critical temperature and the action are necessary to determine the dynamics of the phase transition and  calculate the relevant transition temperatures such as $T_n$,$T_p$, and $T_f$ and the mean bubble separation  $R_* (v_w,\beta)$. 

The hydrodynamics of the phase transition are determined by the pressure and energy density in the symmetric and broken phase where $p_s = -f(0,T)$ and $p_b = -f(\phi_{\text{min}},T)$. The energy density can be computed from the transformation of pressure 
\begin{equation}
	e = T \frac{\partial p}{\partial T} - p,
	\label{eq:energy_density}
\end{equation}
which can be evaluated in both the symmetric and the broken phase. From Eq.~\ref{eq:toy_model}, the pressure in the symmetric and broken phases are  
\begin{eqnarray}
	p_s = -f(0,T) = \frac{1}{3} a T^4 - \Delta V, \\
	p_b  = -f(\Phi_{\text{min}},T),
	\label{eq:pressure} 
\end{eqnarray}
where the pressure in the broken phase has additional dependence on temperature arising from $\phi_{\text{min}}$. The speed of sound may be found from the pressure using Eq.~\ref{eq:cs2} in both the symmetric and broken phase. The temperature dependence from the minimum of the scalar field will result in a speed sound that is function of the model parameters and its form will depend on the overall shape of the potential.  
\begin{figure}
	\centering
	\includegraphics[width=0.6\textwidth]{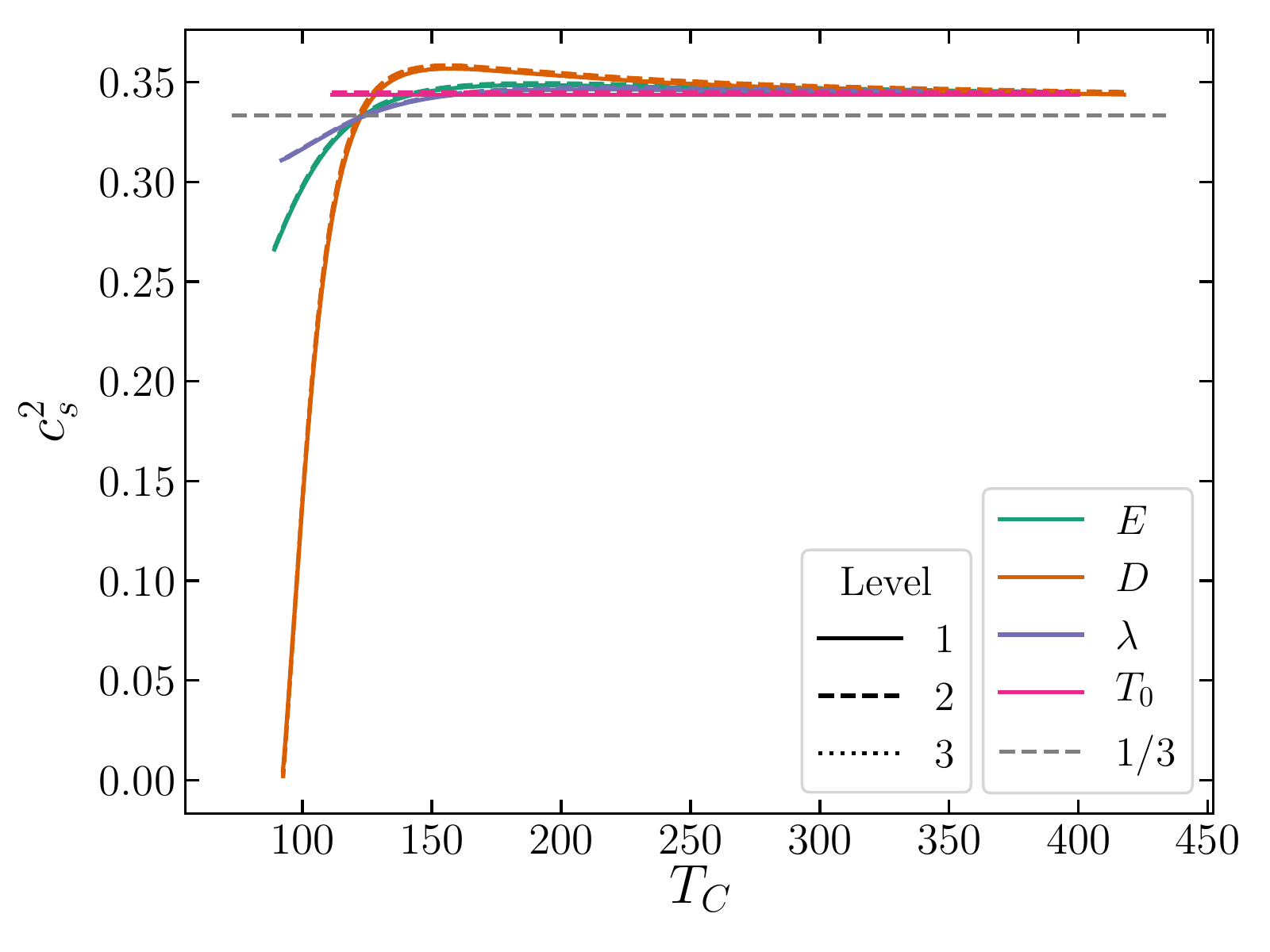}
	\caption{Speed of sound computed in the different levels of diligence.}
	\label{fig:toy-cs2b}
\end{figure} 

\subsection{Results for toy model}
Here we show the different levels of diligence in calculating the thermal parameters and the gravitation wave spectrum in the toy model. The analysis involves individual scans over the different model parameters $(E,D,\lambda,T_0)$ while holding the others fixed. A full analysis of the toy model should involve a randomized scan over all of the parameters but we perform the scan this way in hopes to see any trends in varying the different model parameters. In Eq.~\ref{eq:toy-TC}, the critical temperature is a function of all four model parameters. For this reason, $T_C$ will be used as a basis for each scan. The first step in the beyond the bag calculations is to compute the speed of sound in the symmetric and broken phase.  For the toy model, we only consider detonation, $v_w = 0.92$, where the speed of sound in the symmetric phase is set to $c_s^2 = 1/3$ and the degrees of freedom consist of only the standard model sector. The speed of sound in the broken phase may be found through Eq.~\ref{eq:cs2}. The transition temperatures for the different levels of diligence are $T_n$ (\ref{eq:Tn}), $T_p$ (\ref{eq:LISA_Tp}), and $T_f$ (\ref{eq:Tf}). Example calculations for various phase transition quantities used in the high diligence calculations such as the false vacuum fraction, mean bubble separation, lifetime of the source, and $T_f$ in the toy model may be found in Ref.~\cite{Guo:2020grp}.

In Fig.~\ref{fig:toy-cs2b}, we calculate the speed of sound in the broken phase for each level of diligence. The gray dashed line corresponds to $c_s^2 = 1/3$.  This involves first calculating the speed of sound as a function temperature using Eq.~\ref{eq:cs2} and then evaluating it at $(T_n,T_p,T_f$) computed in the different levels of diligence. We note that in computing the strength of the phase transition only the highest diligence level will involve this calculation. This is merely to show the level of variance in computing the speed of sound at different temperature stages.  For specific values and range chosen, there is only minor change to the speed of sound computed in the different levels however how much variance is present is strongly model dependent.  We do notice that the speed of sound can have a significant deviation away from $c_s^2 = 1/3$ in the bag model.  The strongest deviation is caused by varying the barrier term, $E$, and the quadratic multiplicity term, $D$ as seen in the green and purple curves. The speed of sound can go as low as $c_s^2 \sim 0.22$ and as high as $c_s^2 \sim 0.36$. Varying the zero temperature mass term, $T_0$, did not have any noticeable impact on the speed of sound while the quartic coupling term, $\lambda$, had a mild impact on the speed of sound. This is likely due to the temperature independence of the terms that involve $T_0$ and $\lambda$. The parameters $D$ and $E$ on the other hand, multiply $T^2$ and $T$ respectively and will result in a change in the temperature dependence. The speed of sound in the broken phase is related to the temperature derivatives of the pressure which is evaluated at $p_b(T) = V_{\rm eff}(\phi_{\rm min}(T),T)$ and hence $D$ and $E$ will impact the minimum at finite temperature. The smallest speed of sound in the broken phase corresponds to small $E$ and large $D$. 
\begin{figure}
	\centering
	\includegraphics[width=0.49\textwidth]{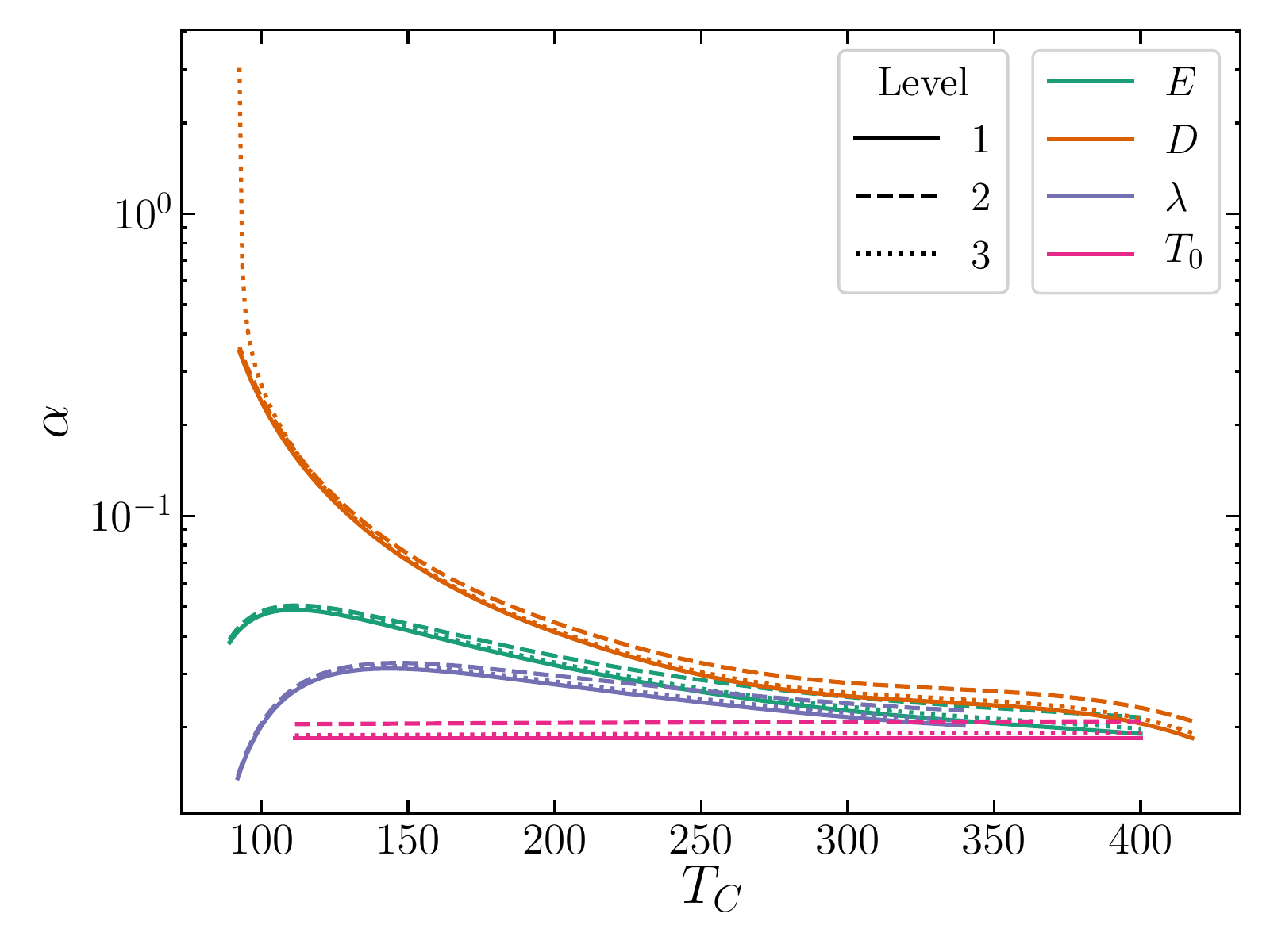}
	\includegraphics[width=0.49\textwidth]{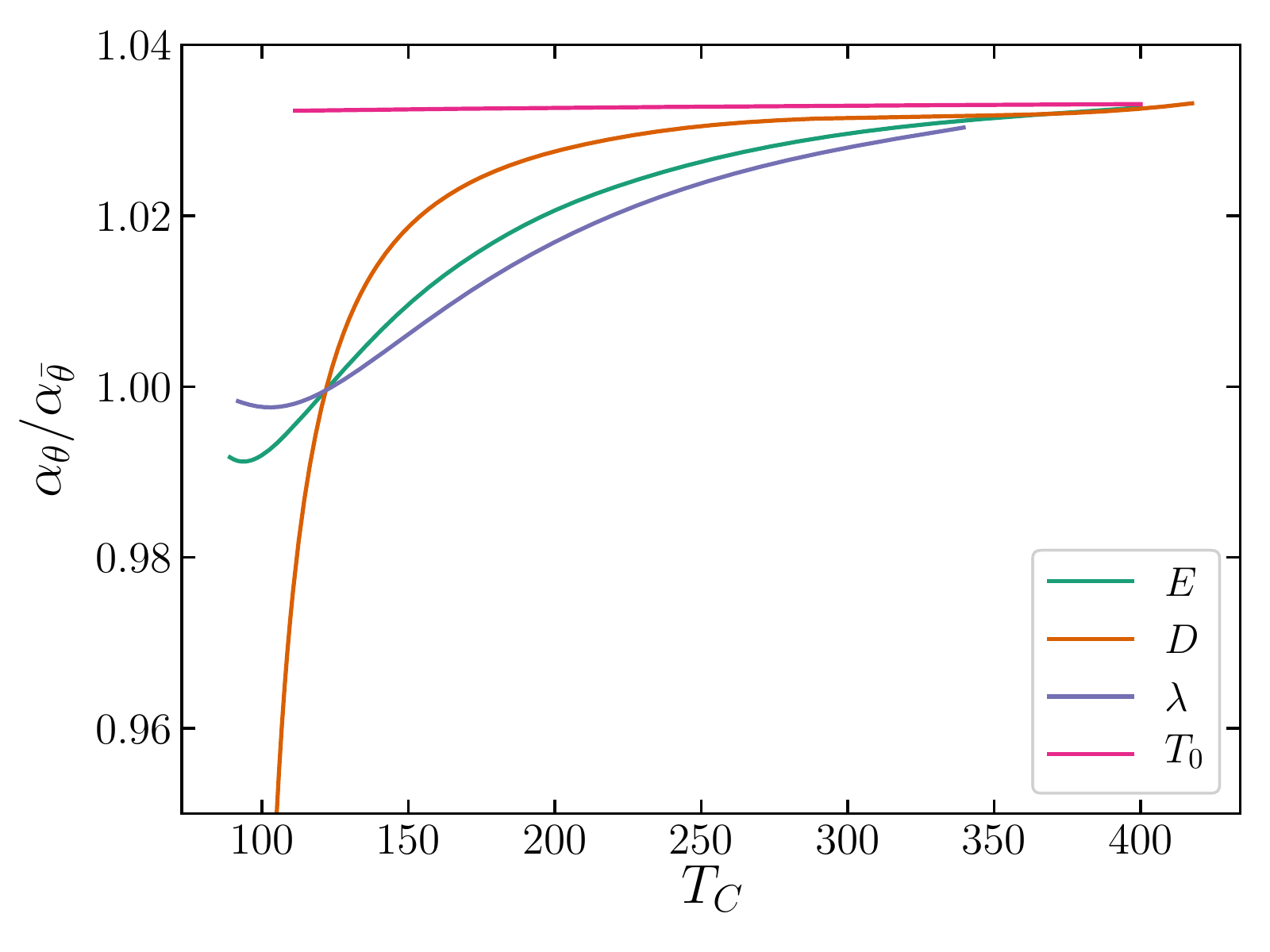}
	\caption{Left: The strength of phase transition computed at the different levels of diligence. Right: The ratio of strength of the phase transition computed at $T_f$ for the bag model $\alpha_\theta$ and the beyond the bag model $\alpha_{\bar{\theta}}$.}
	\label{fig:toy-alpha-compare}
\end{figure} 
 
We show in Fig.~\ref{fig:toy-alpha-compare} the phase transition strength computed in the different level of diligence (left) and the comparison between $\alpha_\theta$ computed in the bag model versus $\alpha_{\bar{\theta}}$ computed in the beyond the bag model at $T_f$ for both quantities (right). We see in the left figure that going higher in the level of diligence results in an increase in the phase transition strength compared to lowest diligence.  This can be attributed to more vacuum energy being released at $T_p$ compared to $T_n$.  On the right, to better compare the difference between the bag model and the beyond the bag model, we compute the ratio of $\alpha_\theta$ and $\alpha_{\bar{\theta}}$ computed at the same temperature, $T_f$. For $T_c < 100$, $\alpha_{\theta}$ is less than  $\alpha_{\bar{\theta}}$ which is the result of $c_s^2 < 1/3$ as seen in Fig.~\ref{fig:toy-cs2b}. This has to do with the $( 1 + c_s^{-2})$ factor in $\alpha_{\bar{\theta}}$. When $T_c > 100$, we see that the opposite is true when $c_s^2 > 1/3$. Similarly, the largest deviations are due to the parameters $D$ and $E$. 

\begin{figure}
	\centering
	\includegraphics[width=0.6\textwidth]{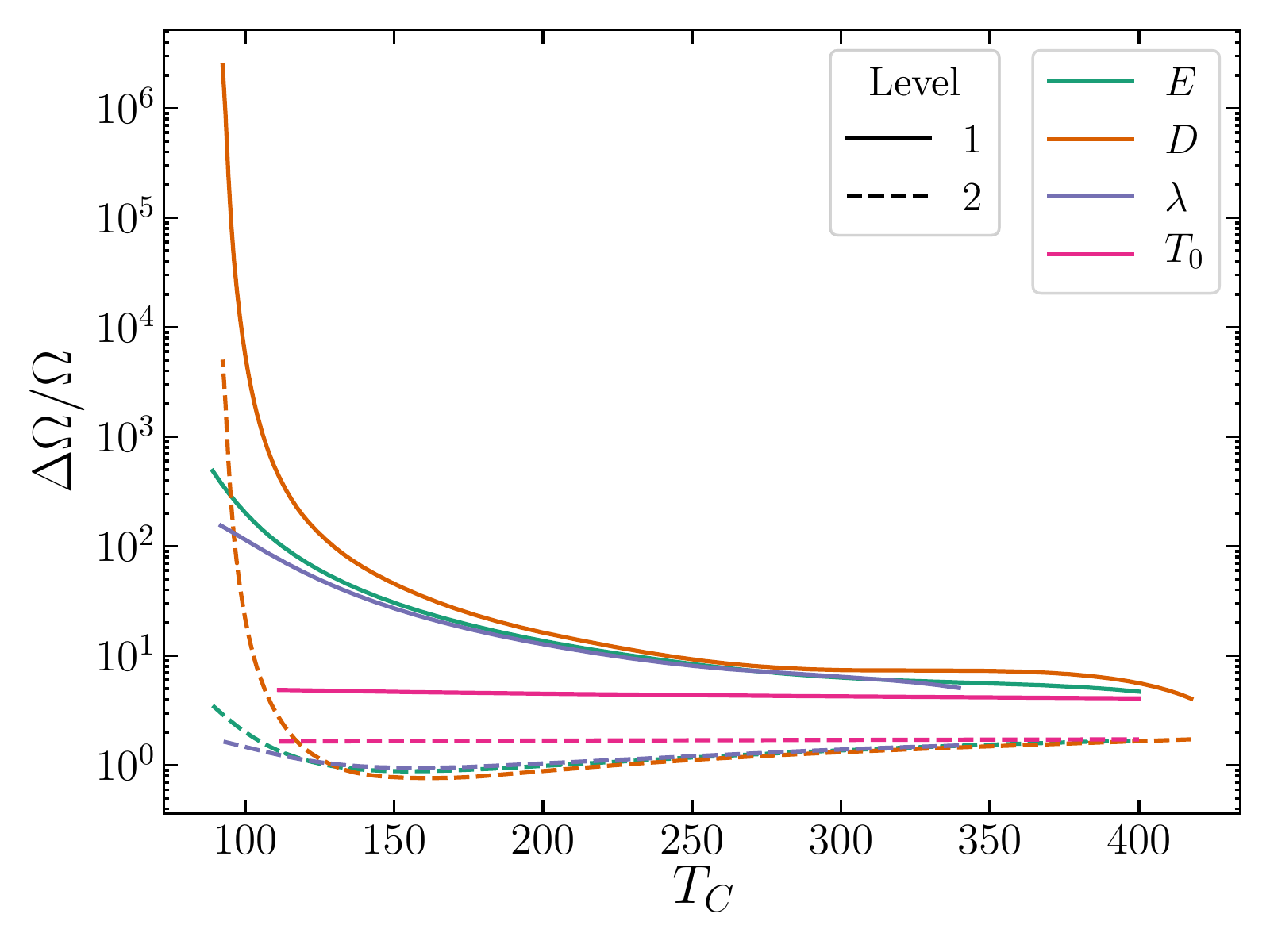}
	\caption{Error of the gravitational spectrum computed at the different levels of diligence.}
	\label{fig:toy-spectrum}
\end{figure} 
The error in the gravitational wave spectrum of the toy model for different scans in the model parameters is shown in the left of Fig.~\ref{fig:toy-spectrum}.  The lowest and modest diligence peak gravitational wave energy density $\Omega_{GW}$ is calculated using Eq.~\ref{eq:omega1} and  Eq.~\ref{eq:omega-suppression-ellis} respectively. The comparison in error is computed with respect to the highest diligence in Eq.~\ref{eq:omega_suppressed}. The lowest diligence level has error in the range $\Delta \Omega / \Omega \sim [10^1,10^3]$ for all parameter scans. The modest diligence level is closest to the highest diligence with error in the range $\Delta \Omega / \Omega \sim [10^0, 10^1]$ for the different scans.  The highest error occurs for $T_C \sim 100$.  This is related to beyond the bag effects which exhibited suppression for the scans in $(E,D)$.  
\bibliographystyle{utphys}

\begin{thebibliography}{100}

\bibitem{Weir:2017wfa}
D.~J. Weir, ``{Gravitational waves from a first order electroweak phase
  transition: a brief review},''
  \href{http://dx.doi.org/10.1098/rsta.2017.0126}{{\em Phil. Trans. Roy. Soc.
  Lond. A} {\bfseries 376} no.~2114, (2018) 20170126},
  \href{http://arxiv.org/abs/1705.01783}{{\ttfamily arXiv:1705.01783
  [hep-ph]}}.

\bibitem{Mazumdar:2018dfl}
A.~Mazumdar and G.~White, ``{Review of cosmic phase transitions: their
  significance and experimental signatures},''
  \href{http://dx.doi.org/10.1088/1361-6633/ab1f55}{{\em Rept. Prog. Phys.}
  {\bfseries 82} no.~7, (2019) 076901},
  \href{http://arxiv.org/abs/1811.01948}{{\ttfamily arXiv:1811.01948
  [hep-ph]}}.

\bibitem{Caprini:2019egz}
C.~Caprini {\em et~al.}, ``{Detecting gravitational waves from cosmological
  phase transitions with LISA: an update},''
  \href{http://dx.doi.org/10.1088/1475-7516/2020/03/024}{{\em JCAP} {\bfseries
  03} (2020) 024}, \href{http://arxiv.org/abs/1910.13125}{{\ttfamily
  arXiv:1910.13125 [astro-ph.CO]}}.

\bibitem{Kajantie:1995kf}
K.~Kajantie, M.~Laine, K.~Rummukainen, and M.~E. Shaposhnikov, ``{The
  Electroweak phase transition: A Nonperturbative analysis},''
  \href{http://dx.doi.org/10.1016/0550-3213(96)00052-1}{{\em Nucl. Phys. B}
  {\bfseries 466} (1996) 189--258},
  \href{http://arxiv.org/abs/hep-lat/9510020}{{\ttfamily
  arXiv:hep-lat/9510020}}.

\bibitem{Kajantie:1996mn}
K.~Kajantie, M.~Laine, K.~Rummukainen, and M.~E. Shaposhnikov, ``{Is there a
  hot electroweak phase transition at m(H) larger or equal to m(W)?},''
  \href{http://dx.doi.org/10.1103/PhysRevLett.77.2887}{{\em Phys. Rev. Lett.}
  {\bfseries 77} (1996) 2887--2890},
  \href{http://arxiv.org/abs/hep-ph/9605288}{{\ttfamily arXiv:hep-ph/9605288}}.

\bibitem{Kajantie:1996qd}
K.~Kajantie, M.~Laine, K.~Rummukainen, and M.~E. Shaposhnikov, ``{A
  Nonperturbative analysis of the finite T phase transition in SU(2) x U(1)
  electroweak theory},''
  \href{http://dx.doi.org/10.1016/S0550-3213(97)00164-8}{{\em Nucl. Phys. B}
  {\bfseries 493} (1997) 413--438},
  \href{http://arxiv.org/abs/hep-lat/9612006}{{\ttfamily
  arXiv:hep-lat/9612006}}.

\bibitem{Csikor:1998eu}
F.~Csikor, Z.~Fodor, and J.~Heitger, ``{Endpoint of the hot electroweak phase
  transition},'' \href{http://dx.doi.org/10.1103/PhysRevLett.82.21}{{\em Phys.
  Rev. Lett.} {\bfseries 82} (1999) 21--24},
  \href{http://arxiv.org/abs/hep-ph/9809291}{{\ttfamily arXiv:hep-ph/9809291}}.

\bibitem{Bazavov:2011nk}
A.~Bazavov {\em et~al.}, ``{The chiral and deconfinement aspects of the QCD
  transition},'' \href{http://dx.doi.org/10.1103/PhysRevD.85.054503}{{\em Phys.
  Rev. D} {\bfseries 85} (2012) 054503},
  \href{http://arxiv.org/abs/1111.1710}{{\ttfamily arXiv:1111.1710 [hep-lat]}}.

\bibitem{Gupta:2011wh}
S.~Gupta, X.~Luo, B.~Mohanty, H.~G. Ritter, and N.~Xu, ``{Scale for the Phase
  Diagram of Quantum Chromodynamics},''
  \href{http://dx.doi.org/10.1126/science.1204621}{{\em Science} {\bfseries
  332} (2011) 1525--1528}, \href{http://arxiv.org/abs/1105.3934}{{\ttfamily
  arXiv:1105.3934 [hep-ph]}}.

\bibitem{Grojean:2004xa}
C.~Grojean, G.~Servant, and J.~D. Wells, ``{First-order electroweak phase
  transition in the standard model with a low cutoff},''
  \href{http://dx.doi.org/10.1103/PhysRevD.71.036001}{{\em Phys. Rev. D}
  {\bfseries 71} (2005) 036001},
  \href{http://arxiv.org/abs/hep-ph/0407019}{{\ttfamily arXiv:hep-ph/0407019}}.

\bibitem{Grojean:2006bp}
C.~Grojean and G.~Servant, ``{Gravitational Waves from Phase Transitions at the
  Electroweak Scale and Beyond},''
  \href{http://dx.doi.org/10.1103/PhysRevD.75.043507}{{\em Phys. Rev. D}
  {\bfseries 75} (2007) 043507},
  \href{http://arxiv.org/abs/hep-ph/0607107}{{\ttfamily arXiv:hep-ph/0607107}}.

\bibitem{Delaunay:2007wb}
C.~Delaunay, C.~Grojean, and J.~D. Wells, ``{Dynamics of Non-renormalizable
  Electroweak Symmetry Breaking},''
  \href{http://dx.doi.org/10.1088/1126-6708/2008/04/029}{{\em JHEP} {\bfseries
  04} (2008) 029}, \href{http://arxiv.org/abs/0711.2511}{{\ttfamily
  arXiv:0711.2511 [hep-ph]}}.

\bibitem{Cline:2009sn}
J.~M. Cline, G.~Laporte, H.~Yamashita, and S.~Kraml, ``{Electroweak Phase
  Transition and LHC Signatures in the Singlet Majoron Model},''
  \href{http://dx.doi.org/10.1088/1126-6708/2009/07/040}{{\em JHEP} {\bfseries
  07} (2009) 040}, \href{http://arxiv.org/abs/0905.2559}{{\ttfamily
  arXiv:0905.2559 [hep-ph]}}.

\bibitem{Carena:2011jy}
M.~Carena, N.~R. Shah, and C.~E.~M. Wagner, ``{Light Dark Matter and the
  Electroweak Phase Transition in the NMSSM},''
  \href{http://dx.doi.org/10.1103/PhysRevD.85.036003}{{\em Phys. Rev. D}
  {\bfseries 85} (2012) 036003},
  \href{http://arxiv.org/abs/1110.4378}{{\ttfamily arXiv:1110.4378 [hep-ph]}}.

\bibitem{Gil:2012ya}
G.~Gil, P.~Chankowski, and M.~Krawczyk, ``{Inert Dark Matter and Strong
  Electroweak Phase Transition},''
  \href{http://dx.doi.org/10.1016/j.physletb.2012.09.052}{{\em Phys. Lett. B}
  {\bfseries 717} (2012) 396--402},
  \href{http://arxiv.org/abs/1207.0084}{{\ttfamily arXiv:1207.0084 [hep-ph]}}.

\bibitem{Carena:2012np}
M.~Carena, G.~Nardini, M.~Quiros, and C.~E. Wagner, ``{MSSM Electroweak
  Baryogenesis and LHC Data},''
  \href{http://dx.doi.org/10.1007/JHEP02(2013)001}{{\em JHEP} {\bfseries 02}
  (2013) 001}, \href{http://arxiv.org/abs/1207.6330}{{\ttfamily arXiv:1207.6330
  [hep-ph]}}.

\bibitem{Fairbairn:2013uta}
M.~Fairbairn and R.~Hogan, ``{Singlet Fermionic Dark Matter and the Electroweak
  Phase Transition},'' \href{http://dx.doi.org/10.1007/JHEP09(2013)022}{{\em
  JHEP} {\bfseries 09} (2013) 022},
  \href{http://arxiv.org/abs/1305.3452}{{\ttfamily arXiv:1305.3452 [hep-ph]}}.

\bibitem{Profumo:2014opa}
S.~Profumo, M.~J. Ramsey-Musolf, C.~L. Wainwright, and P.~Winslow,
  ``{Singlet-catalyzed electroweak phase transitions and precision Higgs boson
  studies},'' \href{http://dx.doi.org/10.1103/PhysRevD.91.035018}{{\em Phys.
  Rev. D} {\bfseries 91} no.~3, (2015) 035018},
  \href{http://arxiv.org/abs/1407.5342}{{\ttfamily arXiv:1407.5342 [hep-ph]}}.

\bibitem{Kozaczuk:2014kva}
J.~Kozaczuk, S.~Profumo, L.~S. Haskins, and C.~L. Wainwright, ``{Cosmological
  Phase Transitions and their Properties in the NMSSM},''
  \href{http://dx.doi.org/10.1007/JHEP01(2015)144}{{\em JHEP} {\bfseries 01}
  (2015) 144}, \href{http://arxiv.org/abs/1407.4134}{{\ttfamily arXiv:1407.4134
  [hep-ph]}}.

\bibitem{Vaskonen:2016yiu}
V.~Vaskonen, ``{Electroweak baryogenesis and gravitational waves from a real
  scalar singlet},'' \href{http://dx.doi.org/10.1103/PhysRevD.95.123515}{{\em
  Phys. Rev. D} {\bfseries 95} no.~12, (2017) 123515},
  \href{http://arxiv.org/abs/1611.02073}{{\ttfamily arXiv:1611.02073
  [hep-ph]}}.

\bibitem{Baldes:2016rqn}
I.~Baldes, T.~Konstandin, and G.~Servant, ``{A first-order electroweak phase
  transition from varying Yukawas},''
  \href{http://dx.doi.org/10.1016/j.physletb.2018.10.015}{{\em Phys. Lett. B}
  {\bfseries 786} (2018) 373--377},
  \href{http://arxiv.org/abs/1604.04526}{{\ttfamily arXiv:1604.04526
  [hep-ph]}}.

\bibitem{Dorsch:2016nrg}
G.~Dorsch, S.~Huber, T.~Konstandin, and J.~No, ``{A Second Higgs Doublet in the
  Early Universe: Baryogenesis and Gravitational Waves},''
  \href{http://dx.doi.org/10.1088/1475-7516/2017/05/052}{{\em JCAP} {\bfseries
  05} (2017) 052}, \href{http://arxiv.org/abs/1611.05874}{{\ttfamily
  arXiv:1611.05874 [hep-ph]}}.

\bibitem{Chiang:2017nmu}
C.-W. Chiang, M.~J. Ramsey-Musolf, and E.~Senaha, ``{Standard Model with a
  Complex Scalar Singlet: Cosmological Implications and Theoretical
  Considerations},'' \href{http://dx.doi.org/10.1103/PhysRevD.97.015005}{{\em
  Phys. Rev. D} {\bfseries 97} no.~1, (2018) 015005},
  \href{http://arxiv.org/abs/1707.09960}{{\ttfamily arXiv:1707.09960
  [hep-ph]}}.

\bibitem{Cao:2017oez}
Q.-H. Cao, F.~P. Huang, K.-P. Xie, and X.~Zhang, ``{Testing the electroweak
  phase transition in scalar extension models at lepton colliders},''
  \href{http://dx.doi.org/10.1088/1674-1137/42/2/023103}{{\em Chin. Phys. C}
  {\bfseries 42} no.~2, (2018) 023103},
  \href{http://arxiv.org/abs/1708.04737}{{\ttfamily arXiv:1708.04737
  [hep-ph]}}.

\bibitem{vonHarling:2017yew}
B.~von Harling and G.~Servant, ``{QCD-induced Electroweak Phase Transition},''
  \href{http://dx.doi.org/10.1007/JHEP01(2018)159}{{\em JHEP} {\bfseries 01}
  (2018) 159}, \href{http://arxiv.org/abs/1711.11554}{{\ttfamily
  arXiv:1711.11554 [hep-ph]}}.

\bibitem{Bian:2017wfv}
L.~Bian, H.-K. Guo, and J.~Shu, ``{Gravitational Waves, baryon asymmetry of the
  universe and electric dipole moment in the CP-violating NMSSM},''
  \href{http://dx.doi.org/10.1088/1674-1137/42/9/093106}{{\em Chin. Phys. C}
  {\bfseries 42} no.~9, (2018) 093106},
  \href{http://arxiv.org/abs/1704.02488}{{\ttfamily arXiv:1704.02488
  [hep-ph]}}. [Erratum: Chin.Phys.C 43, 129101 (2019)].

\bibitem{Chala:2018ari}
M.~Chala, C.~Krause, and G.~Nardini, ``{Signals of the electroweak phase
  transition at colliders and gravitational wave observatories},''
  \href{http://dx.doi.org/10.1007/JHEP07(2018)062}{{\em JHEP} {\bfseries 07}
  (2018) 062}, \href{http://arxiv.org/abs/1802.02168}{{\ttfamily
  arXiv:1802.02168 [hep-ph]}}.

\bibitem{Alves:2018jsw}
A.~Alves, T.~Ghosh, H.-K. Guo, K.~Sinha, and D.~Vagie, ``{Collider and
  Gravitational Wave Complementarity in Exploring the Singlet Extension of the
  Standard Model},'' \href{http://dx.doi.org/10.1007/JHEP04(2019)052}{{\em
  JHEP} {\bfseries 04} (2019) 052},
  \href{http://arxiv.org/abs/1812.09333}{{\ttfamily arXiv:1812.09333
  [hep-ph]}}.

\bibitem{Angelescu:2018dkk}
A.~Angelescu and P.~Huang, ``{Multistep Strongly First Order Phase Transitions
  from New Fermions at the TeV Scale},''
  \href{http://dx.doi.org/10.1103/PhysRevD.99.055023}{{\em Phys. Rev. D}
  {\bfseries 99} no.~5, (2019) 055023},
  \href{http://arxiv.org/abs/1812.08293}{{\ttfamily arXiv:1812.08293
  [hep-ph]}}.

\bibitem{Beniwal:2018hyi}
A.~Beniwal, M.~Lewicki, M.~White, and A.~G. Williams, ``{Gravitational waves
  and electroweak baryogenesis in a global study of the extended scalar singlet
  model},'' \href{http://dx.doi.org/10.1007/JHEP02(2019)183}{{\em JHEP}
  {\bfseries 02} (2019) 183}, \href{http://arxiv.org/abs/1810.02380}{{\ttfamily
  arXiv:1810.02380 [hep-ph]}}.

\bibitem{Bruggisser:2018mrt}
S.~Bruggisser, B.~Von~Harling, O.~Matsedonskyi, and G.~Servant, ``{Electroweak
  Phase Transition and Baryogenesis in Composite Higgs Models},''
  \href{http://dx.doi.org/10.1007/JHEP12(2018)099}{{\em JHEP} {\bfseries 12}
  (2018) 099}, \href{http://arxiv.org/abs/1804.07314}{{\ttfamily
  arXiv:1804.07314 [hep-ph]}}.

\bibitem{Athron:2019teq}
P.~Athron, C.~Balazs, A.~Fowlie, G.~Pozzo, G.~White, and Y.~Zhang, ``{Strong
  first-order phase transitions in the NMSSM \textemdash{} a comprehensive
  survey},'' \href{http://dx.doi.org/10.1007/JHEP11(2019)151}{{\em JHEP}
  {\bfseries 11} (2019) 151}, \href{http://arxiv.org/abs/1908.11847}{{\ttfamily
  arXiv:1908.11847 [hep-ph]}}.

\bibitem{Kainulainen:2019kyp}
K.~Kainulainen, V.~Keus, L.~Niemi, K.~Rummukainen, T.~V. Tenkanen, and
  V.~Vaskonen, ``{On the validity of perturbative studies of the electroweak
  phase transition in the Two Higgs Doublet model},''
  \href{http://dx.doi.org/10.1007/JHEP06(2019)075}{{\em JHEP} {\bfseries 06}
  (2019) 075}, \href{http://arxiv.org/abs/1904.01329}{{\ttfamily
  arXiv:1904.01329 [hep-ph]}}.

\bibitem{Ellis:2019flb}
S.~A.~R. Ellis, S.~Ipek, and G.~White, ``{Electroweak Baryogenesis from
  Temperature-Varying Couplings},''
  \href{http://dx.doi.org/10.1007/JHEP08(2019)002}{{\em JHEP} {\bfseries 08}
  (2019) 002}, \href{http://arxiv.org/abs/1905.11994}{{\ttfamily
  arXiv:1905.11994 [hep-ph]}}.

\bibitem{Ellis:2019tjf}
J.~Ellis, M.~Fairbairn, M.~Lewicki, V.~Vaskonen, and A.~Wickens,
  ``{Intergalactic Magnetic Fields from First-Order Phase Transitions},''
  \href{http://dx.doi.org/10.1088/1475-7516/2019/09/019}{{\em JCAP} {\bfseries
  09} (2019) 019}, \href{http://arxiv.org/abs/1907.04315}{{\ttfamily
  arXiv:1907.04315 [astro-ph.CO]}}.

\bibitem{Papaefstathiou:2020iag}
A.~Papaefstathiou and G.~White, ``{The Electro-Weak Phase Transition at
  Colliders: Confronting Theoretical Uncertainties and Complementary
  Channels},'' \href{http://arxiv.org/abs/2010.00597}{{\ttfamily
  arXiv:2010.00597 [hep-ph]}}.

\bibitem{Postma:2020toi}
M.~Postma and G.~White, ``{Cosmological phase transitions: is effective field
  theory just a toy?},'' \href{http://arxiv.org/abs/2012.03953}{{\ttfamily
  arXiv:2012.03953 [hep-ph]}}.

\bibitem{Alves:2020bpi}
A.~Alves, D.~Gon\c{c}alves, T.~Ghosh, H.-K. Guo, and K.~Sinha, ``{Di-Higgs
  Blind Spots in Gravitational Wave Signals},''
  \href{http://arxiv.org/abs/2007.15654}{{\ttfamily arXiv:2007.15654
  [hep-ph]}}.

\bibitem{Baum:2020vfl}
S.~Baum, M.~Carena, N.~R. Shah, C.~E.~M. Wagner, and Y.~Wang, ``{Nucleation is
  More than Critical}: {A Case Study of the Electroweak Phase Transition in the
  NMSSM},'' \href{http://arxiv.org/abs/2009.10743}{{\ttfamily arXiv:2009.10743
  [hep-ph]}}.

\bibitem{Paul:2020wbz}
A.~Paul, U.~Mukhopadhyay, and D.~Majumdar, ``{Gravitational Wave Signatures
  from Domain Wall and Strong First-Order Phase Transitions in a Two Complex
  Scalar extension of the Standard Model},''
  \href{http://arxiv.org/abs/2010.03439}{{\ttfamily arXiv:2010.03439
  [hep-ph]}}.

\bibitem{Zhou:2020idp}
L.~Bian, H.-K. Guo, Y.~Wu, and R.~Zhou, ``{Gravitational wave and collider
  searches for electroweak symmetry breaking patterns},''
  \href{http://dx.doi.org/10.1103/PhysRevD.101.035011}{{\em Phys. Rev. D}
  {\bfseries 101} no.~3, (2020) 035011},
  \href{http://arxiv.org/abs/1906.11664}{{\ttfamily arXiv:1906.11664
  [hep-ph]}}.

\bibitem{Zhang:2021alu}
Z.~Zhang, C.~Cai, X.-M. Jiang, Y.-L. Tang, Z.-H. Yu, and H.-H. Zhang, ``{Phase
  transition gravitational waves from pseudo-Nambu-Goldstone dark matter and
  two Higgs doublets},'' \href{http://arxiv.org/abs/2102.01588}{{\ttfamily
  arXiv:2102.01588 [hep-ph]}}.

\bibitem{Cline:2021iff}
J.~M. Cline, A.~Friedlander, D.-M. He, K.~Kainulainen, B.~Laurent, and
  D.~Tucker-Smith, ``{Baryogenesis and gravity waves from a UV-completed
  electroweak phase transition},''
  \href{http://arxiv.org/abs/2102.12490}{{\ttfamily arXiv:2102.12490
  [hep-ph]}}.

\bibitem{Davoudiasl:2019ugw}
H.~Davoudiasl, ``{LIGO/Virgo Black Holes from a First Order Quark Confinement
  Phase Transition},''
  \href{http://dx.doi.org/10.1103/PhysRevLett.123.101102}{{\em Phys. Rev.
  Lett.} {\bfseries 123} no.~10, (2019) 101102},
  \href{http://arxiv.org/abs/1902.07805}{{\ttfamily arXiv:1902.07805
  [hep-ph]}}.

\bibitem{Schwarz:2009ii}
D.~J. Schwarz and M.~Stuke, ``{Lepton asymmetry and the cosmic QCD
  transition},'' \href{http://dx.doi.org/10.1088/1475-7516/2009/11/025}{{\em
  JCAP} {\bfseries 11} (2009) 025},
  \href{http://arxiv.org/abs/0906.3434}{{\ttfamily arXiv:0906.3434 [hep-ph]}}.
  [Erratum: JCAP 10, E01 (2010)].

\bibitem{Caprini:2010xv}
C.~Caprini, R.~Durrer, and X.~Siemens, ``{Detection of gravitational waves from
  the QCD phase transition with pulsar timing arrays},''
  \href{http://dx.doi.org/10.1103/PhysRevD.82.063511}{{\em Phys. Rev. D}
  {\bfseries 82} (2010) 063511},
  \href{http://arxiv.org/abs/1007.1218}{{\ttfamily arXiv:1007.1218
  [astro-ph.CO]}}.

\bibitem{Barenboim:2017dfq}
G.~Barenboim and W.-I. Park, ``{A full picture of large lepton number
  asymmetries of the Universe},''
  \href{http://dx.doi.org/10.1088/1475-7516/2017/04/048}{{\em JCAP} {\bfseries
  04} (2017) 048}, \href{http://arxiv.org/abs/1703.08258}{{\ttfamily
  arXiv:1703.08258 [hep-ph]}}.

\bibitem{Schwaller:2015tja}
P.~Schwaller, ``{Gravitational Waves from a Dark Phase Transition},''
  \href{http://dx.doi.org/10.1103/PhysRevLett.115.181101}{{\em Phys. Rev.
  Lett.} {\bfseries 115} no.~18, (2015) 181101},
  \href{http://arxiv.org/abs/1504.07263}{{\ttfamily arXiv:1504.07263
  [hep-ph]}}.

\bibitem{Chao:2017vrq}
W.~Chao, H.-K. Guo, and J.~Shu, ``{Gravitational Wave Signals of Electroweak
  Phase Transition Triggered by Dark Matter},''
  \href{http://dx.doi.org/10.1088/1475-7516/2017/09/009}{{\em JCAP} {\bfseries
  09} (2017) 009}, \href{http://arxiv.org/abs/1702.02698}{{\ttfamily
  arXiv:1702.02698 [hep-ph]}}.

\bibitem{Croon:2018erz}
D.~Croon, V.~Sanz, and G.~White, ``{Model Discrimination in Gravitational Wave
  spectra from Dark Phase Transitions},''
  \href{http://dx.doi.org/10.1007/JHEP08(2018)203}{{\em JHEP} {\bfseries 08}
  (2018) 203}, \href{http://arxiv.org/abs/1806.02332}{{\ttfamily
  arXiv:1806.02332 [hep-ph]}}.

\bibitem{Archer-Smith:2019gzq}
P.~Archer-Smith, D.~Linthorne, and D.~Stolarski, ``{Gravitational Wave Signals
  from Multiple Hidden Sectors},''
  \href{http://dx.doi.org/10.1103/PhysRevD.101.095016}{{\em Phys. Rev. D}
  {\bfseries 101} no.~9, (2020) 095016},
  \href{http://arxiv.org/abs/1910.02083}{{\ttfamily arXiv:1910.02083
  [hep-ph]}}.

\bibitem{Pandey:2020hoq}
M.~Pandey and A.~Paul, ``{Gravitational Wave Emissions from First Order Phase
  Transitions with Two Component FIMP Dark Matter},''
  \href{http://arxiv.org/abs/2003.08828}{{\ttfamily arXiv:2003.08828
  [hep-ph]}}.

\bibitem{Mohamadnejad:2019vzg}
A.~Mohamadnejad, ``{Gravitational waves from scale-invariant vector dark matter
  model: Probing below the neutrino-floor},''
  \href{http://dx.doi.org/10.1140/epjc/s10052-020-7756-6}{{\em Eur. Phys. J. C}
  {\bfseries 80} no.~3, (2020) 197},
  \href{http://arxiv.org/abs/1907.08899}{{\ttfamily arXiv:1907.08899
  [hep-ph]}}.

\bibitem{Helmboldt:2019pan}
A.~J. Helmboldt, J.~Kubo, and S.~van~der Woude, ``{Observational prospects for
  gravitational waves from hidden or dark chiral phase transitions},''
  \href{http://dx.doi.org/10.1103/PhysRevD.100.055025}{{\em Phys. Rev. D}
  {\bfseries 100} no.~5, (2019) 055025},
  \href{http://arxiv.org/abs/1904.07891}{{\ttfamily arXiv:1904.07891
  [hep-ph]}}.

\bibitem{Croon:2019iuh}
D.~Croon, R.~Houtz, and V.~Sanz, ``{Dynamical Axions and Gravitational
  Waves},'' \href{http://dx.doi.org/10.1007/JHEP07(2019)146}{{\em JHEP}
  {\bfseries 07} (2019) 146}, \href{http://arxiv.org/abs/1904.10967}{{\ttfamily
  arXiv:1904.10967 [hep-ph]}}.

\bibitem{Fairbairn:2019xog}
M.~Fairbairn, E.~Hardy, and A.~Wickens, ``{Hearing without seeing:
  gravitational waves from hot and cold hidden sectors},''
  \href{http://dx.doi.org/10.1007/JHEP07(2019)044}{{\em JHEP} {\bfseries 07}
  (2019) 044}, \href{http://arxiv.org/abs/1901.11038}{{\ttfamily
  arXiv:1901.11038 [hep-ph]}}.

\bibitem{Breitbach:2018ddu}
M.~Breitbach, J.~Kopp, E.~Madge, T.~Opferkuch, and P.~Schwaller, ``{Dark, Cold,
  and Noisy: Constraining Secluded Hidden Sectors with Gravitational Waves},''
  \href{http://dx.doi.org/10.1088/1475-7516/2019/07/007}{{\em JCAP} {\bfseries
  07} (2019) 007}, \href{http://arxiv.org/abs/1811.11175}{{\ttfamily
  arXiv:1811.11175 [hep-ph]}}.

\bibitem{Bhoonah:2020oov}
A.~Bhoonah, J.~Bramante, S.~Nerval, and N.~Song, ``{Gravitational Waves From
  Dark Sectors, Oscillating Inflatons, and Mass Boosted Dark Matter},''
  \href{http://arxiv.org/abs/2008.12306}{{\ttfamily arXiv:2008.12306
  [hep-ph]}}.

\bibitem{Hall:2019rld}
E.~Hall, T.~Konstandin, R.~McGehee, and H.~Murayama, ``{Asymmetric Matters from
  a Dark First-Order Phase Transition},''
  \href{http://arxiv.org/abs/1911.12342}{{\ttfamily arXiv:1911.12342
  [hep-ph]}}.

\bibitem{Li:2020eun}
M.~Li, Q.-S. Yan, Y.~Zhang, and Z.~Zhao, ``{Prospects of gravitational waves in
  the minimal left-right symmetric model},''
  \href{http://arxiv.org/abs/2012.13686}{{\ttfamily arXiv:2012.13686
  [hep-ph]}}.

\bibitem{Ghosh:2020ipy}
T.~Ghosh, H.-K. Guo, T.~Han, and H.~Liu, ``{Electroweak Phase Transition with
  an SU(2) Dark Sector},'' \href{http://arxiv.org/abs/2012.09758}{{\ttfamily
  arXiv:2012.09758 [hep-ph]}}.

\bibitem{Huang:2021rrk}
F.~Huang, V.~Sanz, J.~Shu, and X.~Xue, ``{LIGO as a probe of Dark Sectors},''
  \href{http://arxiv.org/abs/2102.03155}{{\ttfamily arXiv:2102.03155
  [hep-ph]}}.

\bibitem{Huang:2020mso}
W.-C. Huang, M.~Reichert, F.~Sannino, and Z.-W. Wang, ``{Testing the Dark
  Confined Landscape: From Lattice to Gravitational Waves},''
  \href{http://arxiv.org/abs/2012.11614}{{\ttfamily arXiv:2012.11614
  [hep-ph]}}.

\bibitem{Bigazzi:2020avc}
F.~Bigazzi, A.~Caddeo, A.~L. Cotrone, and A.~Paredes, ``{Dark Holograms and
  Gravitational Waves},'' \href{http://arxiv.org/abs/2011.08757}{{\ttfamily
  arXiv:2011.08757 [hep-ph]}}.

\bibitem{Azatov:2021ifm}
A.~Azatov, M.~Vanvlasselaer, and W.~Yin, ``{Dark Matter production from
  relativistic bubble walls},''
  \href{http://arxiv.org/abs/2101.05721}{{\ttfamily arXiv:2101.05721
  [hep-ph]}}.

\bibitem{Chao:2017ilw}
W.~Chao, W.-F. Cui, H.-K. Guo, and J.~Shu, ``{Gravitational wave imprint of new
  symmetry breaking},'' \href{http://dx.doi.org/10.1088/1674-1137/abb4cb}{{\em
  Chin. Phys. C} {\bfseries 44} no.~12, (2020) 123102},
  \href{http://arxiv.org/abs/1707.09759}{{\ttfamily arXiv:1707.09759
  [hep-ph]}}.

\bibitem{Croon:2018kqn}
D.~Croon, T.~E. Gonzalo, and G.~White, ``{Gravitational Waves from a Pati-Salam
  Phase Transition},'' \href{http://dx.doi.org/10.1007/JHEP02(2019)083}{{\em
  JHEP} {\bfseries 02} (2019) 083},
  \href{http://arxiv.org/abs/1812.02747}{{\ttfamily arXiv:1812.02747
  [hep-ph]}}.

\bibitem{Haba:2019qol}
N.~Haba and T.~Yamada, ``{Gravitational waves from phase transition in minimal
  SUSY $U(1)_{B-L}$ model},''
  \href{http://dx.doi.org/10.1103/PhysRevD.101.075027}{{\em Phys. Rev. D}
  {\bfseries 101} no.~7, (2020) 075027},
  \href{http://arxiv.org/abs/1911.01292}{{\ttfamily arXiv:1911.01292
  [hep-ph]}}.

\bibitem{Greljo:2019xan}
A.~Greljo, T.~Opferkuch, and B.~A. Stefanek, ``{Gravitational Imprints of
  Flavor Hierarchies},''
  \href{http://dx.doi.org/10.1103/PhysRevLett.124.171802}{{\em Phys. Rev.
  Lett.} {\bfseries 124} no.~17, (2020) 171802},
  \href{http://arxiv.org/abs/1910.02014}{{\ttfamily arXiv:1910.02014
  [hep-ph]}}.

\bibitem{Brdar:2019fur}
V.~Brdar, L.~Graf, A.~J. Helmboldt, and X.-J. Xu, ``{Gravitational Waves as a
  Probe of Left-Right Symmetry Breaking},''
  \href{http://dx.doi.org/10.1088/1475-7516/2019/12/027}{{\em JCAP} {\bfseries
  12} (2019) 027}, \href{http://arxiv.org/abs/1909.02018}{{\ttfamily
  arXiv:1909.02018 [hep-ph]}}.

\bibitem{Okada:2020vvb}
N.~Okada, O.~Seto, and H.~Uchida, ``{Gravitational waves from breaking of an
  extra $U(1)$ in $SO(10)$ grand unification},''
  \href{http://arxiv.org/abs/2006.01406}{{\ttfamily arXiv:2006.01406
  [hep-ph]}}.

\bibitem{Huang:2020bbe}
W.-C. Huang, F.~Sannino, and Z.-W. Wang, ``{Gravitational Waves from Pati-Salam
  Dynamics},'' \href{http://dx.doi.org/10.1103/PhysRevD.102.095025}{{\em Phys.
  Rev. D} {\bfseries 102} no.~9, (2020) 095025},
  \href{http://arxiv.org/abs/2004.02332}{{\ttfamily arXiv:2004.02332
  [hep-ph]}}.

\bibitem{Marzola:2017jzl}
L.~Marzola, A.~Racioppi, and V.~Vaskonen, ``{Phase transition and gravitational
  wave phenomenology of scalar conformal extensions of the Standard Model},''
  \href{http://dx.doi.org/10.1140/epjc/s10052-017-4996-1}{{\em Eur. Phys. J. C}
  {\bfseries 77} no.~7, (2017) 484},
  \href{http://arxiv.org/abs/1704.01034}{{\ttfamily arXiv:1704.01034
  [hep-ph]}}.

\bibitem{Ellis:2019oqb}
J.~Ellis, M.~Lewicki, J.~M. No, and V.~Vaskonen, ``{Gravitational wave energy
  budget in strongly supercooled phase transitions},''
  \href{http://dx.doi.org/10.1088/1475-7516/2019/06/024}{{\em JCAP} {\bfseries
  06} (2019) 024}, \href{http://arxiv.org/abs/1903.09642}{{\ttfamily
  arXiv:1903.09642 [hep-ph]}}.

\bibitem{Bian:2019szo}
L.~Bian, W.~Cheng, H.-K. Guo, and Y.~Zhang, ``{Gravitational waves triggered by
  $B-L$ charged hidden scalar and leptogenesis},''
  \href{http://arxiv.org/abs/1907.13589}{{\ttfamily arXiv:1907.13589
  [hep-ph]}}.

\bibitem{Okada:2018xdh}
N.~Okada and O.~Seto, ``{Probing the seesaw scale with gravitational waves},''
  \href{http://dx.doi.org/10.1103/PhysRevD.98.063532}{{\em Phys. Rev. D}
  {\bfseries 98} no.~6, (2018) 063532},
  \href{http://arxiv.org/abs/1807.00336}{{\ttfamily arXiv:1807.00336
  [hep-ph]}}.

\bibitem{Hasegawa:2019amx}
T.~Hasegawa, N.~Okada, and O.~Seto, ``{Gravitational waves from the minimal
  gauged $U(1)_{B-L}$ model},''
  \href{http://dx.doi.org/10.1103/PhysRevD.99.095039}{{\em Phys. Rev. D}
  {\bfseries 99} no.~9, (2019) 095039},
  \href{http://arxiv.org/abs/1904.03020}{{\ttfamily arXiv:1904.03020
  [hep-ph]}}.

\bibitem{Dev:2016feu}
P.~S.~B. Dev and A.~Mazumdar, ``{Probing the Scale of New Physics by Advanced
  LIGO/VIRGO},'' \href{http://dx.doi.org/10.1103/PhysRevD.93.104001}{{\em Phys.
  Rev. D} {\bfseries 93} no.~10, (2016) 104001},
  \href{http://arxiv.org/abs/1602.04203}{{\ttfamily arXiv:1602.04203
  [hep-ph]}}.

\bibitem{Aggarwal:2020olq}
N.~Aggarwal {\em et~al.}, ``{Challenges and Opportunities of Gravitational Wave
  Searches at MHz to GHz Frequencies},''
  \href{http://arxiv.org/abs/2011.12414}{{\ttfamily arXiv:2011.12414 [gr-qc]}}.

\bibitem{Ackley:2020atn}
K.~Ackley {\em et~al.}, ``{Neutron Star Extreme Matter Observatory: A
  kilohertz-band gravitational-wave detector in the global network},''
  \href{http://dx.doi.org/10.1017/pasa.2020.39}{{\em Publ. Astron. Soc.
  Austral.} {\bfseries 37} (2020) e047},
  \href{http://arxiv.org/abs/2007.03128}{{\ttfamily arXiv:2007.03128
  [astro-ph.HE]}}.

\bibitem{Hindmarsh:2017gnf}
M.~Hindmarsh, S.~J. Huber, K.~Rummukainen, and D.~J. Weir, ``{Shape of the
  acoustic gravitational wave power spectrum from a first order phase
  transition},'' \href{http://dx.doi.org/10.1103/PhysRevD.96.103520}{{\em Phys.
  Rev. D} {\bfseries 96} no.~10, (2017) 103520},
  \href{http://arxiv.org/abs/1704.05871}{{\ttfamily arXiv:1704.05871
  [astro-ph.CO]}}. [Erratum: Phys.Rev.D 101, 089902 (2020)].

\bibitem{Bodeker:2017cim}
D.~Bodeker and G.~D. Moore, ``{Electroweak Bubble Wall Speed Limit},''
  \href{http://dx.doi.org/10.1088/1475-7516/2017/05/025}{{\em JCAP} {\bfseries
  05} (2017) 025}, \href{http://arxiv.org/abs/1703.08215}{{\ttfamily
  arXiv:1703.08215 [hep-ph]}}.

\bibitem{Hindmarsh:2013xza}
M.~Hindmarsh, S.~J. Huber, K.~Rummukainen, and D.~J. Weir, ``{Gravitational
  waves from the sound of a first order phase transition},''
  \href{http://dx.doi.org/10.1103/PhysRevLett.112.041301}{{\em Phys. Rev.
  Lett.} {\bfseries 112} (2014) 041301},
  \href{http://arxiv.org/abs/1304.2433}{{\ttfamily arXiv:1304.2433 [hep-ph]}}.

\bibitem{Hindmarsh:2016lnk}
M.~Hindmarsh, ``{Sound shell model for acoustic gravitational wave production
  at a first-order phase transition in the early Universe},''
  \href{http://dx.doi.org/10.1103/PhysRevLett.120.071301}{{\em Phys. Rev.
  Lett.} {\bfseries 120} no.~7, (2018) 071301},
  \href{http://arxiv.org/abs/1608.04735}{{\ttfamily arXiv:1608.04735
  [astro-ph.CO]}}.

\bibitem{Hindmarsh:2015qta}
M.~Hindmarsh, S.~J. Huber, K.~Rummukainen, and D.~J. Weir, ``{Numerical
  simulations of acoustically generated gravitational waves at a first order
  phase transition},'' \href{http://dx.doi.org/10.1103/PhysRevD.92.123009}{{\em
  Phys. Rev. D} {\bfseries 92} no.~12, (2015) 123009},
  \href{http://arxiv.org/abs/1504.03291}{{\ttfamily arXiv:1504.03291
  [astro-ph.CO]}}.

\bibitem{Caprini:2015zlo}
C.~Caprini {\em et~al.}, ``{Science with the space-based interferometer eLISA.
  II: Gravitational waves from cosmological phase transitions},''
  \href{http://dx.doi.org/10.1088/1475-7516/2016/04/001}{{\em JCAP} {\bfseries
  04} (2016) 001}, \href{http://arxiv.org/abs/1512.06239}{{\ttfamily
  arXiv:1512.06239 [astro-ph.CO]}}.

\bibitem{Kahniashvili:2008pf}
T.~Kahniashvili, A.~Kosowsky, G.~Gogoberidze, and Y.~Maravin, ``{Detectability
  of Gravitational Waves from Phase Transitions},''
  \href{http://dx.doi.org/10.1103/PhysRevD.78.043003}{{\em Phys. Rev. D}
  {\bfseries 78} (2008) 043003},
  \href{http://arxiv.org/abs/0806.0293}{{\ttfamily arXiv:0806.0293
  [astro-ph]}}.

\bibitem{Kahniashvili:2008pe}
T.~Kahniashvili, L.~Campanelli, G.~Gogoberidze, Y.~Maravin, and B.~Ratra,
  ``{Gravitational Radiation from Primordial Helical Inverse Cascade MHD
  Turbulence},'' \href{http://dx.doi.org/10.1103/PhysRevD.78.123006}{{\em Phys.
  Rev. D} {\bfseries 78} (2008) 123006},
  \href{http://arxiv.org/abs/0809.1899}{{\ttfamily arXiv:0809.1899
  [astro-ph]}}. [Erratum: Phys.Rev.D 79, 109901 (2009)].

\bibitem{Kahniashvili:2009mf}
T.~Kahniashvili, L.~Kisslinger, and T.~Stevens, ``{Gravitational Radiation
  Generated by Magnetic Fields in Cosmological Phase Transitions},''
  \href{http://dx.doi.org/10.1103/PhysRevD.81.023004}{{\em Phys. Rev. D}
  {\bfseries 81} (2010) 023004},
  \href{http://arxiv.org/abs/0905.0643}{{\ttfamily arXiv:0905.0643
  [astro-ph.CO]}}.

\bibitem{Caprini:2009yp}
C.~Caprini, R.~Durrer, and G.~Servant, ``{The stochastic gravitational wave
  background from turbulence and magnetic fields generated by a first-order
  phase transition},''
  \href{http://dx.doi.org/10.1088/1475-7516/2009/12/024}{{\em JCAP} {\bfseries
  12} (2009) 024}, \href{http://arxiv.org/abs/0909.0622}{{\ttfamily
  arXiv:0909.0622 [astro-ph.CO]}}.

\bibitem{Kisslinger:2015hua}
L.~Kisslinger and T.~Kahniashvili, ``{Polarized Gravitational Waves from
  Cosmological Phase Transitions},''
  \href{http://dx.doi.org/10.1103/PhysRevD.92.043006}{{\em Phys. Rev. D}
  {\bfseries 92} no.~4, (2015) 043006},
  \href{http://arxiv.org/abs/1505.03680}{{\ttfamily arXiv:1505.03680
  [astro-ph.CO]}}.

\bibitem{Pol:2019yex}
A.~Roper~Pol, S.~Mandal, A.~Brandenburg, T.~Kahniashvili, and A.~Kosowsky,
  ``{Numerical simulations of gravitational waves from early-universe
  turbulence},'' \href{http://dx.doi.org/10.1103/PhysRevD.102.083512}{{\em
  Phys. Rev. D} {\bfseries 102} no.~8, (2020) 083512},
  \href{http://arxiv.org/abs/1903.08585}{{\ttfamily arXiv:1903.08585
  [astro-ph.CO]}}.

\bibitem{Croon:2020cgk}
D.~Croon, O.~Gould, P.~Schicho, T.~V. Tenkanen, and G.~White, ``{Theoretical
  uncertainties for cosmological first-order phase transitions},''
  \href{http://arxiv.org/abs/2009.10080}{{\ttfamily arXiv:2009.10080
  [hep-ph]}}.

\bibitem{Kajantie:1995dw}
K.~Kajantie, M.~Laine, K.~Rummukainen, and M.~E. Shaposhnikov, ``{Generic rules
  for high temperature dimensional reduction and their application to the
  standard model},'' \href{http://dx.doi.org/10.1016/0550-3213(95)00549-8}{{\em
  Nucl. Phys. B} {\bfseries 458} (1996) 90--136},
  \href{http://arxiv.org/abs/hep-ph/9508379}{{\ttfamily arXiv:hep-ph/9508379}}.

\bibitem{Curtin:2016urg}
D.~Curtin, P.~Meade, and H.~Ramani, ``{Thermal Resummation and Phase
  Transitions},'' \href{http://dx.doi.org/10.1140/epjc/s10052-018-6268-0}{{\em
  Eur. Phys. J. C} {\bfseries 78} no.~9, (2018) 787},
  \href{http://arxiv.org/abs/1612.00466}{{\ttfamily arXiv:1612.00466
  [hep-ph]}}.

\bibitem{Schicho:2021gca}
P.~M. Schicho, T.~V.~I. Tenkanen, and J.~\"Osterman, ``{Robust approach to
  thermal resummation: Standard Model meets a singlet},''
  \href{http://arxiv.org/abs/2102.11145}{{\ttfamily arXiv:2102.11145
  [hep-ph]}}.

\bibitem{Gould:2021dzl}
O.~Gould, ``{Real scalar phase transitions: a nonperturbative analysis},''
  \href{http://arxiv.org/abs/2101.05528}{{\ttfamily arXiv:2101.05528
  [hep-ph]}}.

\bibitem{Patel:2011th}
H.~H. Patel and M.~J. Ramsey-Musolf, ``{Baryon Washout, Electroweak Phase
  Transition, and Perturbation Theory},''
  \href{http://dx.doi.org/10.1007/JHEP07(2011)029}{{\em JHEP} {\bfseries 07}
  (2011) 029}, \href{http://arxiv.org/abs/1101.4665}{{\ttfamily arXiv:1101.4665
  [hep-ph]}}.

\bibitem{Garbrecht:2015yza}
B.~Garbrecht and P.~Millington, ``{Self-consistent solitons for vacuum decay in
  radiatively generated potentials},''
  \href{http://dx.doi.org/10.1103/PhysRevD.92.125022}{{\em Phys. Rev. D}
  {\bfseries 92} (2015) 125022},
  \href{http://arxiv.org/abs/1509.08480}{{\ttfamily arXiv:1509.08480
  [hep-ph]}}.

\bibitem{Ai:2020sru}
W.-Y. Ai, J.~S. Cruz, B.~Garbrecht, and C.~Tamarit, ``{Gradient effects on
  false vacuum decay in gauge theory},''
  \href{http://dx.doi.org/10.1103/PhysRevD.102.085001}{{\em Phys. Rev. D}
  {\bfseries 102} no.~8, (2020) 085001},
  \href{http://arxiv.org/abs/2006.04886}{{\ttfamily arXiv:2006.04886
  [hep-th]}}.

\bibitem{Niemi:2020hto}
L.~Niemi, M.~Ramsey-Musolf, T.~V. Tenkanen, and D.~J. Weir, ``{Thermodynamics
  of a two-step electroweak phase transition},''
  \href{http://arxiv.org/abs/2005.11332}{{\ttfamily arXiv:2005.11332
  [hep-ph]}}.

\bibitem{Ellis:2018mja}
J.~Ellis, M.~Lewicki, and J.~M. No, ``{On the Maximal Strength of a First-Order
  Electroweak Phase Transition and its Gravitational Wave Signal},''
  \href{http://dx.doi.org/10.1088/1475-7516/2019/04/003}{{\em JCAP} {\bfseries
  04} (2019) 003}, \href{http://arxiv.org/abs/1809.08242}{{\ttfamily
  arXiv:1809.08242 [hep-ph]}}.

\bibitem{Guo:2020grp}
H.-K. Guo, K.~Sinha, D.~Vagie, and G.~White, ``{Phase Transitions in an
  Expanding Universe: Stochastic Gravitational Waves in Standard and
  Non-Standard Histories},'' \href{http://arxiv.org/abs/2007.08537}{{\ttfamily
  arXiv:2007.08537 [hep-ph]}}.

\bibitem{Giese:2020rtr}
F.~Giese, T.~Konstandin, and J.~van~de Vis, ``{Model-independent energy budget
  of cosmological first-order phase transitions\textemdash{}A sound argument to
  go beyond the bag model},''
  \href{http://dx.doi.org/10.1088/1475-7516/2020/07/057}{{\em JCAP} {\bfseries
  07} no.~07, (2020) 057}, \href{http://arxiv.org/abs/2004.06995}{{\ttfamily
  arXiv:2004.06995 [astro-ph.CO]}}.

\bibitem{Giese:2020znk}
F.~Giese, T.~Konstandin, K.~Schmitz, and J.~Van De~Vis, ``{Model-independent
  energy budget for LISA},'' \href{http://arxiv.org/abs/2010.09744}{{\ttfamily
  arXiv:2010.09744 [astro-ph.CO]}}.

\bibitem{Cutting:2019zws}
D.~Cutting, M.~Hindmarsh, and D.~J. Weir, ``{Vorticity, kinetic energy, and
  suppressed gravitational wave production in strong first order phase
  transitions},'' \href{http://dx.doi.org/10.1103/PhysRevLett.125.021302}{{\em
  Phys. Rev. Lett.} {\bfseries 125} no.~2, (2020) 021302},
  \href{http://arxiv.org/abs/1906.00480}{{\ttfamily arXiv:1906.00480
  [hep-ph]}}.

\bibitem{Quiros:1999jp}
M.~Quiros, ``{Finite temperature field theory and phase transitions},'' in {\em
  {ICTP Summer School in High-Energy Physics and Cosmology}}, pp.~187--259.
\newblock 1, 1999.
\newblock \href{http://arxiv.org/abs/hep-ph/9901312}{{\ttfamily
  arXiv:hep-ph/9901312}}.

\bibitem{Wainwright:2011kj}
C.~L. Wainwright, ``{CosmoTransitions: Computing Cosmological Phase Transition
  Temperatures and Bubble Profiles with Multiple Fields},''
  \href{http://dx.doi.org/10.1016/j.cpc.2012.04.004}{{\em Comput. Phys.
  Commun.} {\bfseries 183} (2012) 2006--2013},
\href{http://arxiv.org/abs/1109.4189}{{\ttfamily arXiv:1109.4189 [hep-ph]}}.

\bibitem{Ramsey-Musolf:2019lsf}
M.~J. Ramsey-Musolf, ``{The electroweak phase transition: a collider target},''
  \href{http://dx.doi.org/10.1007/JHEP09(2020)179}{{\em JHEP} {\bfseries 09}
  (2020) 179}, \href{http://arxiv.org/abs/1912.07189}{{\ttfamily
  arXiv:1912.07189 [hep-ph]}}.

\bibitem{Espinosa:2010hh}
J.~R. Espinosa, T.~Konstandin, J.~M. No, and G.~Servant, ``{Energy Budget of
  Cosmological First-order Phase Transitions},''
  \href{http://dx.doi.org/10.1088/1475-7516/2010/06/028}{{\em JCAP} {\bfseries
  1006} (2010) 028},
\href{http://arxiv.org/abs/1004.4187}{{\ttfamily arXiv:1004.4187 [hep-ph]}}.

\bibitem{Cutting:2018tjt}
D.~Cutting, M.~Hindmarsh, and D.~J. Weir, ``{Gravitational waves from vacuum
  first-order phase transitions: from the envelope to the lattice},''
  \href{http://dx.doi.org/10.1103/PhysRevD.97.123513}{{\em Phys. Rev. D}
  {\bfseries 97} no.~12, (2018) 123513},
  \href{http://arxiv.org/abs/1802.05712}{{\ttfamily arXiv:1802.05712
  [astro-ph.CO]}}.

\bibitem{Guth:1981uk}
A.~H. Guth and E.~J. Weinberg, ``{Cosmological Consequences of a First Order
  Phase Transition in the SU(5) Grand Unified Model},''
  \href{http://dx.doi.org/10.1103/PhysRevD.23.876}{{\em Phys. Rev. D}
  {\bfseries 23} (1981) 876}.

\bibitem{Scherrer:1984fd}
R.~J. Scherrer and M.~S. Turner, ``{Decaying Particles Do Not Heat Up the
  Universe},'' \href{http://dx.doi.org/10.1103/PhysRevD.31.681}{{\em Phys. Rev.
  D} {\bfseries 31} (1985) 681}.

\bibitem{OConnell:2006rsp}
D.~O'Connell, M.~J. Ramsey-Musolf, and M.~B. Wise, ``{Minimal Extension of the
  Standard Model Scalar Sector},''
  \href{http://dx.doi.org/10.1103/PhysRevD.75.037701}{{\em Phys. Rev. D}
  {\bfseries 75} (2007) 037701},
  \href{http://arxiv.org/abs/hep-ph/0611014}{{\ttfamily arXiv:hep-ph/0611014}}.

\bibitem{Elias_Mir__2014}
J.~Elias-Miró, J.~R. Espinosa, and T.~Konstandin, ``Taming infrared
  divergences in the effective potential,''
  \href{http://dx.doi.org/10.1007/jhep08(2014)034}{{\em Journal of High Energy
  Physics} {\bfseries 2014} no.~8, (Aug, 2014) }.
  \url{http://dx.doi.org/10.1007/JHEP08(2014)034}.

\bibitem{Barger:2007im}
V.~Barger, P.~Langacker, M.~McCaskey, M.~J. Ramsey-Musolf, and G.~Shaughnessy,
  ``{LHC Phenomenology of an Extended Standard Model with a Real Scalar
  Singlet},'' \href{http://dx.doi.org/10.1103/PhysRevD.77.035005}{{\em Phys.
  Rev. D} {\bfseries 77} (2008) 035005},
  \href{http://arxiv.org/abs/0706.4311}{{\ttfamily arXiv:0706.4311 [hep-ph]}}.

\bibitem{Profumo:2007wc}
S.~Profumo, M.~J. Ramsey-Musolf, and G.~Shaughnessy, ``{Singlet Higgs
  phenomenology and the electroweak phase transition},''
  \href{http://dx.doi.org/10.1088/1126-6708/2007/08/010}{{\em JHEP} {\bfseries
  08} (2007) 010}, \href{http://arxiv.org/abs/0705.2425}{{\ttfamily
  arXiv:0705.2425 [hep-ph]}}.

\bibitem{Alves:2019igs}
A.~Alves, D.~Gon\c{c}alves, T.~Ghosh, H.-K. Guo, and K.~Sinha, ``{Di-Higgs
  Production in the $4b$ Channel and Gravitational Wave Complementarity},''
  \href{http://dx.doi.org/10.1007/JHEP03(2020)053}{{\em JHEP} {\bfseries 03}
  (2020) 053}, \href{http://arxiv.org/abs/1909.05268}{{\ttfamily
  arXiv:1909.05268 [hep-ph]}}.

\bibitem{Alves:2018oct}
A.~Alves, T.~Ghosh, H.-K. Guo, and K.~Sinha, ``{Resonant Di-Higgs Production at
  Gravitational Wave Benchmarks: A Collider Study using Machine Learning},''
  \href{http://dx.doi.org/10.1007/JHEP12(2018)070}{{\em JHEP} {\bfseries 12}
  (2018) 070}, \href{http://arxiv.org/abs/1808.08974}{{\ttfamily
  arXiv:1808.08974 [hep-ph]}}.

\bibitem{Liu:2021jyc}
W.~Liu and K.-P. Xie, ``{Probing electroweak phase transition with multi-TeV
  muon colliders and gravitational waves},''
  \href{http://arxiv.org/abs/2101.10469}{{\ttfamily arXiv:2101.10469
  [hep-ph]}}.

\bibitem{Chiang:2018gsn}
C.-W. Chiang, Y.-T. Li, and E.~Senaha, ``{Revisiting electroweak phase
  transition in the standard model with a real singlet scalar},''
  \href{http://dx.doi.org/10.1016/j.physletb.2018.12.017}{{\em Phys. Lett. B}
  {\bfseries 789} (2019) 154--159},
  \href{http://arxiv.org/abs/1808.01098}{{\ttfamily arXiv:1808.01098
  [hep-ph]}}.

\bibitem{Fuyuto:2014yia}
K.~Fuyuto and E.~Senaha, ``{Improved sphaleron decoupling condition and the
  Higgs coupling constants in the real singlet-extended standard model},''
  \href{http://dx.doi.org/10.1103/PhysRevD.90.015015}{{\em Phys. Rev. D}
  {\bfseries 90} no.~1, (2014) 015015},
  \href{http://arxiv.org/abs/1406.0433}{{\ttfamily arXiv:1406.0433 [hep-ph]}}.

\bibitem{Athron:2019nbd}
P.~Athron, C.~Bal\'azs, M.~Bardsley, A.~Fowlie, D.~Harries, and G.~White,
  ``{BubbleProfiler: finding the field profile and action for cosmological
  phase transitions},'' \href{http://dx.doi.org/10.1016/j.cpc.2019.05.017}{{\em
  Comput. Phys. Commun.} {\bfseries 244} (2019) 448--468},
  \href{http://arxiv.org/abs/1901.03714}{{\ttfamily arXiv:1901.03714
  [hep-ph]}}.

\bibitem{Wang:2020nzm}
X.~Wang, F.~P. Huang, and X.~Zhang, ``{The energy budget and the gravitational
  wave spectra beyond the bag model},''
  \href{http://arxiv.org/abs/2010.13770}{{\ttfamily arXiv:2010.13770
  [astro-ph.CO]}}.

\bibitem{Dine:1992wr}
M.~Dine, R.~G. Leigh, P.~Y. Huet, A.~D. Linde, and D.~A. Linde, ``{Towards the
  theory of the electroweak phase transition},''
  \href{http://dx.doi.org/10.1103/PhysRevD.46.550}{{\em Phys.\ Rev.\ D}
  {\bfseries 46} (1992) 550--571},
  \href{http://arxiv.org/abs/hep-ph/9203203}{{\ttfamily arXiv:hep-ph/9203203}}.

\end{thebibliography}
\providecommand{\href}[2]{#2}\begingroup\raggedright\endgroup

\end{document}